\documentclass[twoside,headsepline,11pt,DIV=14,BCOR=10mm,numbers=noenddot]{scrbook}

\usepackage{savesym}
\usepackage[english]{babel}
\usepackage[style=english]{csquotes}
\usepackage[style=numeric, backref=true, urldate=iso,backend=biber]{biblatex}
\usepackage{hyperref}
\usepackage{todonotes}
\usepackage{booktabs}
\usepackage{scrhack}
\usepackage{makeidx}
\makeindex
\usepackage[totoc]{idxlayout}
\usepackage[symbols,nogroupskip,sort=none]{glossaries-extra}
\usepackage[T1]{fontenc}
\usepackage[
    type={CC},
    modifier={by-nc-sa},
    version={4.0},
]{doclicense}
\usepackage{xpatch}
\usepackage{hyperxmp}

\usepackage{amsmath}
\savesymbol{iint}
\usepackage{amsthm}
\usepackage{mathtools}
\usepackage{thmtools}

\usepackage{textcomp}
\usepackage{xspace}
\usepackage{newpxmath}
\restoresymbol{AMS}{iint}
\usepackage{fontawesome}
\usepackage{tcolorbox}
\usepackage{scrlayer-scrpage}
\usepackage[onehalfspacing]{setspace}
\usepackage{epigraph}
\usepackage{enumitem}
\usepackage{diagbox}
\usepackage[largesc]{newpxtext}
\usepackage{pifont}

\usepackage{prftree}
\usepackage{cmll}

\usepackage{algorithm}
\usepackage{algpseudocode}

\usepackage{tikz}
\usetikzlibrary{shapes.multipart, positioning, cd, automata, arrows, shapes.geometric,decorations.markings,petri}
\usepackage{transparent}
\usepackage{caption}
\usepackage{graphicx}

\newcommand{\N}{\ensuremath{\mathbb{N}}}

\makeindex
\newcommand{\lindex}[1]{%
  \lowercase{\def\temp{#1}}%
  \expandafter\index{\temp|textit}%
}
\newcommand{\reflindex}[1]{%
  \lowercase{\def\temp{#1}}%
  \expandafter\index{\temp}%
}
\newcommand{\define}[2][]{\ifthenelse{\equal{#1}{}}{\emph{#2}\lindex{#2}}{\emph{#2}\index{#1|textit}}}

\newcommand{\reftodef}[2][]{\ifthenelse{\equal{#1}{}}{#2\reflindex{#2}}{#2\index{#1}}}

\definecolor{maincolor}{rgb}{0,0.31,0.62}
\colorlet{maincolorDark}{maincolor!80!black}
\colorlet{maincolorMedium}{maincolor!50!white}
\colorlet{maincolorLight}{maincolor!20!white}
\colorlet{maincolorExtraLight}{maincolor!10!white}
\colorlet{maincolorExtraExtraLight}{maincolor!5!white}
\definecolor{cec1d24}{RGB}{236,29,36}
\definecolor{cffffff}{RGB}{255,255,255}

\hypersetup{
    colorlinks,
    citecolor=maincolor,
    filecolor=black,
    linkcolor=maincolor,
    urlcolor=red!60!black
}

\definecolor{typecolor}{HTML}{0101FD}
\definecolor{termcolor}{HTML}{007D9A}
\definecolor{contextcolor}{HTML}{888888}
\definecolor{universecolor}{HTML}{A139B7}

\newcommand{\fragment}[1]{\ensuremath{\mathsf{#1}}}
\newcommand{\fragmenttxt}[1]{\textsf{#1}}
\renewcommand{\phi}{\varphi}
\renewcommand{\epsilon}{\varepsilon}

\DeclareMathOperator{\petritrans}{\tikz{\draw[opacity=0](0,-0.1)--(0,0.1);\draw [-stealth, thick](0,0) -- (0.5,0);\draw[draw=black, fill=black] (0.15,-0.05) rectangle ++(0.1,0.1);}}
\DeclareMathOperator{\linking}{\tikz{\draw[opacity=0](0,-0.1)--(0,0.1);\draw [very thick](0,0) -- (0.5,0);}}

\DeclareMathOperator{\bigO}{\mathcal{O}}
\newcommand{\class}[1]{\ensuremath{\mathsf{#1}}\xspace}

\newcommand{\NP}{\class{NP}}
\newcommand{\PSPACE}{\class{PSPACE}}

\newcommand{\problem}[1]{\ensuremath{\mathsf{#1}}\xspace}

\newcommand{\Cat}[1]{{\ensuremath{\mathcal{#1}}}}

\newcommand{\bnfsep}{\ |\ }

\useosf

\let\originalepigraph\epigraph 
\renewcommand\epigraph[2]{\originalepigraph{\textit{#1}}{#2}}

\addto\captionsenglish{
  \renewcommand{\contentsname}%
    {Contents\strut}%
}

\let\olditemize=\itemize \let\endolditemize=\enditemize \renewenvironment{itemize}{\olditemize \itemsep0em}{\endolditemize}

\DeclareMathAlphabet{\mathcal}{OMS}{cmsy}{m}{n}

\setkomafont{disposition}{\normalfont}

\setkomafont{chapter}{\Huge\textbf}

\addtokomafont{part}{\thispagestyle{empty}\fontsize{40}{48} \selectfont \textsc}
\addtokomafont{partnumber}{\Huge\textsc}

\makeatletter

\setkomafont{section}{\LARGE\bfseries}
\setkomafont{subsection}{\Large\bfseries}
\setkomafont{subsubsection}{\large\bfseries}
\makeatother

\theoremstyle{remark}
\newtheorem*{remark}{Remark}

\declaretheoremstyle[
  headfont=\normalfont\bfseries,
  numbered=yes,
  bodyfont=\itshape,
  spaceabove=3pt,
  spacebelow=3pt,
  notefont=\bfseries,
]{thmtoolsThmStyle}

\declaretheoremstyle[
  headfont=\normalfont\itshape,
  numbered=yes,
  bodyfont=\normalfont,
  spaceabove=3pt,
  spacebelow=3pt,
  notefont=\itshape,
]{thmtoolsExStyle}

\declaretheoremstyle[
  headfont=\normalfont\bfseries,
  numbered=yes,
  bodyfont=\normalfont,
  spaceabove=3pt,
  spacebelow=3pt,
  qed={$\lrcorner$},
  notefont=\bfseries,
]{thmtoolsDefStyle}

\declaretheoremstyle[
  headfont=\large\normalfont\bfseries,
  numbered=yes,
  bodyfont=\normalfont,
  spaceabove=3pt,
  spacebelow=3pt,
  mdframed={
    skipabove=10pt,
    skipbelow=10pt,
    hidealllines=true,
    backgroundcolor={maincolorExtraExtraLight},
    innerleftmargin=8pt,
    innerrightmargin=8pt},
  postheadhook = {\hspace*{\parindent}},
]{thmtoolsProbStyle}

\declaretheorem[
  style=thmtoolsThmStyle,
  title=Theorem,
  refname={theorem,theorems},
  Refname={Theorem,Theorems},
  numberwithin = chapter,
]{theorem}
\declaretheorem[
  style=thmtoolsThmStyle,
  title=Lemma,
  refname={lemma,lemmas},
  Refname={Lemma,Lemmas},
  sibling=theorem
]{lemma}
\declaretheorem[
  style=thmtoolsThmStyle,
  title=Proposition,
  refname={proposition,propositions},
  Refname={Proposition,Propositions},
  sibling=theorem
]{proposition}
\declaretheorem[
  style=thmtoolsThmStyle,
  title=Corollary,
  refname={corollary,corollaries},
  Refname={Corollary,Corollaries},
  sibling=theorem
]{corollary}

\declaretheorem[
  style=thmtoolsExStyle,
  title=Example,
  refname={example,examples},
  Refname={Example,Examples},
  sibling=theorem
]{example}

\declaretheorem[
  style=thmtoolsDefStyle,
  title=Definition,
  refname={definition,definitions},
  Refname={Definition,Definitions},
  sibling=theorem
]{definition}

\declaretheorem[
  style=thmtoolsProbStyle,
  title=Problem,
  refname={problem,problems},
  Refname={Problem,Problems},
  sibling=theorem,
]{defproblem}

\tcbuselibrary{theorems, skins, breakable}

\newtcbtheorem[number within=chapter]{fdefinition}{Definition}{
  enhanced,
  colback=maincolorExtraExtraLight,
  coltitle=black,
  boxrule=0pt,
  fonttitle=\large\bfseries,
  description delimiters parenthesis,
  sharpish corners,
  attach boxed title to top,
  boxed title style={colframe=maincolorMedium, colback=maincolorLight, sharpish corners, boxrule=0mm},
  toptitle=0.7mm,
  enlarge top by=0cm,
  bottomtitle=0.5mm,
  breakable}{def}

\newtcbtheorem[number within=chapter, use counter from=fdefinition]{ftheorem}{Theorem}{
  enhanced,
  colback=maincolorExtraExtraLight,
  coltitle=black,
  boxrule=0pt,
  fonttitle=\large\bfseries,
  description delimiters parenthesis,
  sharpish corners,
  attach boxed title to top,
  boxed title style={colframe=maincolorMedium, colback=maincolorLight, sharpish corners, boxrule=0mm},
  toptitle=0.7mm,
  enlarge top by=0cm,
  bottomtitle=0.5mm,
  breakable}{thm}

\newtcbtheorem[number within=chapter, use counter from=fdefinition]{fexample}{Example}{
  enhanced,
  coltitle=maincolor,
  colback=maincolorExtraExtraLight,
  colframe=maincolorExtraLight,
  fonttitle=\large\bfseries,
  description delimiters parenthesis,
  sharpish corners,
  detach title,
  before upper={\tcbtitle\par},
  enlarge top by=0.4cm,
  breakable}{example}

  \newtcbtheorem[number within=chapter]{fproblem}{Problem}{
    enhanced,
    coltitle=maincolor,
    colback=maincolorExtraExtraLight,
    colframe=maincolorExtraLight,
    fonttitle=\large,
    description delimiters parenthesis,
    sharpish corners,
    detach title,
    before upper={\tcbtitle\par},
    enlarge top by=0.4cm,
    breakable}{problem}

  \newtcbtheorem[number within=chapter, use counter from=fdefinition]{fproposition}{Proposition}{
  enhanced,
  coltitle=maincolor,
  colback=maincolorExtraExtraLight,
  colframe=maincolorExtraExtraLight,
  description delimiters parenthesis,
  fonttitle=\bfseries,
  sharpish corners,
  detach title,
  before upper={\tcbtitle\quad},
  enlarge top by=0.4cm,
  breakable}{proposition}

  \newtcbtheorem[number within=chapter, use counter from=fdefinition]{flemma}{Lemma}{
    enhanced,
    coltitle=maincolor,
    colback=maincolorExtraExtraLight,
    description delimiters parenthesis,
    fonttitle=\bfseries,
    sharpish corners,
    colframe=maincolorExtraExtraLight,
    detach title,
    before upper={\tcbtitle\quad},
    enlarge top by=0.4cm,
    breakable}{lemma}

  \newtcbtheorem[number within=chapter, use counter from=fdefinition]{fcorollary}{Corollary}{
    enhanced,
    coltitle=maincolor,
    colback=maincolorExtraExtraLight,
    description delimiters parenthesis,
    fonttitle=\bfseries,
    sharpish corners,
    colframe=maincolorExtraExtraLight,
    detach title,
    before upper={\tcbtitle\quad},
    enlarge top by=0.4cm,
    breakable}{corollary}

  \newtcbtheorem[number within=chapter]{customNumberDefinition}{Definition}{
  enhanced,
  colback=maincolorExtraExtraLight,
  coltitle=black,
  colframe=maincolorMedium,
  fonttitle=\large\bfseries,
  separator sign none,
  description delimiters none,
  sharpish corners,
  attach boxed title to top,
  boxed title style={colframe=maincolorMedium, colback=maincolorLight, sharpish corners, boxrule=0.5mm},
  toptitle=0.7mm,
  enlarge top by=0.4cm,
  bottomtitle=0.7mm,
  breakable}{def}

\newtcbtheorem[number within=chapter, use counter from=fdefinition]{customNumberTheorem}{Theorem}{
  enhanced,
  colback=maincolorExtraExtraLight,
  coltitle=black,
  colframe=maincolorMedium,
  fonttitle=\large\bfseries,
  separator sign none,
  description delimiters none,
  sharpish corners,
  attach boxed title to top,
  boxed title style={colframe=maincolorMedium, colback=maincolorLight, sharpish corners, boxrule=0.5mm},
  toptitle=0.7mm,
  enlarge top by=0.4cm,
  bottomtitle=0.7mm,
  breakable}{thm}

\newtcbtheorem[number within=chapter, use counter from=fdefinition]{customNumberExample}{Example}{
  enhanced,
  coltitle=maincolor,
  colback=maincolorExtraExtraLight,
  colframe=maincolorExtraLight,
  fonttitle=\large\bfseries,
  description delimiters parenthesis,
  sharpish corners,
  detach title,
  before upper={\tcbtitle\par},
  enlarge top by=0.4cm,
  breakable}{example}

\newtcbtheorem[number within=chapter, use counter from=fdefinition]{customNumberProposition}{Proposition}{
  enhanced,
  coltitle=maincolor,
  colback=maincolorExtraExtraLight,
  fonttitle=\bfseries,
  sharpish corners,
  colframe=maincolorExtraExtraLight,
  detach title,
  before upper={\tcbtitle\quad},
  enlarge top by=0.4cm,
  breakable}{proposition}

\clearpairofpagestyles
\setkomafont{pageheadfoot}{\normalfont}

\ihead{\textsc{\headmark}}
\automark[section]{chapter}
\ohead{\pagemark}

\newcommand\Accept{\textbf{accept}}
\newcommand\Reject{\textbf{reject}}

\algnewcommand\algorithmicOutput{\textbf{Output:}}
\algnewcommand\Output{\item[\algorithmicOutput]}
\algnewcommand\algorithmicInput{\textbf{Input:}}
\algnewcommand\Input{\item[\algorithmicInput]}

\DeclareMathOperator{\incr}{\ensuremath{\uparrow}}
\DeclareMathOperator{\decr}{\ensuremath{\downarrow}}
\DeclareMathOperator{\val}{\ensuremath{\text{val}}}
\newcommand{\frownsmile}{%
  \mathrel{%
    \vcenter{\offinterlineskip
      \ialign{##\cr$\frown$\cr\noalign{\kern-0.5pt}$\smile$\cr}%
    }%
  }%
}
\newcommand{\smilefrown}{%
  \mathrel{%
    \vcenter{\offinterlineskip
      \ialign{##\cr$\smile$\cr\noalign{\kern0pt}$\frown$\cr}%
    }%
  }%
}
\newcommand{\strictfrown}{%
  \mathrel{%
    \vcenter{\offinterlineskip
      \ialign{##\cr$\frown$\cr\noalign{\kern0pt}\phantom{$\frown$}\cr}%
    }%
  }%
}
\newcommand{\strictsmile}{%
  \mathrel{%
    \vcenter{\offinterlineskip
      \ialign{##\cr$\smile$\cr\noalign{\kern0pt}\phantom{$\frown$}\cr}%
    }%
  }%
}
\glsxtrnewsymbol[description={classical or intuitionistic logic formulas}]{clform}{\ensuremath{\phi, \psi, \dots}}
\glsxtrnewsymbol[description={linear logic formulas}]{llform}{\ensuremath{A, B, C, \dots}}
\glsxtrnewsymbol[description={sequents}]{sequents}{\ensuremath{\Gamma, \Delta, \dots}}
\glsxtrnewsymbol[description={vector of variables}]{vecvar}{\ensuremath{\overline{x}}}
\glsxtrnewsymbol[description={implies}]{implies}{\ensuremath{\implies}}
\glsxtrnewsymbol[description={follows from}]{follows}{\ensuremath{\impliedby}}
\glsxtrnewsymbol[description={constant symbols (zero-ary connectives) of linear logic}]{constant}{\ensuremath{0, 1, \bot, \top}}
\glsxtrnewsymbol[description={unary sentential connective of negation}]{senneg}{\ensuremath{\neg}}
\glsxtrnewsymbol[description={unary linear logic connective of negation}]{linneg}{\ensuremath{\cdot^\bot}}
\glsxtrnewsymbol[description={unary linear logic connectives of exponentiation}]{linexp}{\ensuremath{!, ?}}
\glsxtrnewsymbol[description={binary sentential connectives}]{senbin}{\ensuremath{\wedge, \vee, \rightarrow}}
\glsxtrnewsymbol[description={binary linear logic connectives}]{linbin}{\ensuremath{\oplus, \with, \otimes, \parr, \multimap, \multimapdot}}
\glsxtrnewsymbol[description={equality by definition}]{eq}{\ensuremath{\coloneq}}
\glsxtrnewsymbol[description={semantic equivalence}]{semeq}{\ensuremath{\equiv}}
\glsxtrnewsymbol[description={an isomorphism}]{iso}{\ensuremath{\simeq}}
\glsxtrnewsymbol[description={congruence relation induced by $\lesssim$}]{congrrel}{\ensuremath{\sim}}
\glsxtrnewsymbol[description={syntactical consequence}]{syncon}{\ensuremath{\vdash}}
\glsxtrnewsymbol[description={sematical consequence}]{semcon}{\ensuremath{\models}}
\glsxtrnewsymbol[description={the empty set}]{empty}{\ensuremath{\emptyset}}
\glsxtrnewsymbol[description={inclusion}]{incl}{\ensuremath{\subseteq}}
\glsxtrnewsymbol[description={proper inclusion}]{pincl}{\ensuremath{\subsetneq}}
\glsxtrnewsymbol[description={disjoint union}]{disjointu}{\ensuremath{\uplus}}
\glsxtrnewsymbol[description={order relation on the natural numbers}]{ordernat}{\ensuremath{\leq}}
\glsxtrnewsymbol[description={order relation on a lattice}]{orderlatt}{\ensuremath{\preccurlyeq}}
\glsxtrnewsymbol[description={preorder relation on the exponential modalities}]{orderexp}{\ensuremath{\lesssim}}
\glsxtrnewsymbol[description={placeholder in category theory}]{placeholder}{\ensuremath{-}}
\glsxtrnewsymbol[description={generalized Routley-Meyer star operator}]{starop}{\ensuremath{\cdot^\star}}

\glsxtrnewsymbol[description={the class of elementary funcitons}]{elemfunc}{\ensuremath{\mathscr{F}}}

\glsxtrnewsymbol[description={assignments}]{assign}{\ensuremath{\mathfrak{I}, \mathfrak{K}}}

\glsxtrnewsymbol[description={linear Kripke frame}]{kripke}{\ensuremath{\mathcal{K}}}

\glsxtrnewsymbol[description={a logic}]{alogic}{\ensuremath{\mathcal{L}}}

\glsxtrnewsymbol[description={the natural numbers (with 0)}]{nat}{\ensuremath{\mathbb{N}}}

\glsxtrnewsymbol[description={the powerset of a set $A$}]{power}{\ensuremath{\mathcal{P}(A)}}

\glsxtrnewsymbol[description={logical theory}]{logtheo}{\ensuremath{\mathbb{T}}}

\glsxtrnewsymbol[description={the integers}]{integ}{\ensuremath{\mathbb{Z}}}
\glsxtrnewsymbol[description={the positive integers}]{integ+}{\ensuremath{\mathbb{Z}_+}}

\title{Computational~Complexity of~Deciding~Provability~in Linear~Logic~and~Its~Fragments}
\author{Florian Chudigiewitsch}
\date{\oldstylenums{14}. September \oldstylenums{2021}}
\publishers{Institut für Theoretische Informatik\\[1ex] Gottfried Wilhelm Leibniz Universität Hannover}

\begin{document}

\makeatletter

\begin{titlepage}
    \begin{center}
        \ \\[-9ex]
        \includegraphics[height=10ex]{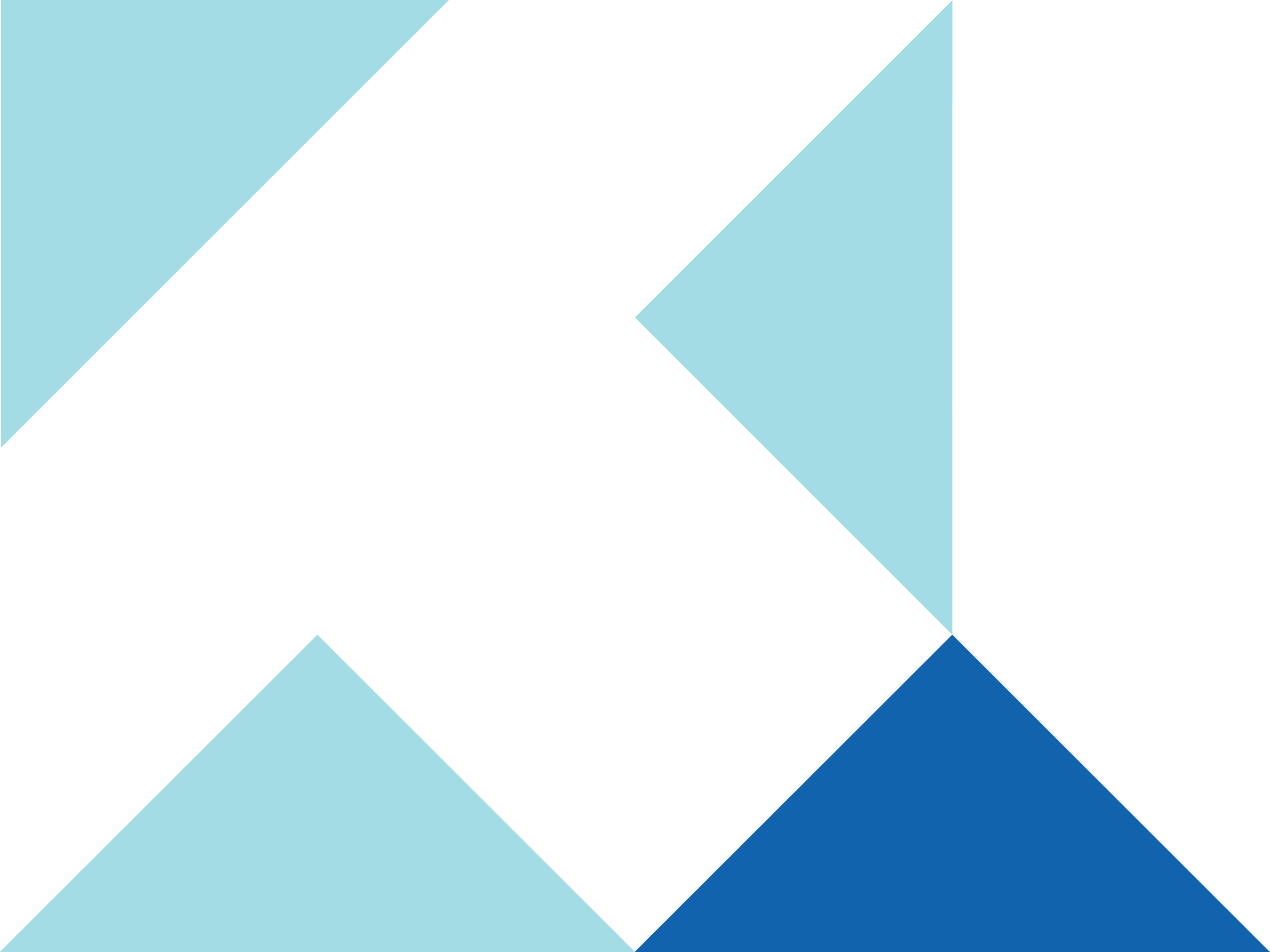}\hfill\includegraphics[height=10ex]{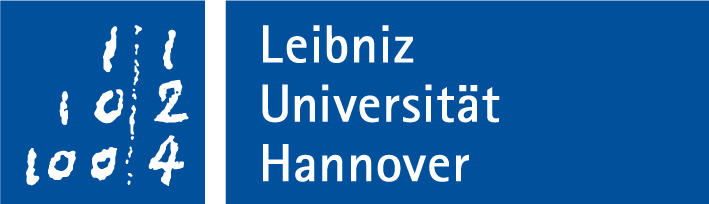}\\[7ex]
        {\fontsize{32}{30} \bfseries \@title \par}~\\[7ex]
        {\Large \scshape Masterarbeit}\\[2ex]
        im Studiengang Informatik\\[5ex]
        eingereicht von\\[2ex]
        {\Large  \@author}\\[2ex] 
        \@date\\[7ex]
        \begin{tabular}{l l}
            \textbf{Erstprüfer:} & PD Dr.\,rer.\,nat.\,habil. Arne Meier\\
            \textbf{Zweitprüfer:} & Prof.\,Dr.\,rer.\,nat. Heribert Vollmer\\
            \textbf{Betreuer:} & PD Dr.\,rer.\,nat.\,habil. Arne Meier\\
            \textbf{Matrikelnummer:} & \oldstylenums{3216390}\\
        \end{tabular}\\[5ex]
        {\large  \@publishers}\\[6ex]
    \end{center}
    \makebox[0pt][l]{%
    \raisebox{-\totalheight}[0pt][0pt]{%
        {\transparent{0.5}\includegraphics[width=\textwidth]{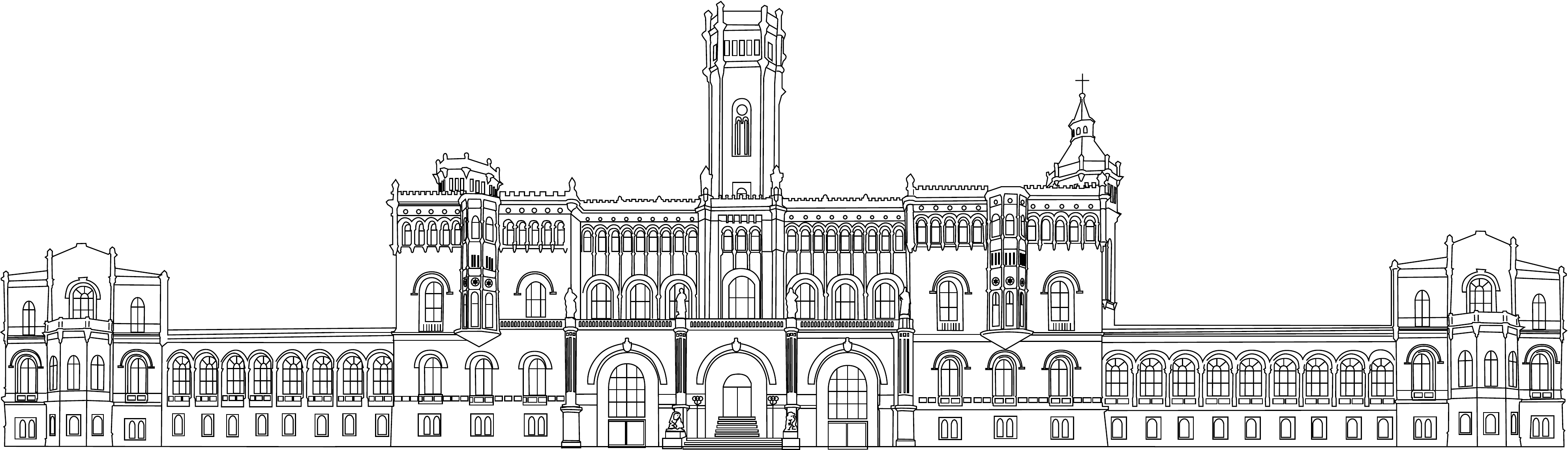}}}}%
\end{titlepage}

\thispagestyle{empty}
\vspace*{\fill}
\doclicenseThis

\makeatother

\frontmatter


\cleardoublepage
\thispagestyle{empty}
\phantomsection\addcontentsline{toc}{chapter}{Abstract}
\begin{center}
    {\huge \textbf{Abstract}}
\end{center}
Linear logic was conceived in 1987 by Girard and, in contrast to classical logic, restricts the usage of the structural inference rules of weakening and contraction. With this, atoms of the logic are no longer interpreted as truth, but as information or resources. This interpretation makes linear logic a useful tool for formalisation in mathematics and computer science. Linear logic has, for example, found applications in proof theory, quantum logic, and the theory of programming languages. A central problem of the logic is the question whether a given list of formulas is provable with the calculus. In the research regarding the complexity of this problem, some results were achieved, but other questions are still open. To present these questions and give new perspectives, this thesis consists of three main parts which build on each other:
\begin{itemize}
    \item We present the syntax, proof theory, and various approaches to a semantics for linear logic. Here already, we will meet some open research questions.
    \item We present the current state of the complexity-theoretic characterization of the most important fragments of linear logic. Here, further research problems are presented and it becomes apparent that until now, the results have all made use of different approaches.
    \item We prove an original complexity characterization of a fragment of the logic and present ideas for a new, structural approach to the examination of provability in linear logic.
\end{itemize}

\cleardoublepage
\thispagestyle{empty}
\phantomsection\addcontentsline{toc}{chapter}{Kurzfassung}
\begin{center}
    {\huge \textbf{Kurzfassung}}
\end{center}
Die Lineare Logik wurde erstmals 1987 von Girard definiert und schränkt im Vergleich zur klassischen Logik die Benutzbarkeit der strukturellen Inferenzregeln der Abschwächung und der Kontraktion ein. Durch diese Einschränkung werden Atome der Logik nicht mehr als Wahrheitswert, sondern als Information oder Ressource interpretiert. Diese Interpretation macht die Logik zu einem nützlichen Formalisierungswerkzeug der Mathematik und Informatik. Unter anderem hat die Lineare Logik in der Beweistheorie, Quantenlogik und Theorie von Programmiersprachen Anwendung gefunden. Ein zentrales Problem der Logik ist die Frage, ob eine gegebene Liste von Formeln im Kalkül der Logik beweisbar ist. Die Untersuchung der Komplexität dieses Problems hat zwar schon einige Resultate hervorgebracht, allerdings sind immer noch viele Fragen offen. Um diese Fragen erläutern und neue Perspektiven geben zu können, befasst sich diese Arbeit mit drei aufeinander aufbauenden Hauptthemen:
\begin{itemize}
    \item Wir stellen die Syntax, Beweistheorie und verschiedene Ansätze für eine Semantik der Linearen Logik vor. Schon hier treffen wir auf erste ungelöste Forschungsprobleme.
    \item Wir stellen den aktuellen Stand der komplexitätstheoretischen Charakterisierung der wichtigsten Fragmente der Linearen Logik vor. Hier werden weitere offene Forschungsfragen vorgestellt und es wird deutlich, dass die bisher erzielten Resultate alle auf unterschiedli\-chen Ansätzen fußen.
    \item Wir beweisen eine neue Komplexitätscharakterisierung eines Fragmentes der Logik und geben Ideen für einen neuen, strukturelleren Ansatz für die Betrachtung des Beweisbarkeitsproblems in der Linearen Logik.
\end{itemize}

\tableofcontents


\mainmatter

\chapter{Introduction\strut}

\epigraph{A mathematical proof is evidence that can be checked computationally. And thus was born computer science.}{\textsc{Moshe Vardi}}

If we want to express the sentence ``a mathematician is a device that turns coffee into theorems'' in the language of said mathematician -- that is, mathematical logic -- one way we would probably think of is classical propositional sequents:
\[\text{mathematician}\colon \text{coffee} \vdash \text{theorems}.\]
But in classical logic, the following derivation is possible:

\begin{displaymath}
    \prftree[r]{(contraction)}
        {\prftree[r]{(weakening)}
        {\text{coffee} \vdash \text{theorems}}
        {\text{coffee}, \text{coffee} \vdash \text{coffee} \wedge \text{theorems}}}
        {\text{coffee} \vdash \text{coffee} \wedge \text{theorems}}
\end{displaymath}

Where does the extra coffee come from? It becomes apparent that this expression does not capture how much coffee the mathematician needs and how many theorems they can produce from it\footnote{Anecdotal evidence from writing this thesis suggests that the ratio is quite high.}. This is because we use classical logic to reason about \emph{truth}, and truth can be, once established, used as many times as we like. But in this case, we want to reason about \emph{resources}, which are consumed when they are transformed into something else.

This is where linear logic enters the picture: By restricting the use of the weakening and contraction rules of classical logic, the formulas used in a proof can neither be copied arbitrarily nor be dropped. Because we restrict certain structural rules of classical logic, linear logic falls under the umbrella of substructural logics. Linear logic was conceived by~\textcite{Gir_87} when he studied \reftodef[coherence space]{coherence spaces}, structures that are used in domain theory, and extensions of the $\lambda$-calculus. In turn, coherence spaces now provide one of the many different approaches to give a semantics for linear logic.

The resource interpretation makes linear logic a prime candidate for many applications in computer science: Linear type systems in programming languages are used to track shared resources, and in quantum computing we can make use of the fact that formulas may not be copied arbitrarily to formalize the no-cloning theorem. When viewing these resources as parallel acting agents with or without interaction, we can use linear logic for the verification of cryptographic protocols, and parallel computation. We will see at the end of Chapter~\ref{ch:linlog} how this is done exactly after we described how linear logic works in detail.

When we use logics in a practical setting, a problem that often arises is the following: Given a list of formulas, is this list derivable through the use of the inference rules from the calculus of the logic? For example, the formula may express a specification, and we want to prove that this specification does not lead to any inconsistencies. It is a natural question to ask for the complexity of this problem. We will see that compared to classical logic, the provability problem becomes much harder, so it makes sense to also examine fragments of the logic. The complexity-theoretic lens can also reveal structure in the problems, which can in the best case be leveraged in the construction of more efficient algorithms.

\subsection*{Structure of the Thesis}

In Chapter~\ref{ch:prelim}, we will review basic notions in the fields of complexity theory, logic and proof theory, and category theory, which form the cornerstone for the theoretical examination of linear logic. In Chapter~\ref{ch:linlog}, we will describe linear logic and its fragments, with their syntax, proof theory, and semantics. In Chapter~\ref{ch:classic-comp}, we present the current state of the art regarding the complexity of deciding provability in linear logic and its fragments. In Chapter~\ref{ch:new}, we establish new complexity results and provide ideas for a new approach to the examination of the complexity of linear logic. We conclude the thesis in Chapter~\ref{ch:conclude} with the discussion of the findings and an outlook of the further research directions in this area.

\chapter{Preliminaries\strut }\label{ch:prelim}

In this chapter, we will establish the notation used throughout the thesis. We will also point out which familiarities are assumed in an informal manner and refer to standard literature for each topic. We will furthermore review advanced notions more formally.

\section{Mathematical Logic}

Since this thesis examines the complexity properties of logical decision problems, we will assume that the reader has some background in logic, such as provided in the introductory books by~\textcite{Shoenfield1967-SHOML-2} or~\textcite{Hinman2007-HINFOM}. Intuitionistic logic and constructivism are presented in detail in the book by~\textcite{Beeson85}. We will review some key parts of classical and intuitionistic logic, proof theory, and the finite model property from finite model theory.

\subsection{Classical Logic}

The term ``\reftodef{classical logic}'' refers to the class of logics which is arguably studied and used to the greatest extend in mathematics and computer science. This class forms the cornerstone of modern mathematical logic and is treated in almost every introductory book on logic, such as the two books named above. Because of that, we will not give a formal introduction to classical logic, but rather informally remark some properties which are interesting for our treatment of linear logic.

\begin{description}
    \item[Law of excluded middle] The \reftodef{law of excluded middle} (LEM) states that for a formula $\varphi$, either $\varphi$ or its negation $\neg\varphi$ holds. It is semantically equivalent to double negation elimination $\neg\neg\varphi\rightarrow\varphi$. Both properties make the logics who possess them non-constructive, meaning that one can prove existential propositions without constructing a witness explicitly. The implications of this are discussed further in Section~\ref{intuitionisticlog}.
    \item[De Morgan duality] The \reftodef[de Morgan duality!classical]{de Morgan duality} of classical logic permits us to express conjunctions and disjunctions purely in terms of each other via negation. In propositional logic, these dualities are denoted as
    \[\neg(\phi \vee \psi) \iff (\neg\phi)\wedge(\neg\psi)\text{\ \ \ \ \ \ and\ \ \ \ \ \ }\neg(\phi\wedge\psi) \iff (\neg\phi) \vee (\neg\psi).\] 
\end{description}

The two classical logics which are of utmost importance are \reftodef{propositional logic} and \reftodef{first-order logic}. We distinguish between two aspects of a logic: its \reftodef{syntax}, which determines the well-formed expressions of our language, and its \reftodef{semantics}, which defines the behavior of the logic.

Syntactically, we conceive a logic to be defined over a formal language with an alphabet that provides logical and non-logical symbols. In the case of propositional logic, the logical symbols typically include the logical connectives $\{\,\wedge, \neg\,\}$, from which all other connectives can be derived, parentheses, and an infinite set of variables. First-order logic extends these symbols by the \reftodef[quantifier]{quantifiers} $\forall$ and $\exists$. The non-logical symbols consist of predicate and function symbols which are provided by a so-called signature. We then define some formulation rules which state -- most often inductively -- how well-formed formulas may be constructed.

A \reftodef{deductive system} is a way to show in a purely syntactical way that one formula is a logical consequence of another formula. One example is sequent calculi, which will play a central part in this thesis. The rules for a classical first-order sequent calculus are given in Definition~\ref{lk}. In sequent calculi, we examine syntactical objects which are called sequents, denoted as $\Gamma \vdash \Delta$, where $\Gamma$ and $\Delta$ are lists of formulas called \reftodef[cedent]{cedents}.

Semantics is concerned with the interpretation of the language. Most often, an interpretation is a specification of a structure (also called model), which consists of a domain (also called a universe) and a function that assigns to each function symbol a function and to each predicate symbol a relation. The standard semantics for a sequent $\Gamma\vdash\Delta$ asserts for example that when every $\gamma\in\Gamma$ is true, then at least one $\delta\in\Delta$ will be true. When we have access to de Morgan dualities in a logic, we can often bring $\Gamma$ to the right-hand side by negating it: $\vdash \neg\Gamma, \Delta$. This one-sided calculus has the advantage that we have to define far fewer inference rules. We can also view classical propositional logic from an algebraic standpoint. The class of algebras which correspond to propositional logic are called \reftodef[Boolean algebra]{Boolean algebras}.

\subsection{Intuitionistic Logic}\label{intuitionisticlog}

\epigraph{there's a really great joke about non-constructive proofs}{\href{https://twitter.com/qntm/status/1279802274584330244}{@qntm on Twitter}}

When we restrict sequents to only have one formula on the right-hand side (e.\,g. $\Gamma\vdash \phi$, where $\Gamma$ is a list of formulas and $\phi$ is a formula), we arrive at \reftodef{intuitionistic logic}. In this non-classical logic, the law of excluded middle is not a tautology\footnote{It can of course always be assumed for a proof if we desire so.}. Instead, we have the law of contradiction $(\phi\rightarrow \psi)\rightarrow((\phi\rightarrow \neg\psi)\rightarrow \neg\phi)$. In this logic, we are no longer concerned with truth and falsity, but rather notions of proofs and refutations: just because we have no proof of $\phi$, it does not mean that we automatically refute it\footnote{Amazingly, we can refute the refutation of LEM, thus $\neg\neg(\phi\vee\neg\phi)$ is a tautology.}. This leads to the Brouwer-Heyting-Kolmogorov interpretation. For example, to prove the formula $\Phi\equiv\varphi\wedge\psi$, we have to give a proof for both $\varphi$ and $\psi$. To prove $\Psi\equiv\varphi\rightarrow \psi$, we have to give a function that converts a proof for $\varphi$ into a proof for $\psi$. The algebras used for intuitionistic logic are called \reftodef[Heyting algebra]{Heyting algebras}.

There are rather profound advantages and disadvantages which accompany the use of intuitionistic logic over classical logic. On the one hand, proofs by contradiction are now limited to proving negative statements, since $\neg\neg\phi\rightarrow\phi$ no longer holds. On the other hand, intuitionistic logic is constructive, which means that a proof of the existence of an object always gives rise to an algorithmic procedure constructing exactly this object. This property is used in many modern proof assistants, which help mathematicians and logicians to prove statements about systems which are too complex to reason manually about them, computationally.

Thus, the history around classical and intuitionistic logic is mainly defined by the heated discussions around the question of which logic is the true logic of mathematics. It was no help that the mathematicians who disagreed most vehemently on this topic were two of the most famous mathematicians of their era: While David Hilbert was a strong advocate of classical logic, Luitzen Egbertus Jan Brouwer stood on the side of constructivism. For example, Hilbert wrote -- rather dramatically -- that ``taking the principle of excluded middle from the mathematician would be the same, say, as proscribing the telescope to the astronomer or to the boxer the use of his fists. To prohibit existence statements and the principle of excluded middle is tantamount to relinquishing the science of mathematics altogether.''~\cite{Heijenoort}. The disagreement between the parties was so severe that it became known as the ``Grundlagenstreit''. Eventually, Hilbert had Brouwer removed as editor from the leading mathematical journal of the time, the ``Mathematische Annalen'', and his view succeeded in becoming the mathematical standard. But until today, there are discussions about which system to use. Since a solution to this controversy will likely not be achieved in this thesis, we will not waste any resources to argue for one logic or the other, but rather adopt the pragmatical standpoint that we use the one that is most useful to us in a given situation.

We can embed intuitionistic logic into classical logic via various translations. The definition and proof of correctness of one such translation, the \reftodef{double negation translation}, can be found in Buss~\cite[Chapter 5]{Buss1998-BUSHOP}.

\subsection{Proof Theory}

Around the beginning of the 20th century, logicians examined the foundations of mathematics much more closely than before and they wanted to talk about mathematics itself with formal methods. One outcome of this was modern proof theory: The idea is to view proofs as mathematical objects which we can examine using mathematical machinery. Through time, the field grew rapidly, establishing subfields such as structural proof theory, provability logic, proof mining, automated theorem proving, and proof complexity. In this section, we will review some basic notions of proof theory that will be used in this thesis. For this, we will mainly employ the books by~\textcite{takeuti2013proof} and by~\textcite{troelstra_schwichtenberg_2000}. Another reference that gives a broad overview over the field is the handbook by~\textcite{Buss1998-BUSHOP}. We set the stage by introducing the system we will be examining in this section: the classical first-order sequent calculus.

\begin{definition}[Classical first order sequent calculus]\label{lk}
    The sequent calculus of classical first order logic ($\fragment{LK}$) consists of the following rules:

    \textbf{Axiom}
    \begin{displaymath}
        \prftree[r]{(id)}
            {}
            {\vdash A, \neg A}
    \end{displaymath}

    \textbf{Structural rules}
    \begin{center}
        \begin{minipage}[b]{0.3\textwidth}
            \begin{displaymath}
                \prftree[r]{(weakening)}
                    {\vdash \Gamma}
                    {\vdash \Gamma, A}
            \end{displaymath}
        \end{minipage}
        \begin{minipage}[b]{0.3\textwidth}
            \begin{displaymath}
                \prftree[r]{(contraction)}
                    {\vdash \Gamma, A, A}
                    {\vdash \Gamma, A}
            \end{displaymath}
        \end{minipage}
    \end{center}
    \begin{center}
        \begin{minipage}[b]{0.3\textwidth}
            \begin{displaymath}
                \prftree[r]{(exchange)}
                    {\vdash \Gamma}
                    {\vdash \Gamma'}
            \end{displaymath}
        \end{minipage}
        \begin{minipage}[b]{0.3\textwidth}
            \begin{displaymath}
                \prftree[r]{(cut)}
                    {\vdash \Gamma, A}
                    {\vdash \neg A, \Delta}
                    {\vdash \Gamma, \Delta}
            \end{displaymath}
        \end{minipage}
    \end{center}
    \phantom{xxxxx}where $\Gamma'$ is a permutation of $\Gamma$.\\[0.5ex]

    \textbf{Logical rules}
    \begin{center}
        \begin{minipage}[b]{0.3\textwidth}
            \begin{displaymath}
                \prftree[r]{($\wedge$)}
                    {\vdash \Gamma, A}
                    {\vdash \Gamma, B}
                    {\vdash \Gamma, A\wedge B}
            \end{displaymath}
        \end{minipage}
        \begin{minipage}[b]{0.3\textwidth}
            \begin{displaymath}
                \prftree[r]{($\vee_1$)}
                    {\vdash \Gamma, A}
                    {\vdash \Gamma, A\vee B}
            \end{displaymath}
        \end{minipage}
        \begin{minipage}[b]{0.3\textwidth}
            \begin{displaymath}
                \prftree[r]{($\vee_2$)}
                    {\vdash \Gamma, B}
                    {\vdash \Gamma, A\vee B}
            \end{displaymath}
        \end{minipage}
    \end{center}
    \begin{center}
        \begin{minipage}[b]{0.3\textwidth}
            \begin{displaymath}
                \prftree[r]{($\forall$)}
                    {\vdash \Gamma, A}
                    {\vdash \Gamma, \forall x A}
            \end{displaymath}
        \end{minipage}
        \begin{minipage}[b]{0.3\textwidth}
            \begin{displaymath}
                \prftree[r]{($\exists$)}
                    {\vdash \Gamma, A[t/x]}
                    {\vdash \Gamma, \exists x A}
            \end{displaymath}
        \end{minipage}
    \end{center}
    \phantom{xxxxx}where $x$ does not occur in $\Gamma$ and $A[t/x]$ means substitution of $x$ by $t$.
\end{definition}

This calculus was given the name \define[\textsf{LK}]{}\textsf{LK} by Gentzen, which stands for ``Logistisches Kalkül''~\cite{gentzen34, gentzen35}. He developed this system to prove one of the central results of structural proof theory, the cut-elimination theorem. This theorem -- also known as ``Gentzens Hauptsatz'' -- was a breakthrough in modern proof theory because it enables proof search: when trying to prove a formula from axioms, we need to start from the formula and apply the rules above in ``reverse'', deciding what rule to apply and eventually landing at axioms. Observe that the cut rule is the only one in which we use a formula in the premises which does not occur in the conclusion. Intuitively, this corresponds to finding lemmas in the process of proving a theorem. Cut elimination also brings with it a wealth of corollaries, such as the subformula property, an easy way to prove consistency, Craig's interpolation theorem, Herbrand's theorem, and much more.

\begin{proposition}[Cut-elimination theorem]
    If a sequent is provable in \fragment{LK}, then it is provable in \fragment{LK} without a cut.
\end{proposition}

\begin{proof}
    See Takeuti~\cite[Chapter 1, Paragraph 5]{takeuti2013proof}.
\end{proof}

Since we will use it later on, we also state the definition of the subformula property, which is implied by cut-elimination. This property is, for example, a key part in the standard proof of the decidability of \reftodef[\textsf{LK}]{\textsf{LK}}. It holds for first-order as well as propositional logic.

\begin{definition}[Subformula property]\label{subform}
    A calculus has the \define{subformula property} if any sequent that is provable in the calculus can be proven by use of its subformulas only.
\end{definition}

\subsection{Finite Model Theory}

Finite model theory, with its deep connection to descriptive complexity theory, is a prime candidate for the computational examination of logics. Standard literature on finite model theory includes the book by~\textcite{DBLP:books/sp/Libkin04}, which we will use, and the book by~\textcite{DBLP:series/txtcs/GradelKLMSVVW07}. Of special interest for us is the finite model property, because it has direct recursion-theoretic consequences.

\begin{definition}[Finite model property]\label{fmp}
    We say that a class $K$ of sentences has the \define{finite model property} if for every sentence $\varphi$ in $K$, either $\varphi$ is unsatisfiable, or it has a finite model.
\end{definition}

Examples for logics which admit the finite model property are first-order logic restricted to one universal quantification~\cite{DBLP:books/sp/Libkin04}, or modal logic~\cite{DBLP:books/cu/BlackburnRV01}. What makes the finite model property so valuable is the following proposition.

\begin{proposition}
    If a logic $\mathcal{L}$ is finitely axiomatizable and has the finite model property, then it is decidable.
\end{proposition}

Another useful tool that finite model theory provides is that we can separate logic fragments and provide bounds for the complexity by showing that models of a certain complexity can be defined. This will find applications at various points in this thesis.

\section{Complexity Theory}

We assume some familiarity with basic complexity-theoretic concepts such as Turing machines, (un-)decidability, $\mathcal{O}$-notation, time and space complexity, (co-)nondeterminism, reductions, oracle machines, hardness, and completeness. Standard literature which introduces these topics is the book by~\textcite{AB_09}, the book by~\textcite{DBLP:books/daglib/0072413}, and the book by~\textcite{DBLP:books/daglib/0086373}.

In this thesis, we will use the following classes in particular:
\[\class{P} \subseteq \class{NP} \subseteq \class{PSPACE} \subseteq \Sigma_1^0,\]
where $\Sigma_1^0$ denotes the first level of the arithmetical hierarchy, i.\,e. the recursively enumerable sets. We will also briefly mention the subpolynomial time classes \class{AC}, \class{NC}, \class{L}, and \class{NL}. Since it will be directly used in a reduction, we furthermore recall the canonical \class{PSPACE}-complete problem, \textsf{QBF}. The completeness, and especially the further adaptation to provide complete problems for each level of the polynomial hierarchy, is due to~\textcite{WRATHALL197623}.

\begin{defproblem}[\class{QBF}]\label{qbf-problem}
    \begin{description}
        \item[Input:] A formula $\Phi$ of the form $Q_1x_1Q_2x_2\cdots Q_nx_n\varphi(x_1, x_2, \dots, x_n)$, where $Q_i\in \left\{\,\forall, \exists\,\right\}$ and $x_i\in \left\{\,0, 1\,\right\}$ for $1\leq i\leq n$ and $\varphi$ is a unquantified Boolean formula.
        \item[Output:] $\Phi \equiv 1$?
    \end{description} 
\end{defproblem}

The gap between \class{PSPACE} and \class{NP} is large enough to define a hierarchy that conveys some granularity between them. Given a language $A$ and a complexity class $\mathsf{C}$, we write $\mathsf{C}^A$ for the class of languages which can be decided by an algorithm of class $\mathsf{C}$ which may, at any point, query in constant time whether some word is in $A$ or not. Given another class of languages $\mathsf{D}$, we furthermore write $\mathsf{C}^{\mathsf{D}} \coloneq{} \bigcup_{A\in \mathsf{D}} \mathsf{C}^A$. We write $\overline{A}$ for the complement of $A$.

\begin{definition}[Polynomial hierarchy]
    We define the classes of the \define{polynomial hierarchy} as follows:
    \begin{align*}
        \Delta^p_0 = \Sigma^p_0 = \Pi^p_0 &\coloneq \class{P}\\
        \Delta_{k+1}^p &\coloneq \class{P}^{\Sigma^p_k}\\
        \Sigma_{k+1}^p &\coloneq \class{NP}^{\Sigma^p_k}\\
        \Pi_{k+1}^p &\coloneq \left\{\,A\mid \overline{A}\in \Sigma^p_{k+1}\,\right\}\\
        \class{PH} &\coloneq \bigcup_i \left(\Delta^p_k \cup \Sigma^p_k \cup \Pi^p_k\right)
    \end{align*}
    where $i, k\geq 0$.
\end{definition}

We can adapt the problem \class{QBF} to provide complete problems for each level in \class{PH}. For this, let $\Sigma^q_k$ be the class of QBFs with $k$ quantifier alternations beginning with an existential quantifier, and let $\Pi^q_k$ be the class of QBFs with $k$ quantifier alternations beginning with a universal quantifier. Then we have the following result.

\begin{proposition}
    For $k\geq 1$, we have that $\Sigma^q_k$-evaluation is $\Sigma^p_k$-complete, and that $\Pi^q_k$-evaluation is $\Pi^p_k$-complete.
\end{proposition}

A special type of Turing machines are alternating Turing machines. They generalize the notions of nondeterminism and co-nondeterminism in that we can alternate between the two modes of operation in each step. There are many equivalent ways to define this machine model, we go with the definition by~\textcite{AB_09}.

\begin{definition}[Alternating time]\label{at}\define[Turing machine!alternating]{}
    For every $T\colon \mathbb{N} \rightarrow \mathbb{N}$, we say that an alternating Turing machine (ATM) $M$ runs in $T(n)$-time if for every input $x\in \{\,0, 1\,\}^\ast$ with $|x| = n$ and for every possible sequence of transition function choices, $M$ halts after at most $T\left(|x|\right)$ steps.

    We say that a language $L$ is in $\class{ATIME}\left(T(n)\right)$ if there is a constant $c$ and a $c\cdot T(n)$-time ATM~$M$ such that for every $x\in \{\,0, 1\,\}^\ast$, $M$ accepts if and only if $x\in L$. The definition of accepting an input is as follows:
    
    Let $G_{M, x}$ be the directed acyclic \emph{configuration graph} of $M$ on input $x$, where there is an edge from a configuration $C$ to configuration $C'$ if and only if $C'$ can be obtained from $C$ by one step on $M$'s transition function. We label some of the vertices in this graph by ``ACCEPT'' by repeatedly applying the following rules to exhaustion:
    \begin{itemize}
        \item The configuration $C_{\texttt{acc}}$ where the machine is in an accepting state is labeled ``ACCEPT''.
        \item If a configuration $C$ is in a state labeled $\exists$ and there is an edge from $C$ to a configuration $C'$ labeled ``ACCEPT'', then we label $C$ ``ACCEPT''.
        \item If a configuration $C$ is in a state labeled $\forall$ and both the configurations $C', C''$ reachable from it in one step are labeled ``ACCEPT'', then we label $C$ ``ACCEPT''.
    \end{itemize}
    We say that $M$ \emph{accepts} $x$ if at the end of this process the starting configuration $C_\texttt{start}$ is labeled ``ACCEPT''.
\end{definition}

The relations between classical and alternating Turing machines are extensively studied. For us, it suffices that we can imagine that compared to classical Turing machines, these machines are more expressive from a complexity-theoretic standpoint. In particular, the following equality will be used in this thesis, the proof of which can be found in the paper by~\textcite{CKS_81}.

\begin{proposition}\label{atime-pspace}
    $\class{PSPACE} = \class{ATIME}(n^{\mathcal{O}(1)})$.
\end{proposition}

Some decision problems we will encounter will still be way harder than what the classes named above could capture. For them, we will define a class of non-elementary problems, which we will call \class{TOWER}.

\begin{definition}[Elementary functions]\label{efun}
    Let $\text{exp}_0(n) \coloneq n$ and $\text{exp}_{k+1}(n) \coloneq 2^{\text{exp}_k(n)}$. We define the class of \define[elementary function]{elementary functions} as
    \[\mathscr{F}\coloneq \left\{\,f\colon \mathbb{N}\rightarrow\mathbb{N}\mid f\text{ is computable in time }\bigO\!\left(\text{exp}_k(n)\right) \text{ for some fixed }k\in \mathbb{N}\,\right\}.\qedhere\]
\end{definition}

\begin{definition}[\fragmenttxt{TOWER}]\label{tower}
    The complexity class \class{TOWER} is defined as
    \[\class{TOWER} \coloneq \bigcup_{f\in \mathscr{F}} \class{DTIME}\left(\text{exp}_{f(n)}(1)\right).\qedhere\]
\end{definition}

\begin{remark}
    In this thesis, we only need the class \class{TOWER}, but note that~\textcite{Schmitz_2016} provides a natural generalization of this class to a hierarchy of non-elementary complexity classes.
\end{remark}

\section{Category Theory and Categorical Logic}

\subsection{Category Theory}

Category theory was first introduced by Eilenberg and Mac Lane in their paper ``General Theory of Natural Equivalences''~\cite{EM_45}. It was quickly adapted to provide a convenient language to describe algebraic topology and homological algebra.  Later on, it proved to be a valuable tool in the analysis of logic. The definitive reference is the book by~\textcite{Mac_78}, for our purposes, the much more accessible book by~\textcite{Awo_10} suffices. The following definitions stem from this work, if not stated otherwise.

\begin{definition}[Category]\label{cat}
    A \define{category} consists of the following data:
    \begin{itemize}
        \item \define[object]{Objects} $A,B,C,\dots$
        \item \define[morphism]{Morphisms} $f,g,h,\dots$
        \item For each morphism $f$, there are given objects $\text{dom}(f),\ \text{cod}(f)$, called the \define{domain} and \define{codomain} of $f$. We write $f\colon A \rightarrow B$ to indicate that $A = \text{dom}(f)$ and $B = \text{cod}(f)$.
        \item Given morphisms $f\colon A \rightarrow B$ and $g\colon B\rightarrow C$, that is, with $\text{cod}(f) = \text{dom}(g)$
        there is given a morphism $g \circ f\colon A\rightarrow C$, called the \define{composite} of $f$ and $g$.
        \item For each object $A$, there is given a morphism $1_A\colon A\rightarrow A$, called the \define[morphism!identity]{identity morphism} of $A$.
    \end{itemize}

    These data are required to satisfy the following laws:
    \begin{itemize}
        \item Associativity:
        \[h\circ(g\circ f) = (h\circ g)\circ f\]
        for all $f\colon A\rightarrow B, g\colon B\rightarrow C, h\colon C\rightarrow D$.
        \item Unit:
        \[f\circ 1_A = f = 1_B \circ f\]
        for all $f\colon A\rightarrow B$.\qedhere
    \end{itemize}
\end{definition}

Categories, due to their general structure, arise throughout mathematics and computer science: Standard examples include the category \textbf{Set} of sets with functions as morphisms, the category \textbf{Grp} of groups with group homomorphisms as morphisms, and the category \textbf{Vect}$_k$ of vector spaces over the field $k$ with $k$-linear maps as morphisms. An example for the application of categories in logic is that for a given logical deduction system, we can associate a category of proofs, whose objects are the formulas and the morphisms are sequents in the system. Transitivity then corresponds to chaining together proofs, while the identity morphism can be interpreted as the initial sequent. We denote $C\in \mathcal{C}_0$ for an object of category $\mathcal{C}$ and $f\in \mathcal{C}_1$ for a morphism of category $\mathcal{C}$.

One of the limitations of set theory is that we cannot build the set of all sets, since this would enable us to derive Russell's paradox. Similar problems arise in other formal systems if the concept of ``self-inclusion'' is not handled carefully. What if we want to speak of categories as a category? We need some nomenclature to differentiate categories of different ``sizes''. A category $\mathcal{C}$ is called \define[category!small]{small} if its collection of objects and its collection of morphisms are both sets. Otherwise it is called \define[category!large]{large}. Furthermore, a category $\mathcal{C}$ is called \define[category!locally small]{locally small}, if for all $A, B\in \mathcal{C}_0$, the collection Hom$_\mathcal{C}(A, B) \coloneq \left\{\,f\in \mathcal{C}_1\mid f\colon A\rightarrow B\,\right\}$ is a set (called the hom-set). Now, we can construct the category \textbf{Cat} of small categories, which is itself a large category.

\begin{remark}
    A central idea in category theory is the commutativity of diagrams. Commutative diagrams play a similar role in category theory as equations in algebra~\cite{BC_02}. A diagram
    \begin{center}
        \begin{tikzcd}
            A \ar[r, "f"] \ar[d, "g", swap] & C\\
            B \ar[ru, "h", swap] & \\
        \end{tikzcd}
    \end{center}
    \emph{commutes} if $f = h\circ g$, so intuitively, it does not matter whether we go directly from $A$ to $C$ via $f$ or take a detour over $B$ via $g$ and $h$.
\end{remark}

Morphisms often play the roles of certain mappings, which vary depending on which category we consider. But we also want to talk about mappings \emph{between} categories which uphold the general structure we expect from a category. These mappings are called functors.

\begin{definition}[Functor]\label{functor}
    A \define{functor}
    \[F\colon \Cat{C}\rightarrow \Cat{D}\]
    between categories \Cat{C} and \Cat{D} is a mapping from objects to objects and morphisms to morphisms, in such a way that
    \begin{enumerate}[label=(\alph*)]
        \item $F(f\colon A\rightarrow B) = F(f) \colon F(A)\rightarrow F(B),$
        \item $F(1_A) = 1_{F(A)},$
        \item $F(g\circ f) = F(g)\circ F(f).$\qedhere
    \end{enumerate}
\end{definition}

A special kind of functor is the one that maps a category onto itself, e.\,g. $F\colon \mathcal{C}\rightarrow \mathcal{C}$ for a category $\mathcal{C}$. We call these functors \emph{endofunctors}. Since functors are only another mathematical object, nothing keeps us from defining a category in which the objects are functors. This gives us the ability to relate functors to one another in the language of category theory itself. The morphisms of this category are called natural transformations and are defined in the obvious way.

\begin{definition}[Natural transformation]\label{nat_trans}
    For categories $\mathcal{C}$, $\mathcal{D}$ and functors $F, G\colon \mathcal{C} \rightarrow \mathcal{D}$, a \define{natural transformation} $\vartheta\colon F \rightarrow G$ is a family of morphisms in $\mathcal{D}$
    \[(\vartheta_A\colon F(A)\rightarrow G(A))_{A\in \mathcal{C}_0}\]
    such that, for any $f\colon A\rightarrow A'$ in $\mathcal{C}$, one has $\vartheta_{A'}\circ F(f) = G(f)\circ \vartheta_A$, that is, the following diagram commutes:
    \begin{center}
        \begin{tikzcd}
            F(A) \ar[r, "\vartheta_A"] \ar[d, "F(f)", swap] & G(A) \ar[d, "G(f)"]\\
            F(A') \ar[r, "\vartheta_{A'}"] & G(A').\\
        \end{tikzcd}
    \end{center}
    Given such a natural transformation $\vartheta\colon F\rightarrow G$, the $\mathcal{D}$-morphism $\vartheta_A\colon F(A)\rightarrow G(A)$ is called the \define{component of $\vartheta$} at $A$. When all components of $\vartheta$ are isomorphisms, we call $\vartheta$ a \define{natural isomorphism} and denote it as $\vartheta\colon F \overset{\simeq}{\rightarrow} G.$
\end{definition}

We can now establish an equivalence relation between categories: Two categories $\mathcal{C}, \mathcal{D}$ are \emph{equivalent}, in symbols $\mathcal{C} \simeq \mathcal{D}$, if we have two functors $F\colon \mathcal{C} \rightarrow \mathcal{D}$ and $G\colon \mathcal{D}\rightarrow \mathcal{C}$ with natural isomorphisms $\alpha\colon F\circ G \overset{\simeq}{\rightarrow} 1_\mathcal{D}$ and $\beta\colon G\circ F\overset{\simeq}{\rightarrow} 1_\mathcal{C}$.

\begin{remark}
    Natural transformations are also a starting point for higher category theory: looking back at \textbf{Cat}, it has as morphisms sets of functors, but they themselves form a category, with natural transformations as morphisms. This leads to the definition of \textbf{Cat} as a 2-category. We now have as objects small categories, and two levels of morphisms: as 1-morphisms, we have functors, and as 2-morphisms, we have natural transformations. This process can be continued indefinitely, leading to the generalization of category theory to $\infty$-category theory.
\end{remark}

For the next definition, we need two further notions: First, the \define[category!opposite]{opposite category} $\mathcal{C}^\text{op}$ for a category $\mathcal{C}$ is the category that has the same objects as $\mathcal{C}$ and a morphism $g\colon B\rightarrow A$ for every morphism $f\colon A\rightarrow B$ in $\mathcal{C}$. Intuitively speaking, $\mathcal{C}^\text{op}$ is constructed from a category $\mathcal{C}$ by reversing all morphisms. The composition operation is the same as in $\mathcal{C}$.

Second, we need the notion of product categories, which are defined in the obvious way: For two categories $\mathcal{C}, \mathcal{D}$, the \define[category!product]{product category} $\mathcal{C}\times \mathcal{D}$ is the category whose objects are ordered pairs $(A, B)$ for $A\in \mathcal{C}_0$ and $B\in\mathcal{D}_0$, and whose morphisms are ordered pairs $((A\rightarrow A'), (B\rightarrow B'))$ for $A,A'\in \mathcal{C}_0$ and $B,B'\in\mathcal{D}_0$. Composition of morphisms is defined componentwise by composition in $\mathcal{C}$ and $\mathcal{D}$. With this nomenclature at hand, we can define a specific class of functors, the hom-functors. The definition we use is taken from~\textcite{homF}.

\begin{definition}[Hom-functors]\label{homfun}
    Given a locally small category $\mathcal{C}$, its \define[functor!hom-functor]{hom-functor} is the functor
    \[\text{hom}\colon \mathcal{C}^\text{op}\times \mathcal{C} \rightarrow \textbf{Set}\]
    which sends
    \begin{itemize}
        \item an object $(A, A') \in (\mathcal{C}^\text{op} \times \mathcal{C})_0$ to the hom-set Hom$_\mathcal{C}(A, A')$ in \textbf{Set}, the set of morphisms $q\colon A\rightarrow A'$ in $\mathcal{C}$.
        \item a morphism $(A, A') \rightarrow (B, B')$, i.\,e. a pair of morphisms
        \begin{center}
            \begin{tikzcd}
                A \ar[d, "f^\text{op}", swap] & A' \ar[d, "g"]\\
                B & B'\\
            \end{tikzcd}
        \end{center}
        in $\mathcal{C}$ to the mapping of sets Hom$_\mathcal{C}(A, A') \rightarrow \text{Hom}_\mathcal{C}(B, B')$ defined as
        \[(q\colon A\rightarrow A') \mapsto \begin{tikzcd}
            A  \ar[r, "q"] & A' \ar[d, "g"]\\
            B \ar[u, "f"] & B'\\
        \end{tikzcd}.\]
    \end{itemize}\vspace{-4em}\qedhere
\end{definition}

A special relation that functors can have is that of adjointness. The notion of adjoint functors is one of the high points of category theory and is used in many areas of modern algebra and applied category theory. Some of these applications are presented in the book by~\textcite{Awo_10}, Chapter 9 or the book by~\textcite{fong_spivak_2019}, Section 3.4. Again, the definition we use is taken from~\textcite{adjointF}.

\begin{definition}[Adjoint functors]\label{adjointfun}
    Given categories $\mathcal{C}$ and $\mathcal{D}$ and functors $L\colon \mathcal{C} \rightarrow \mathcal{D}$, and $R\colon \mathcal{D} \rightarrow \mathcal{C}$, $L$ and $R$ are called a \emph{pair of} \define[functor!adjoint]{adjoint functors}, with $L$ being the \define{left adjoint} and $R$ being the \define{right adjoint}, if there exists a natural isomorphism between the hom-functors of the form
    \[\text{Hom}_\mathcal{D}(L(-),-) \simeq \text{Hom}_\mathcal{C}(-, R(-)).\qedhere\]
\end{definition}

There are many different but equivalent ways to define adjoint functors, which is due to the fact that they are a ubiquitous structure in category theory. We will give two notable examples for adjunctions to gain a little bit of intuition for them.

\begin{example}[Adjoint functors]\label{adjoint}
    We present a simple example for adjoint functors from order theory and a more advanced example from logic.
    \begin{enumerate}
        \item Take two preordered sets $(A, \leq)$ and $(B, \leq)$ as a degenerate example of a category, in that there is at most one morphism between two objects, which is the order relation. A Galois connection between these sets consists of two monotone functions $f\colon A\rightarrow B$ and $g\colon B\rightarrow A$, s.\,t.
        \[\forall a\in A, b\in B\ (f(a)\leq b\iff a\leq g(b)).\]
        A Galois connection between two such sets forms an adjunction.
        \item Observe that $\text{Form}(\overline{x}) \coloneq \{\,\varphi(\overline{x})\mid \phi(\overline{x}) \text{ has at most }\overline{x}\text{ free}\,\}$ for lists of variables $\overline{x}$ and first-order formulas $\phi$ is a preordered set under the entailment relation of first-order logic and define the functor $*\colon \text{Form}(\overline{x}) \rightarrow \text{Form}(\overline{x}, y)$, taking every $\phi(\overline{x})$ to itself. We notice that for every $\psi(\overline{x}, y)\in \text{Form}(\overline{x}, y)$, the variable $y$ cannot occur free in $\forall y.\psi(\overline{x}, y)$. We thus have a map $\forall y\colon \text{Form}(\overline{x}, y)\rightarrow \text{Form}(\overline{x})$. The rules for universal introduction and elimination show us that these functors are adjoint:
        \[*\dashv \forall.\]
        When we examine existential quantification in the same manner, we get
        \[\exists \dashv * \dashv \forall.\]
    \end{enumerate}
\end{example}

Another elementary notion of category theory we use in this thesis is that of products. The intuition behind this construction is captured in the notion of cartesian products in the category of sets, we just give the abstract category-theoretic definition.

\begin{definition}[Products]\label{prod}
    Given a category $\mathcal{C}$, a \define[product]{product diagram} for the objects $A$ and $B$ consists of an object $P$ and morphisms
    \begin{center}
        \begin{tikzcd}
            A & P \ar[l, "p_1", swap] \ar[r, "p_2"] & B\\
        \end{tikzcd}
    \end{center}
    satisfying that, given any diagram of the form
    \begin{center}
        \begin{tikzcd}
            A & X \ar[l, "x_1", swap] \ar[r, "x_2"] & B\\
        \end{tikzcd}
    \end{center}
    there exists a unique $u\colon X\rightarrow P$, making the following diagram commute.
    \begin{center}
        \begin{tikzcd}
            {} & X \ar[ld, "x_1", swap] \ar[rd, "x_2"] \ar[d, "u", dashed] & {}\\
            A & P \ar[l, "p_1"] \ar[r, "p_2", swap] & B\\
        \end{tikzcd}
    \end{center}
    A \define[product!finite]{finite product} is a product with a finite number of factors.
\end{definition}

The category $\mathcal{C}^\text{op}$ gave us a first glimpse at a very useful concept in category theory: duality. Duality in category theory means that for many definitions and theorems, a dual definition or theorem naturally arises. In the case of products, this materializes itself in the definition of coproducts, whose definition is achieved by simply reversing the morphisms in the definition of products.

\begin{definition}[Coproduct]
    Given a category $\mathcal{C}$, a \define[coproduct]{coproduct diagram} for the objects $A$ and $B$ consists of an object $Q$ and morphisms
    \begin{center}
        \begin{tikzcd}
            A \ar[r, "q_1"] & Q & B \ar[l, "q_2", swap]\\
        \end{tikzcd}
    \end{center}
    satisfying that, given any diagram of the form
    \begin{center}
        \begin{tikzcd}
            A \ar[r, "x_1"] & X & B \ar[l, "x_2", swap]\\
        \end{tikzcd}
    \end{center}
    there exists a unique $u\colon X\rightarrow Q$, making the following diagram commute.
    \begin{displaymath}
        \begin{tikzcd}
            {} & X   & {}\\
            A \ar[ru, "x_1"] \ar[r, "q_1", swap] & P \ar[u, "u", swap, dashed] & B \ar[lu, "x_2", swap] \ar[l, "q_2"]\\
        \end{tikzcd}
    \end{displaymath}
    Dually to products, we have that a \define{finite coproduct} is a coproduct with a finite number of summands.
\end{definition}

The last notion we introduce is that of (co-)monads. Monads play an important role in the theory of functional programming, where they are used to model concepts like non-deterministic computation, probabilistic computation, exceptions, side effects, and interactive input.

\begin{definition}[Monad]\label{monad}
    A \define{monad} on a category $\mathcal{C}$ consists of an endofunctor $T\colon \mathcal{C} \rightarrow \mathcal{C}$ and natural transformations $\eta\colon 1_\mathcal{C} \rightarrow T$, and $\mu\colon T\circ T\rightarrow T$ satisfying
    \begin{align*}
        \mu \circ \mu_T &= \mu\circ T(\mu)\\
        \mu\circ \eta_T &= 1_\mathcal{C} = \mu\circ T(\eta).\qedhere
    \end{align*}
\end{definition}

The definition bears a strong similarity to the one of monoids. A monad is indeed a monoidal monoid in the \reftodef[category!monoidal]{monoidal category} $\mathcal{C}^\mathcal{C}$ with composition as the monoidal product.\footnote{The jargon of category theory, due to its abstract nature, can be quite confusing for people who are new to the field. In fact, an equivalent description of monads lead to the quote from James Iry: ``A monad is just a monoid in the category of endofunctors, what's the problem?''~\cite{james}.} Thus, the two laws above are called the \emph{associativity} and \emph{unit} laws. A related concept to monads are Kleisli categories.

\begin{definition}[Kleisli category]
    Given a category $\mathcal{C}$ and a monad $(T, \eta, \mu)$, its \define[category!Kleisli]{Kleisli category} $\mathcal{C}_T$ is defined as follows:
    \begin{itemize}
        \item the objects are the same as those of $\mathcal{C}$, but written $A_T, B_T, \dots$,
        \item a morphism $f_T\colon A_T \rightarrow B_T$ is a morphism $f\colon A\rightarrow T(B)$ in $\mathcal{C}$,
        \item the identity arrow $1_{A_T}\colon A_T \rightarrow A_T$ is the arrow $\eta_A\colon A\rightarrow T(A)$ in $\mathcal{C}$,
        \item for composition, given $f_T\colon A_T \rightarrow B_T$ and $g_T\colon B_T \rightarrow C_T$, the composite $g_T\circ f_T\colon A_T \rightarrow C_T$ is defined to be
        \[\mu_C\circ T(g_T)\circ f_T\]
        as indicated in the following diagram:
        \begin{displaymath}
            \begin{tikzcd}
                A \ar[r, "g_T\circ f_T"] \ar[d, "f_T", swap] & T(C)\\
                T(B) \ar[r, "T(g_T)", swap] & T(T(C))\ar[u, "\mu_C", swap]
            \end{tikzcd}
        \end{displaymath}
    \end{itemize}\vspace{-2em}\qedhere
\end{definition}

For a short example of how monads and their corresponding Kleisli categories -- in the form of Kleisli triples -- are used in Haskell to model side effects, we refer to~\textcite[Chapter 2.4.3]{DBLP:books/sp/Doberkat15}. Again following the concept of duality, we define the dual of monads.

\begin{definition}[Comonad]
    A \define{comonad} of a category $\mathcal{C}$ is a monad on the category $\mathcal{C}^\text{op}$.
\end{definition}

In particular, the comonad is thus a comonoid in the \reftodef[category!monoidal]{monoidal category} of endofunctors. It consists of an endofunctor $G\colon \mathcal{C}\rightarrow \mathcal{C}$ and natural transformations $\varepsilon\colon G\rightarrow 1$, the \define[counit!in a comonad]{counit}, and $\delta\colon G\rightarrow G^2$, the \define[comultiplication!in a comonad]{comultiplication}, which suffice the equations

\begin{align*}
    \delta_G \circ \delta &= G(\delta) \circ \delta\\
    \varepsilon_G\circ \delta &= 1_G = G(\varepsilon) \circ \delta.\\    
\end{align*}
From a data structure point of view, we can picture a comonad intuitively as a container with a notion of a current value. Also, the three-way adjunction of quantifiers from Example~\ref{adjoint} gives rise to the monad $T = * \circ \exists$ and the comonad $G = * \circ \forall$. Then, $T$ and $G$ are in turn adjoint. For a given category and comonad, we can construct the \define[category!co-Kleisli]{co-Kleisli category} in the obvious way dual to above.

\subsection{Categorical Logic}

The idea behind categorical logic is that we take a \emph{pluralistic} point of view on logic: Instead of viewing mathematics as an edifice built on top of it, logic becomes part of mathematics itself and can as such be examined with methods from category theory. The main advantage of this is that we have a flexible framework for defining logical systems that suit our area, rather than taking the cumbersome way of building enough theory on top of first-order logic. A good introduction into categorical logic which also deals with linear logic is given by~\textcite{Abramsky_2010}.

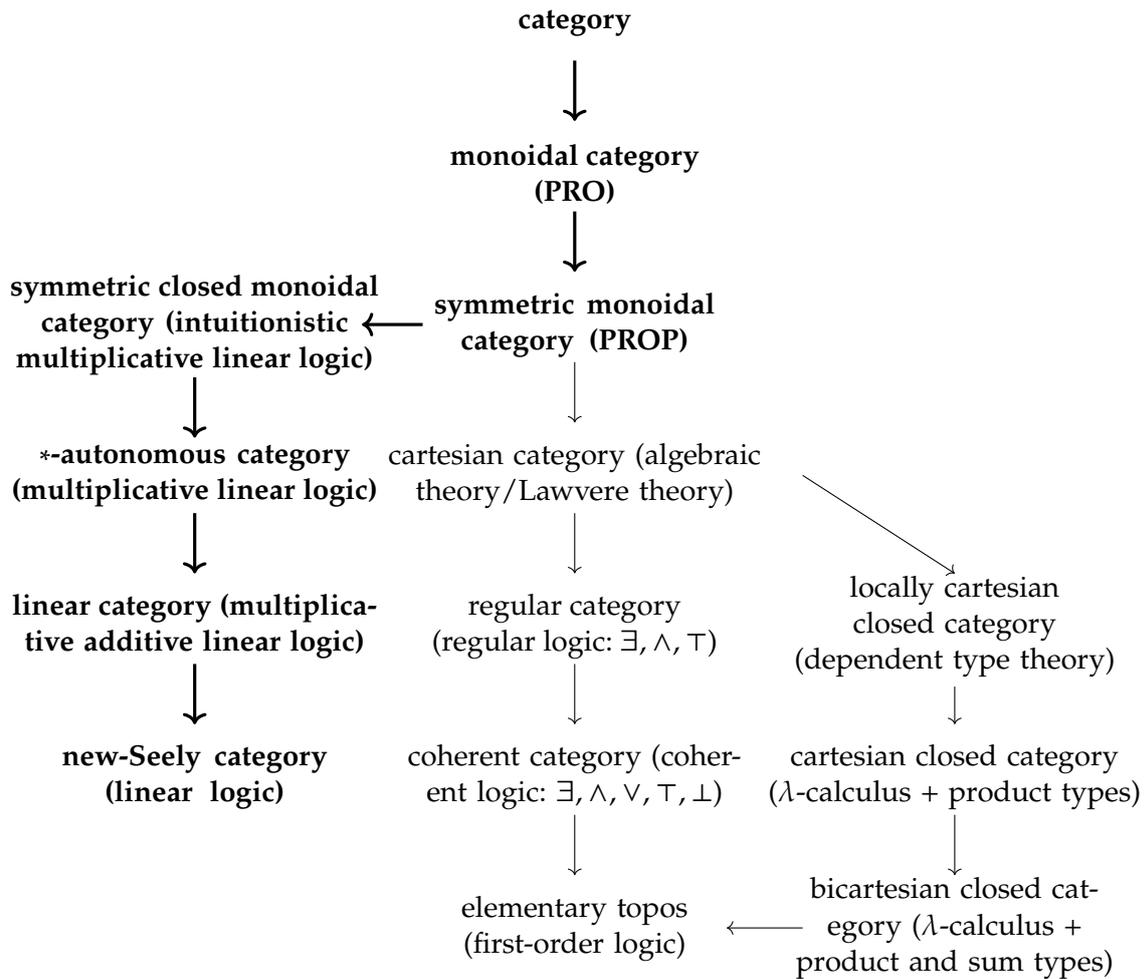
\begin{figure}[t]
    \centering
    \begin{tikzpicture}[every text node part/.style={align=center}]
        \node at (0,0) [] {\textbf{category}};
        \node at (0,-2) [] {\textbf{monoidal category} \\ \textbf{(PRO)}};
        \node at (0,-4) [text width=5cm] {\textbf{symmetric monoidal category (PROP)}};
        \node at (0,-6) [text width=5cm] {cartesian category (algebraic theory/Lawvere theory)};
        \node at (0,-8) [] {regular category \\ (regular logic: $\exists, \wedge, \top$)};
        \node at (0,-10) [text width=5cm] {coherent category (coherent logic: $\exists, \wedge, \vee, \top, \bot$)};
        \node at (0,-12) [text width=5cm] {elementary topos (first-order logic)};

        \node at (5,-8) [text width=5cm] {locally cartesian closed category\\(dependent type theory)};
        \node at (5,-10) [text width=5cm] {cartesian closed category ($\lambda$-calculus + product types)};
        \node at (5,-12) [text width=5cm] {bicartesian closed category ($\lambda$-calculus + product and sum types)};

        \node at (-5,-4) [text width=5cm] {\textbf{symmetric closed monoidal category (intuitionistic multiplicative linear logic)}};
        \node at (-5,-6) [text width=5cm] {\textbf{$*$-autonomous category (multiplicative linear logic)}};
        \node at (-5,-8) [text width=5cm] {\textbf{linear category (multiplicative additive linear logic)}};
        \node at (-5,-10) [text width=5cm] {\textbf{new-Seely category} \\ \textbf{(linear logic)}};

        \draw[->, very thick] (0, -0.5) -- (0, -1.3);
        \draw[->, very thick] (0, -2.5) -- (0, -3.3);
        \draw[->] (0, -4.5) -- (0, -5.3);
        \draw[->] (0, -6.5) -- (0, -7.3);
        \draw[->] (0, -8.5) -- (0, -9.3);
        \draw[->] (0, -10.5) -- (0, -11.3);

        \draw[->, very thick] (-2, -4) -- (-2.8, -4);
        \draw[->, very thick] (-5, -4.7) -- (-5, -5.5);
        \draw[->, very thick] (-5, -6.5) -- (-5, -7.3);
        \draw[->, very thick] (-5, -8.5) -- (-5, -9.3);

        \draw[->] (3, -6) -- (5, -7.3);
        \draw[->] (5, -8.8) -- (5, -9.3);
        \draw[->] (5, -10.5) -- (5, -11.3);
        \draw[->] (3, -12) -- (2, -12);
    \end{tikzpicture}
    \caption{A family tree of categorical logic. The arrows denote inclusions. Adapted from~\textcite{Pat_20}, Figure 1.1.}
    \label{catlogfamtree}
\end{figure}

Categorical logic also helped in the development of algebraic logic, where it was not clear how to deal with quantifiers. Lawvere recognized in the 1960s that quantifiers can be understood as adjoint functors, as presented in Example~\ref{adjoint}, giving a geometric interpretation for first-order logic. Over the years, several more connections to logic were found. Figure~\ref{catlogfamtree} shows an incomplete overview. The path in the middle shows how we can construct a categorical model of classical first-order logic. This structure is called a topos. It is built to be a more general version of sets, where the membership relation is substituted by a so-called subobject classifier. More on the usage of topos theory for the analysis of logic can be read in~\textcite{DBLP:books/daglib/0067551}. The right loop pictures another very important branch, since it marks where the Curry-Howard-Lambek correspondence was discovered. Curry and Howard found the correspondence between proofs in intuitionistic logic and programs of the typed $\lambda$-calculus, which we have briefly discussed above (see also~\textcite{mimram}). Lambek then extended the notion to include also category theory. We will follow the bold path in Section~\ref{sec:catlog}, providing categories successively with enough structure to be a model of linear logic.

\chapter{Foundations of Linear Logic}\label{ch:linlog}

In this chapter, we will present the syntax, proof theory, and semantics of linear logic. For this, we will first think about how the omission of the weakening and contraction rules impact the logic, and define a sequent calculus for linear logic under the consideration that weakening and contraction should not be admissible. We define reasonable fragments of the logic and present a useful structure called proof nets. After this, we will present different approaches to giving a semantics to linear logic fragments. The rest of the chapter is devoted to peculiarities and applications of linear logic.

\section{Syntax and Proof Theory}

In the previous chapter, we have seen that when we impose certain restrictions to the classical sequent calculus, our formal system behaves differently and that this leads to new interpretations of the nature of the objects we are reasoning about. In the case of intuitionistic logic, we no longer consider truth, but provability. Now, what happens if we omit certain structural rules of the classical sequent calculus? Of course, the formal system would again behave differently.

If we omit the weakening and the contraction rule, we can interpret the objects we reason about as \emph{information} or \emph{resources} which cannot be created out of thin air nor dropped as we please. This is exactly what we will do to define linear logic. We will see how this new interpretation of logical primitives leads to interesting challenges in finding a semantics for linear logic which has enough structure to suffice the one induced by the refinement that linear logic provides over classical logics, and at the same time, be modular enough to be able to encompass the rich landscape of fragments of linear logic we can consider. In return, we see that this logic has applications all throughout mathematics and computer science and how the analysis of this logic leads to new insights itself.

Linear logic was introduced in 1987 when Girard published his seminal paper~\cite{Gir_87}. The material in this section stems mostly from this paper, but throughout the years, a number of introductory texts were written~\cite{Be_13, Bra_96, Co_96, Wa_93}, which also helped in developing a more concise and clear presentation of the foundations of linear logic in this thesis. Internet resources used for the presentation include the~\textcite{plato}, the~\textcite{nlab}, and the~\textcite{llwiki}. We will concentrate on propositional linear logic, so when we write linear logic, we always mean propositional linear logic. When adding first-order (or higher order) predicates, we speak of first-order (or higher-order) logic explicitly.

\begin{definition}[Grammar of linear logic]\label{grammarLL}
    The language of \define{linear logic} (\fragment{LL}) is defined by the BNF notation
    \begin{align*}
        A & \Coloneqq p \bnfsep p^{\bot} \bnfsep A^\bot&\\
        &\bnfsep A\otimes A \bnfsep A\oplus A&\text{\emph{tensor} and \emph{plus}}\\
        &\bnfsep A \with A \bnfsep A \parr A&\text{\emph{with} and \emph{par}}\\
        &\bnfsep 1 \bnfsep 0 \bnfsep \top \bnfsep \bot&\text{\emph{units}}\\
        &\bnfsep !A \bnfsep ?A&\text{\emph{of-course} and \emph{why-not}},
    \end{align*}
    where $\cdot^\bot$ is called \emph{negation} and $p$ and $p^{\bot}$ range over the logical atoms.
\end{definition}

The reason behind the ``duplication'' of connectives and the need for the modalities becomes clear when we closer examine the sequent calculus for linear logic given below. For now, just note that we have for each binary connective a neutral element, the correspondence of the neutral elements to the connectives can be seen in Table~\ref{neutralel}.

\begin{table}[ht!]
    \centering
    \caption{Neutral elements}\label{neutralel}
    \begin{tabular}{l l}
        \toprule
        \textbf{Connective} & \textbf{Element}\\ \hline
        $\otimes$ & $1$\\
        $\oplus$ & $0$\\
        $\with$ & $\top$\\
        $\parr$ & $\bot$\\
        \bottomrule
    \end{tabular}
\end{table}

One of the most useful connectives in classical logic is implication since it plays a key part in the deduction process. So it makes sense to define a pendant for linear logic, given in Definition~\ref{implicationLL}. Intuitively, implication in classical logic forces $\psi$ to be true if $\phi$ is true. In intuitionistic logic, where we deal with proofs, the interpretation is that implication is a function which maps a proof of $\phi$ to a proof of $\psi$. In linear logic, where we view primitives as resources, linear implication can be viewed as constructing $\psi$ from $\phi$ \emph{and consuming} $\phi$. This intuition will lead to some major insights later on.

\begin{definition}[Linear implication]\label{implicationLL}
    For two formulas $A$ and $B$, we define the \emph{(multiplicative) linear implication} as $A\multimap B \coloneqq A^{\bot} \parr B$.
\end{definition}

We are now presented with two (equivalent) ways of defining the calculus: We could give a two-sided calculus, with two rules for negation of the form
\[
\prftree[r]{(neg1)}
    {\Gamma\vdash A,\Delta}
    {\Gamma, A^\bot\vdash \Delta}\phantom{xxx}
\prftree[r]{(neg2)}
    {\Gamma, A\vdash \Delta}
    {\Gamma \vdash A^\bot, \Delta}.\]
From these, the de Morgan dualities would follow. The other way is that we construct a one-sided calculus, where we have to define the de Morgan equalities, but have much less inference rules to cover. We will execute the latter idea.

\begin{definition}[Linear negation]\label{negationLL}
    Given an atom $p$ and \fragment{LL}-propositions $A$ and $B$, we define the \define{linear negation} $A^\bot$, as a \define[de Morgan duality!linear]{de Morgan duality} inductively as\vspace{-2em}
    \begin{center}
        \begin{minipage}[b]{0.4\textwidth}
            \begin{align*}
                1^\bot &\coloneq \bot\\
                \top^\bot &\coloneq 0\\
                (p)^\bot &\coloneq p^\bot\\
                (A\otimes B)^\bot &\coloneq A^\bot \parr B^\bot\\
                (A\with B)^\bot &\coloneq A^\bot \oplus B^\bot\\
                (!A)^\bot &\coloneq\ ?A^\bot\\
            \end{align*}
        \end{minipage}
        \begin{minipage}[b]{0.4\textwidth}
            \begin{align*}
                \bot^\bot &\coloneq 1\\
                0^\bot &\coloneq \top\\
                (p^\bot)^\bot &\coloneq p\\
                (A\parr B)^\bot &\coloneq A^\bot \otimes B^\bot\\
                (A\oplus B)^\bot &\coloneq A^\bot \with B^\bot\\
                (?A)^\bot &\coloneq\ !A^\bot.\\
            \end{align*}
        \end{minipage}
    \end{center}\vspace{-4.5em}\qedhere
\end{definition}

\subsection{Sequent Calculus}

We now focus on the heart of linear logic from a proof-theoretical standpoint: the sequent calculus. For clarity, we will present the main ideas of the construction in two-sided form but switch to a one-sided presentation later on. Notice that, when constructing a calculus for linear logic, conjunction and disjunction are each definable in four ways, with two ways for the left and the right rules, respectively. For conjunction, we have

\[\begin{array}{c c}
    \prftree[r]{(L$\wedge$)}{\Gamma, A_i \vdash \Delta}{\Gamma, A_0\wedge A_1 \vdash \Delta} & \prftree[r]{(L$\wedge'$)}{\Gamma, A_0, A_1 \vdash \Delta}{\Gamma, A_0 \wedge A_1\vdash \Delta}\\
    &\\
    \prftree[r]{(R$\wedge$)}{\Gamma, A_0 \vdash \Delta}{\Gamma, A_1 \vdash \Delta}{\Gamma\vdash A_0\wedge A_1, \Delta} & \prftree[r]{(R$\wedge'$)}{\Gamma_0 \vdash A_0, \Delta_0}{\Gamma_1\vdash A_1, \Delta_1}{\Gamma_0, \Gamma_1 \vdash A_0 \wedge A_1, \Delta_0, \Delta_1},\\
\end{array}\]

where $i\in\{\,0, 1\,\}$. So we have to consider the pairs $(\text{L}\wedge,\text{R}\wedge), (\text{L}\wedge,\text{R}\wedge'), (\text{L}\wedge',\text{R}\wedge)$ and $(\text{L}\wedge',\text{R}\wedge')$. But the middle two of these cases lead to versions of weakening and contraction being derivable. Thus, we allow the first and last case to characterize two (distinct!) variants of conjunction. The difference between these variants is that the first one can be regarded as \emph{context-free}, while the second one can be regarded as \emph{context-sharing}. We assign them the symbols $\otimes$ and $\with$ respectively. Executing the same ideas for disjunction, we get the connectives $\parr$ and $\oplus$ with their corresponding inference rules.

Lastly, the modalities $!$ and $?$ are added to provide a guarded re-introduction of weakening and contraction for certain formulas. They behave roughly like the modalities $\Box$ and $\Diamond$ of the modal logic \textsf{S4}. As their notation suggests, they are dual to each other, $!A$ can be thought of as ``$A$ can be used zero, one or many times'' while $?A$ can be thought of as ``$A$ can be obtained zero, one or many times''. We can now present the (already one sided) sequent calculus of linear logic. By the well known process of ``abuse of notation'', $!\Gamma$ (or $?\Gamma$) means that we write $!$ (or $?$) before every formula in $\Gamma$.

\begin{definition}[Sequent calculus of linear logic]\label{seq}
    The sequent calculus of linear logic consists of the following rules:

    \textbf{Axiom}
    \begin{displaymath}
        \prftree[r]{(id)}
            {}
            {\vdash A, A^\bot}
    \end{displaymath}

    \textbf{Structural rules}
    \begin{center}
        \begin{minipage}[b]{0.3\textwidth}
            \begin{displaymath}
                \prftree[r]{(exchange)}
                    {\vdash \Gamma}
                    {\vdash \Gamma'}
            \end{displaymath}
        \end{minipage}
        \begin{minipage}[b]{0.3\textwidth}
            \begin{displaymath}
                \prftree[r]{(cut)}
                    {\vdash \Gamma, A}
                    {\vdash A^{\bot}, \Delta}
                    {\vdash \Gamma, \Delta}
            \end{displaymath}
        \end{minipage}
    \end{center}
    \phantom{xxxxx}where $\Gamma'$ is a permutation of $\Gamma$.\\[0.5ex]

    \textbf{Logical rules}
    \begin{itemize}
    \item Additive rules
        \begin{displaymath}
            \prftree[r]{(truth)}
                {}
                {\vdash \Gamma, \top}
        \end{displaymath}

        \begin{center}
            \begin{minipage}[b]{0.3\textwidth}
                \begin{displaymath}
                    \prftree[r]{($\with$)}
                        {\vdash \Gamma, A}
                        {\vdash \Gamma, B}
                        {\vdash \Gamma, A\with B}
                \end{displaymath}
            \end{minipage}
            \begin{minipage}[b]{0.3\textwidth}
                \begin{displaymath}
                    \prftree[r]{($\oplus_1$)}
                        {\vdash \Gamma, A}
                        {\vdash \Gamma, A\oplus B}
                \end{displaymath}
            \end{minipage}
            \begin{minipage}[b]{0.3\textwidth}
                \begin{displaymath}
                    \prftree[r]{($\oplus_2$)}
                        {\vdash \Gamma, B}
                        {\vdash \Gamma, A\oplus B}
                \end{displaymath}
            \end{minipage}
        \end{center}

    \item Multiplicative rules
        \begin{center}
            \begin{minipage}[b]{0.3\textwidth}
                \begin{displaymath}
                    \prftree[r]{(one)}
                        {}
                        {\vdash 1}
                \end{displaymath}
            \end{minipage}
            \begin{minipage}[b]{0.3\textwidth}
                \begin{displaymath}
                    \prftree[r]{(false)}
                        {\vdash \Gamma}
                        {\vdash \Gamma, \bot}
                \end{displaymath}
            \end{minipage}
        \end{center}
        \begin{center}
            \begin{minipage}[b]{0.3\textwidth}
                \begin{displaymath}
                    \prftree[r]{($\otimes$)}
                        {\vdash \Gamma, A}
                        {\vdash \Gamma, B}
                        {\vdash \Gamma, A\otimes B}
                \end{displaymath}
            \end{minipage}
            \begin{minipage}[b]{0.3\textwidth}
                \begin{displaymath}
                    \prftree[r]{($\parr$)}
                        {\vdash \Gamma, A, B}
                        {\vdash \Gamma, A\parr B}
                \end{displaymath}
            \end{minipage}
        \end{center}
    
    \item Exponential rules
        \begin{center}
            \begin{minipage}[t]{0.3\textwidth}
                \begin{displaymath}
                    \prftree[r]{(of course)}
                        {\vdash\ ?\Gamma, A}
                        {\vdash\ ?\Gamma, !A}
                \end{displaymath}
            \end{minipage}
            \begin{minipage}[t]{0.3\textwidth}
                \begin{displaymath}
                    \prftree[r]{(weakening)}
                        {\vdash \Gamma}
                        {\vdash \Gamma, ?A}
                \end{displaymath}
            \end{minipage}
        \end{center}
        \begin{center}
            \begin{minipage}[t]{0.3\textwidth}
                \begin{displaymath}
                    \prftree[r]{(dereliction)}
                        {\vdash \Gamma, A}
                        {\vdash \Gamma, ?A}
                \end{displaymath}
            \end{minipage}
            \begin{minipage}[t]{0.3\textwidth}
                \begin{displaymath}
                    \prftree[r]{(contraction)}
                        {\vdash \Gamma, ?A, ?A}
                        {\vdash \Gamma, ?A}
                \end{displaymath}
            \end{minipage}
        \end{center}\qedhere
    \end{itemize}
\end{definition}

\begin{remark}
    Since we will not extensively study first-order linear logic in this thesis, we omitted the inference rules for the quantifiers. Nonetheless, their construction is straightforward. They behave akin to the context sharing operators.
\end{remark}

When we now remind ourselves of the controversy around the law of excluded middle in the context of linear logic, we already note a fascinating facette of the calculus we just defined: the main argument can be formulated as whether $\vdash \phi \vee \neg\phi$ should be derivable or not. But in linear logic, we have two versions of disjunction and, lo and behold, $\vdash A\parr A^\bot$ is trivially derivable in linear logic, while $\vdash A\oplus A^\bot$ is not. We have, in some sort, the best of both worlds: The symmetry of classical logic (for example, negation is involutive) together with intuitionistic notions in the additive fragment (for example, additive disjunction suffices the disjunction property). Further examples of provable formulas and equivalences are listed in Table~\ref{appendix:2} and Table~\ref{appendix:3}.

The additive rules provide a notion of weakening, but only in respect to their own connectives. This does not imply weakening for the whole sequent calculus, since it is defined in a ``multiplicative'' way: for a sequent $A_1, \dots, A_n \vdash B_1, \dots, B_m$, the commas on the left represent $\otimes$, the commas on the right represent $\parr$, and $A_1, \dots, A_n \vdash B_1, \dots, B_m$ is derivable if and only if the formula $!(A_1, \dots, A_n) \multimap (B_1, \dots, B_m)$ is.

Apart from the classification of connectives into additive, multiplicative and exponential parts, we can also classify them on the grounds of their \define{polarity}. \textcite{DBLP:journals/logcom/Andreoli92} first viewed proof search in linear logic as a computational task in which we start from the formula we want to prove and read inference rules bottom-up. In this process, formulas may or may not interact with their environment at inferences. Based on this he classified a formula as \define[formula!asynchronous]{asynchronous} (which we will call \define[formula!negative]{negative}) if its top level connective is $\top$, $\with$, $\bot$, $\parr$ or $?$, and as \define[formula!synchronous]{synchronous} (which we will call \define[formula!positive]{positive}) otherwise. Notice how, the formula $(A\parr B)$ simply splits into $A$ and $B$, but that a provable sequent with $\oplus$ as a top connective could evolve into a non-provable sequent. Since in the one-sided sequent, we have only right introduction rules, we can classify derivation steps into (a-)synchronous phases. At the synchronous phase, a formula is selected and becomes the focus of this phase. This proof search technique, called focussing, is very important for proof search in linear logic. It provides a normal form and an abstract model of computation. In conclusion, we arrive at the characterization of connectives given in Table~\ref{classconn}.

\begin{table}[ht!]
    \centering
    \caption{Classification of connectives}\label{classconn}
    \begin{tabular}{l l l l}
        \toprule
        & \textbf{additive} & \textbf{multiplicative} & \textbf{exponential}\\ \hline
        \textbf{positive} & $\oplus, 0$ & $\otimes, 1$ & !\\
        \textbf{negative} & $\with, \top$ & $\parr, \bot$ & ?\\
        \bottomrule
    \end{tabular}
\end{table}

\subsection{Fragments}

The classification of the connectives into additive, multiplicative and exponential connectives also forms the basis of one of the two classes of fragments we will consider in this thesis. The definition of fragments helps us in finding strucutral properties of the logic, and can provide restrictions to the provability problem which make it easier to solve. To avoid confusion, we will now establish a naming system for a selection of fragments we can define. First, we list the allowed connectives for the base fragments in Table~\ref{basicfrag}.

\begin{table}[ht!]
    \centering
    \caption{Base cases of fragments}\label{basicfrag}
    \begin{tabular}{l l}
        \toprule
        \textbf{Name} & \textbf{Connectives}\\ \hline
        \fragment{MLL} & $\cdot^\bot, \otimes, \parr, 1, \top$\\
        \fragment{ALL} & $\cdot^\bot, \oplus, \with, 0, \bot$\\
        \fragment{ELL} & $\cdot^\bot, !, ?$\\
        \bottomrule
    \end{tabular}
\end{table}

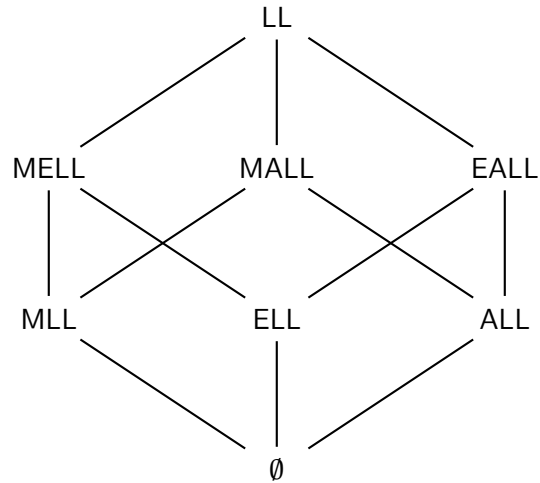
\begin{figure}[t]
    \centering
    \begin{tikzpicture}
        \node at (3,0) [] {$\emptyset$};
    
        \node at (0,2) [] {\fragment{MLL}};
        \node at (3,2) [] {\fragment{ELL}};
        \node at (6,2) [] {\fragment{ALL}};
    
        \node at (0,4) [] {\fragment{MELL}};
        \node at (3,4) [] {\fragment{MALL}};
        \node at (6,4) [] {\fragment{EALL}};

        \node at (3,6) [] {\fragment{LL}};

        \draw [thick, shorten >= 5mm, shorten <= 5mm] (3,0)  -- (0,2);
        \draw [thick, shorten >= 3mm, shorten <= 3mm] (3,0)  -- (3,2);
        \draw [thick, shorten >= 5mm, shorten <= 5mm] (3,0)  -- (6,2);
            
        \draw [thick, shorten >= 3mm, shorten <= 3mm] (0,2)  -- (0,4);
        \draw [thick, shorten >= 5mm, shorten <= 5mm] (0,2)  -- (3,4);
            
        \draw [thick, shorten >= 5mm, shorten <= 5mm] (3,2)  -- (0,4);
        \draw [thick, shorten >= 5mm, shorten <= 5mm] (3,2)  -- (6,4);
            
        \draw [thick, shorten >= 5mm, shorten <= 5mm] (6,2)  -- (3,4);
        \draw [thick, shorten >= 3mm, shorten <= 3mm] (6,2)  -- (6,4);

        \draw [thick, shorten >= 5mm, shorten <= 5mm] (3,6)  -- (0,4);
        \draw [thick, shorten >= 3mm, shorten <= 3mm] (3,6)  -- (3,4);
        \draw [thick, shorten >= 5mm, shorten <= 5mm] (3,6)  -- (6,4);
    \end{tikzpicture}
    \caption{Lattice of linear logic fragments.}
    \label{ll-frag}
\end{figure}

We can of course consider all possible combinations of additives, multiplicatives and exponentials, which is how we obtain the lattice shown in Figure~\ref{ll-frag}. Trivially, the expressiveness of the logic -- and thus its complexity -- increases when we go up the lattice. The complexity of the fragments shown in this lattice will be the main consideration in this thesis.

Nevertheless, we will leave some remarks on other fragments which we do not consider in detail. For this, we extend the notation, in that when we consider a fragment $\mathcal{X}\fragment{LL}$ from the lattice, but without its units, we denote it as $\mathcal{X}\fragment{LL}^-$. When we consider the fragment extended by $n$th-order quantifiers, we denote it as $\mathcal{X}\fragment{LL}_n$. Sometimes, we allow weakening or contraction back into the fragments. If this is the case we write $\mathcal{X}\fragment{LLw}$ and $\mathcal{X}\fragment{LLc}$, respectively. If we demand the fragment to be intuitionistic, we prepend an $\textsf{I}$, i.\,e. we write $\fragment{I}\mathcal{X}\fragment{LL}$. Non-commutative variants ($\textsf{Nc}\mathcal{X}\fragment{LL}$) are also considered in the literature. Lastly, adaptations of linear logic for complexity-theoretic considerations are given by elementary linear logic (\fragment{ElemLL}), light linear logic (\fragment{LLL}), and soft linear logic (\fragment{SLL}). An outlook about their usages is given in Section~\ref{llapplications}.

\subsubsection{Horn Fragments}

The importance of the other class of fragments we consider stems from their extensive usage in logic programming: linear Horn fragments. The fragments extend naturally to linear logic. \textcite{DBLP:journals/apal/Kanovich94} studied their complexity extensively and arrived at a full characterization of the respective lattice. Another positive aspect of the Horn fragments is that Kanovich found a model of the fragments, \reftodef[branching Horn program]{branching Horn programs}. There exists a duality between the Horn fragments and Horn programs, which, like with their classical counterpart, leads to applications in logic programming. The examination of the complexity of these fragments reveals some results which differ from their classical counterpart, whence we present their complexity in this thesis. 

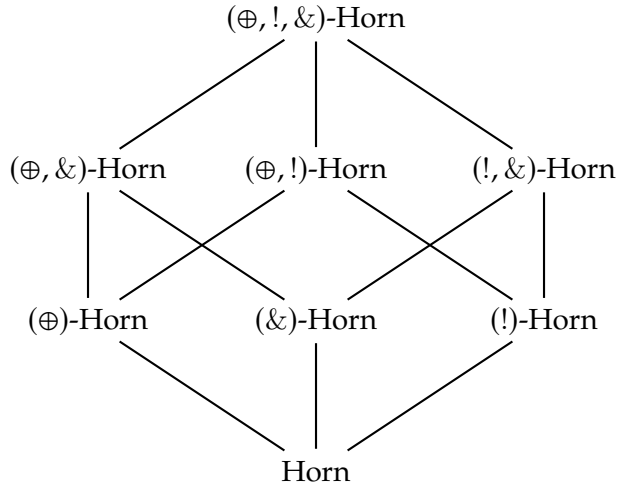
\begin{figure}[t]
    \centering
    \begin{tikzpicture}
        \node at (3,0) [] {Horn};
        
        \node at (0,2) [] {$(\oplus)$-Horn};
        \node at (3,2) [] {$(\with)$-Horn};
        \node at (6,2) [] {$(!)$-Horn};
        
        \node at (0,4) [] {$(\oplus, \with)$-Horn};
        \node at (3,4) [] {$(\oplus, !)$-Horn};
        \node at (6,4) [] {$(!, \with)$-Horn};

        \node at (3,6) [] {$(\oplus, !, \with)$-Horn};

        \draw [thick, shorten >= 5mm, shorten <= 5mm] (3,0)  -- (0,2);
        \draw [thick, shorten >= 3mm, shorten <= 3mm] (3,0)  -- (3,2);
        \draw [thick, shorten >= 5mm, shorten <= 5mm] (3,0)  -- (6,2);
            
        \draw [thick, shorten >= 3mm, shorten <= 3mm] (0,2)  -- (0,4);
        \draw [thick, shorten >= 5mm, shorten <= 5mm] (0,2)  -- (3,4);
            
        \draw [thick, shorten >= 5mm, shorten <= 5mm] (3,2)  -- (0,4);
        \draw [thick, shorten >= 5mm, shorten <= 5mm] (3,2)  -- (6,4);
            
        \draw [thick, shorten >= 5mm, shorten <= 5mm] (6,2)  -- (3,4);
        \draw [thick, shorten >= 3mm, shorten <= 3mm] (6,2)  -- (6,4);

        \draw [thick, shorten >= 5mm, shorten <= 5mm] (3,6)  -- (0,4);
        \draw [thick, shorten >= 3mm, shorten <= 3mm] (3,6)  -- (3,4);
        \draw [thick, shorten >= 5mm, shorten <= 5mm] (3,6)  -- (6,4);
    \end{tikzpicture}
    \caption{Lattice of linear Horn fragments.}
    \label{horn}
\end{figure}

The definition of linear Horn sequents runs analogously to the classical case: We write $W, \Gamma \vdash Z$, where $\Gamma$ is a multiset of linear Horn clauses, and $W$ and $Z$ are simple conjunctions, meaning the multiplicative conjunction of positive literals. When we let simple conjunctions $X$ and $Y$ represent multisets $L$ and $M$, we can model the union of $L$ and $M$ by $X\otimes Y$, the difference $L - M$ by $X - Y$, and containment $L\subseteq M$ by $X\subseteq Y$. When we have $X \subseteq Y$ and $Y\subseteq X$, we write $X = Y$. A linear Horn clause $X\multimap Y$ means that given $X$, $Y$ can be computed, while $X$ is consumed in the process. We will inspect the lattice of Horn fragments depicted in Figure~\ref{horn}.

When considering the computational interpretation of Horn fragments, the $\oplus$-connective can be perceived as non-deterministic branching, the $\with$-connective as non-deterministic choice, and the $!$-modality as reuse of certain resources. We use the following definition.

\begin{definition}[Generalized Horn sequents]
    The eight variants of \define[Horn implication!generalized]{generalized Horn implications} are defined as follows:
    \begin{enumerate}
        \item A Horn implication is a formula of the form $(X\multimap Y)$,
        \item a $(\oplus)$-Horn implication is a formula of the form $(X\multimap (Y_1 \oplus Y_2))$,
        \item and a $(\with)$-Horn implication is a formula of the form $((X_1\multimap Y_1) \with (X_2\multimap Y_2))$.
    \end{enumerate}
    From these, \define[sequent!generalized Horn]{generalized Horn sequents} are defined in the following way.
    \begin{enumerate}
        \item For a multiset $\Gamma$ of Horn implications, a sequent of the form $W, \Gamma \vdash Z$ is called a Horn sequent, and a sequent of the form $W, !\Gamma \vdash Z$ is called an $!$-Horn sequent.
        \item Let $\diamondsuit \in \{\,\oplus, \with\,\}$. For a multiset $\Gamma$ of Horn or $(\diamondsuit)$-Horn implications, a sequent of the form $W, \Gamma\vdash Z$ is called a Horn or an $(\diamondsuit)$-Horn sequent, and a sequent of the form $W, !\Gamma \vdash Z$ is called an $(!, \diamondsuit)$-Horn sequent.
        \item For a multiset $\Gamma$ of generalized Horn implications, a sequent of the form $W, \Gamma\vdash Z$ is called a $(\oplus, \with)$-Horn sequent.\qedhere
    \end{enumerate}
\end{definition}

For simplicity, we will work with Horn fragments extended with the weakening rule in some proofs. The following corollary gives an efficient embedding of the former to the latter.

\begin{corollary}\label{hornweak}
    We can construct two polynomial time algorithms transforming $(\oplus, \with)$-Horn multisets $\Gamma$ into multisets $\Gamma^{+\with}$ and multisets $\Gamma^{-\with}$, respectively, such that
    \begin{itemize}
        \item The multiset $\Gamma^{+\with}$ emerges from $\Gamma$ by replacing every formula $A$ by a certain multiset $A^{+\with}$.
        \item The multiset $\Gamma^{-\with}$ emerges from $\Gamma$ by replacing every formula $A$ by a certain multiset $A^{-\with}$.
        \item If $A$ is either a Horn implication or a $(\with)$-Horn implication, then $A^{+\with}$ consists of $(\with)$-Horn implications, and $A^{-\with}$ consists only of Horn implications.
        \item If $A$ is a $(\oplus)$-Horn implication, then $A^{+\with}$ consists of $(\oplus)$-Horn implications and $(\with)$-Horn implications, and $A^{-\with}$ consists of $(\oplus)$-Horn implications.
        \item The sequent
        \[W, \Gamma \vdash Z\]
        is derivable in linear logic with the weakening rule if and only if the sequent
        \[W, \Gamma^{+\with} \vdash Z\]
        is derivable in linear logic.
        \item The sequent
        \[W, \Gamma \vdash Z\]
        is derivable in linear logic if and only if the sequent
        \[W, \Gamma^{-\with} \vdash Z\]
        is derivable in linear logic with the weakening rule.
    \end{itemize}
\end{corollary}

\subsection{Proof Nets}\label{ssec:proofnets}

Especially in intuitionistic logic, \emph{natural deduction} is often used instead of the sequent calculus. This form of deduction has the advantage that it simplifies proofs drastically. In linear logic, a similar concept was developed, named \emph{proof nets}. When we consider derivations of the form

\begin{displaymath}
    \prftree[r, double]{}
        {\vdash A_1, A_2, \dots,A_n}
        {\vdash (A_1 \parr A_2),\dots, (A_{n-1} \parr A_n)},
\end{displaymath}
it becomes apparent that there are many ways to achieve the derivation which differ only in uninteresting ways, namely the choice of the sequence in which the $(\parr)$ rule is applied. This is due to the fact that in the sequent calculus, we have to choose a linear order on the set of rules we apply, even if they do not interfere with one another.

To abstract away this unnecessary information, Girard conceived proof nets, which can be formalized as directed hypergraphs and enjoy several nice properties for the \fragment{MLL} fragment. We will therefore give a short overview of the concept, which also provides us with a first complexity result regarding the verification of proofs in \fragment{MLL}.

We start by associating a link to every inference rule in \fragment{MLL}. These links will become the hyperedges of the hypergraph that represents the proof net.

\begin{definition}[Proof links]\label{def:prfl}
    We define \define{proof links} as follows.
    \begin{enumerate}
        \item For the axiom rule, we associate the \emph{axiom link}.
        \begin{center}
            \begin{minipage}[b]{0.3\textwidth}
                \begin{displaymath}
                    \prftree[r]{(id)}
                        {}
                        {\vdash A, A^\bot}
                \end{displaymath}
            \end{minipage}
            \begin{minipage}[b]{0.3\textwidth}
                \centering
                \begin{tikzpicture}
                    \node at (0,0) [] {$A$};
                    \node at (1,0) [] {$A^\bot$};
                    \draw  [ultra thick, stealth-stealth] (0, 0.2) -- (0, 0.5) -- (1, 0.5) -- (1, 0.2);
                \end{tikzpicture}
            \end{minipage}
        \end{center}
        \item For the cut rule, we associate the \emph{cut link}.
        \begin{center}
            \begin{minipage}[b]{0.3\textwidth}
                \begin{displaymath}
                    \prftree[r]{(cut)}
                        {\prftree[noline]{\prftree[noline]{\Pi_1}{\vdots}}{\vdash \Gamma, A}}
                        {\prftree[noline]{\prftree[noline]{\Pi_2}{\vdots}}{\vdash \Delta, A^\bot}}
                        {\vdash \Gamma, \Delta}
                \end{displaymath}
            \end{minipage}
            \begin{minipage}[b]{0.3\textwidth}
                \centering
                \begin{tikzpicture}
                    \node at (0,0) [] {$\Gamma$};
                    \node at (1,0) [] {$A$};
                    \node at (2,0) [] {$A^\bot$};
                    \node at (3,0) [] {$\Delta$};
                    \node at (0.5,0.5) [gray] {$N_1$};
                    \node at (2.5,0.5) [gray] {$N_2$};
                    \draw  [>=stealth, >-, ultra thick] (1, -0.2) -- (1, -0.5) -- (1.3, -0.5);
                    \draw  [>=stealth, -<, ultra thick] (1.7, -0.5) -- (2, -0.5) -- (2, -0.2);
                    \draw  [thick, gray, rounded corners=5pt] (0, 0.25) -- (1, 0.25) -- (1, 0.75) -- (0, 0.75) -- cycle;
                    \draw  [thick, gray, rounded corners=5pt] (2, 0.25) -- (3, 0.25) -- (3, 0.75) -- (2, 0.75) -- cycle;
                    \node at (1.5, -0.5) [] {\ding{34}};
                \end{tikzpicture}
            \end{minipage}
        \end{center}
        \item For the $(\otimes)$ rule, we associate the \emph{tensor link}.
        \begin{center}
            \begin{minipage}[b]{0.3\textwidth}
                \begin{displaymath}
                    \prftree[r]{$(\otimes)$}
                        {\prftree[noline]{\prftree[noline]{\Pi_1}{\vdots}}{\vdash \Gamma, A}}
                        {\prftree[noline]{\prftree[noline]{\Pi_2}{\vdots}}{\vdash \Delta, B}}
                        {\vdash \Gamma, \Delta, A\otimes B}
                \end{displaymath}
            \end{minipage}
            \begin{minipage}[b]{0.3\textwidth}
                \centering
                \begin{tikzpicture}
                    \node at (0,0) [] {$\Gamma$};
                    \node at (1,0) [] {$A$};
                    \node at (2,0) [] {$B$};
                    \node at (3,0) [] {$\Delta$};
                    \node at (0.5,0.5) [gray] {$N_1$};
                    \node at (2.5,0.5) [gray] {$N_2$};
                    \node at (1.5, -1) [] {$A\otimes B$};
                    \draw  [>=stealth, >-, ultra thick] (1, -0.2) -- (1, -0.5) -- (1.4, -0.5);
                    \draw  [>=stealth, -<, ultra thick] (1.6, -0.5) -- (2, -0.5) -- (2, -0.2);
                    \draw  [>=stealth, ->, ultra thick] (1.5, -0.6) -- (1.5, -0.85);
                    \draw  [thick, gray, rounded corners=5pt] (0, 0.25) -- (1, 0.25) -- (1, 0.75) -- (0, 0.75) -- cycle;
                    \draw  [thick, gray, rounded corners=5pt] (2, 0.25) -- (3, 0.25) -- (3, 0.75) -- (2, 0.75) -- cycle;
                    \node at (1.5, -0.5) [] {$\otimes$};
                \end{tikzpicture}
            \end{minipage}
        \end{center}\newpage
        \item For the $(\parr)$ rule, we associate the \emph{par link}.
        \begin{center}
            \begin{minipage}[b]{0.3\textwidth}
                \begin{displaymath}
                    \prftree[r]{$(\parr)$}
                        {\prftree[noline]{\prftree[noline]{\Pi}{\vdots}}{\vdash \Gamma, A, B}}
                        {\vdash \Gamma, A\parr B}
                \end{displaymath}
            \end{minipage}
            \begin{minipage}[b]{0.3\textwidth}
                \centering
                \begin{tikzpicture}
                    \node at (0,0) [] {$\Gamma$};
                    \node at (1,0) [] {$A$};
                    \node at (2,0) [] {$B$};
                    \node at (1.5, -1.1) [] {$A\parr B$};

                    \draw  [>=stealth, >-, ultra thick] (1, -0.2) -- (1, -0.5) -- (1.35, -0.5);
                    \draw  [>=stealth, -<, ultra thick] (1.65, -0.5) -- (2, -0.5) -- (2, -0.2);
                    \draw  [>=stealth, ->, ultra thick] (1.5, -0.65) -- (1.5, -0.9);
                    \node at (1.49, -0.51) [] {\fontsize{8}{8}$\parr$};
                    \draw (1.5, -0.5) circle (0.15);

                    \node at (1,0.5) [gray] {$N$};
                    \draw  [thick, gray, rounded corners=5pt] (0, 0.25) -- (2, 0.25) -- (2, 0.75) -- (0, 0.75) -- cycle;
                \end{tikzpicture}
            \end{minipage}
        \end{center}
    \end{enumerate}
    Note that there is no translation of the exchange rule. This is because it has no effect on this graphical representation.
\end{definition}

\begin{figure}[h]
    \centering
    \begin{tikzpicture}
        \node at (0,0) [] {$A^\bot \parr B$};
        \node at (2,0) [] {$A\otimes B^\bot$};
        \node at (1.5,1) [] {$A$};
        \node at (2.5,1) [] {$B^\bot$};
        \node at (4,0) [] {$B$};

        \draw  [>=stealth, >-, ultra thick] (1.5, 0.8) -- (1.5,0.5) -- (1.9,0.5);
        \draw  [>=stealth, -<, ultra thick] (2.1,0.5) -- (2.5,0.5) -- (2.5,0.8);
        \draw  [>=stealth, ->, ultra thick] (2,0.4) -- (2,0.15);
        \node at (2,0.5) [] {$\otimes$};

        \draw  [>=stealth, <->, ultra thick] (2.5, 1.2) -- (2.5, 1.5) -- (4, 1.5) -- (4, 0.2);

        \draw [>=stealth, >-, ultra thick] (0, -0.2) -- (0, -0.5) -- (0.8, -0.5);
        \draw [>=stealth, -<, ultra thick] (1.2, -0.5) -- (2, -0.5) -- (2, -0.2);
        \node at (1,-0.5) [] {\ding{34}};
        
        \node at (8,0) [] {$A$};
        \node at (10,0) [] {$A^\bot$};

        \draw  [>=stealth, <->, ultra thick] (8, 0.2) -- (8, 0.5) -- (10, 0.5) -- (10, 0.2);

        \draw [>=stealth, >-, ultra thick] (8, -0.2) -- (8, -0.5) -- (8.8, -0.5);
        \draw [>=stealth, -<, ultra thick] (9.2, -0.5) -- (10, -0.5) -- (10, -0.2);
        \node at (9,-0.5) [] {\ding{34}};

    \end{tikzpicture}
    \caption{Examples of proof structures.}
    \label{proofstruct}
\end{figure}

\begin{figure}[t]
    \centering
    \begin{tikzpicture}
        \node at (0,0) [] {$\rightsquigarrow$};
        \node at (0,2.5) [] {$\rightsquigarrow$};
        \node at (0,4.5) [] {$\rightsquigarrow$};
        \node at (0,6) [] {$\rightsquigarrow$};

        \node at (-3,6) [] {$A$};
        \node at (-2,6) [] {$A^\bot$};
        \draw  [>=stealth, <->, ultra thick] (-3, 6.2) -- (-3, 6.5) -- (-2, 6.5) -- (-2, 6.2);

        \node at (1.5,6) [] {$A$};
        \node at (2.5,6) [] {$A^\bot$};
        \node at (2,6.5) [gray] {$N$};
        \draw  [thick, gray, rounded corners=5pt] (1.5, 6.25) -- (2.5, 6.25) -- (2.5, 6.75) -- (1.5, 6.75) -- cycle;

        \node at (-4,4.5) [] {$\Gamma$};
        \node at (-3,4.5) [] {$A$};
        \node at (-2,4.5) [] {$A^\bot$};
        \node at (-1,4.5) [] {$\Delta$};
        \node at (-3.5,5) [gray] {$N_1$};
        \node at (-1.5,5) [gray] {$N_2$};
        \draw  [>=stealth, >-, ultra thick] (-3, 4.3) -- (-3, 4) -- (-2.7, 4);
        \draw  [>=stealth, -<, ultra thick] (-2.3, 4) -- (-2, 4) -- (-2, 4.3);
        \draw  [thick, gray, rounded corners=5pt] (-4, 4.75) -- (-3, 4.75) -- (-3, 5.25) -- (-4, 5.25) -- cycle;
        \draw  [thick, gray, rounded corners=5pt] (-2, 4.75) -- (-1, 4.75) -- (-1, 5.25) -- (-2, 5.25) -- cycle;
        \node at (-2.5, 4) [] {\ding{34}};

        \node at (1.5,4.5) [] {$\Gamma$};
        \node at (2.5,4.5) [] {$\Delta$};
        \node at (2,5) [gray] {$N$};
        \draw  [thick, gray, rounded corners=5pt] (1.5, 4.75) -- (2.5, 4.75) -- (2.5, 5.25) -- (1.5, 5.25) -- cycle;

        \node at (-4,2.5) [] {$\Gamma$};
        \node at (-3,2.5) [] {$A$};
        \node at (-2,2.5) [] {$B$};
        \node at (-1,2.5) [] {$\Delta$};
        \node at (-3.5,3) [gray] {$N_1$};
        \node at (-1.5,3) [gray] {$N_2$};
        \node at (-2.5, 1.5) [] {$A\otimes B$};
        \draw  [>=stealth, >-, ultra thick] (-3, 2.3) -- (-3, 2) -- (-2.6, 2);
        \draw  [>=stealth, -<, ultra thick] (-2.4, 2) -- (-2, 2) -- (-2, 2.3);
        \draw  [>=stealth, ->, ultra thick] (-2.5, 1.9) -- (-2.5, 1.65);
        \draw  [thick, gray, rounded corners=5pt] (-4, 2.75) -- (-3, 2.75) -- (-3, 3.25) -- (-4, 3.25) -- cycle;
        \draw  [thick, gray, rounded corners=5pt] (-2, 2.75) -- (-1, 2.75) -- (-1, 3.25) -- (-2, 3.25) -- cycle;
        \node at (-2.5, 2) [] {$\otimes$};

        \node at (1,2.5) [] {$\Gamma$};
        \node at (2,2.5) [] {$A\otimes B$};
        \node at (3,2.5) [] {$\Delta$};
        \node at (2,3) [gray] {$N$};
        \draw  [thick, gray, rounded corners=5pt] (1, 2.75) -- (3, 2.75) -- (3, 3.25) -- (1, 3.25) -- cycle;

        \node at (-4,0) [] {$\Gamma$};
        \node at (-3,0) [] {$A$};
        \node at (-2,0) [] {$B$};
        \node at (-2.5, -1.1) [] {$A\parr B$};
        \draw  [>=stealth, >-, ultra thick] (-3, -0.2) -- (-3, -0.5) -- (-2.65, -0.5);
        \draw  [>=stealth, -<, ultra thick] (-2.35, -0.5) -- (-2, -0.5) -- (-2, -0.2);
        \draw  [>=stealth, ->, ultra thick] (-2.5, -0.65) -- (-2.5, -0.9);
        \node at (-2.51, -0.51) [] {\fontsize{8}{8}$\parr$};
        \draw (-2.5, -0.5) circle (0.15);
        \node at (-3,0.5) [gray] {$N$};
        \draw  [thick, gray, rounded corners=5pt] (-4, 0.25) -- (-2, 0.25) -- (-2, 0.75) -- (-4, 0.75) -- cycle;

        \node at (1,0) [] {$\Gamma$};
        \node at (2,0) [] {$A\parr B$};
        \node at (3,0) [] {$\Delta$};
        \node at (2,0.5) [gray] {$N$};
        \draw  [thick, gray, rounded corners=5pt] (1, 0.25) -- (3, 0.25) -- (3, 0.75) -- (1, 0.75) -- cycle;

    \end{tikzpicture}
    \caption{The reduction rules for the proof net verification.}
    \label{proofnet}
\end{figure}
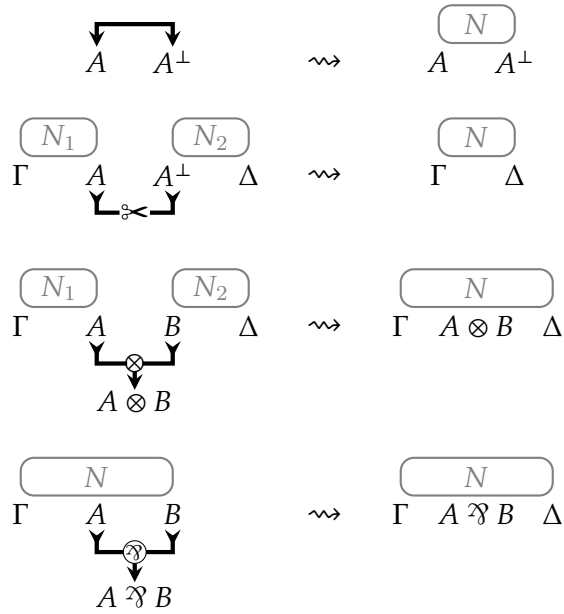

When we apply these links for all rules in our proof inductively, we get a proof structure. Examples of this are depicted in Figure~\ref{proofstruct}. Observe that while the left proof structure is a valid proof of linear logic, namely the elimination of linear implication, the right is not, since we would derive the empty sequent with it. Thus, proof nets need to suffice further correctness criteria to be able to soundly represent linear logic proofs. Over time, various approaches were found to verify the correctness of proof nets. The first one was given by~\textcite{Gir_87}, called \emph{long trip criterion}, but this method required exponential runtime. Danos and Regnier, after finding another criterion called \emph{acyclic connectedness}~\cite{DR89}, which also needed exponential time, developed a criterion that can be na\"{i}vely checked in quadratic time, called \emph{contractibility}. \textcite{GMM_96} used this approach to develop a graph reduction procedure. With this procedure, a hypergraph can be verified to be a valid proof net if it contracts to a singleton node $N$ with the reduction rules given in Figure~\ref{proofnet}.

Quadratic runtime can be na\"{i}vely shown from the fact that for each hyperlink in the graph, we go through the graph to find a fitting reduction candidate to reduce. Later, linear time algorithms for this reduction procedure were found by~\textcite{MO_06} and by~\textcite{G11}. This gives us a first complexity result for linear logic: that the correctness of a proof in \fragment{MLL} can be checked in linear time.

While being useful as a graphical representation of proofs, proof nets for \fragment{MLL} especially have further nice properties regarding cut-elimination which make them a very useful tool for working with \fragment{MLL}. The most prominent properties are that cut-elimination has the Church-Rosser property and that it can be performed very efficiently and locally. Unfortunately, there exists no canonical generalization of proof nets to more general fragments of linear logic at the moment of writing. Especially when also considering the additive fragment, the question of how to present proof net like structures is an active area of research (cf.~\cite{HG_05, HH_16}).

Proof nets also find applications in the research of proof complexity. They were, for example, used by~\textcite{DBLP:phd/hal/Aubert13} to examine sub-polynomial complexity classes such as \class{AC}, \class{NC}, \class{L}, and \class{NL}, tying the proof net representation of linear logic to Boolean circuits.

Effective representations of proof nets is also a field of current research that brings together different areas of mathematics and computer science. An example of this is the paper by~\textcite{DBLP:journals/jar/Acclavio19}, in which he gives an alternative 2-dimensional syntax to proof nets called proof diagrams which are inspired by string diagrams widely used in category theory.

\section{Semantics}

The study of the syntactical part of linear logic has brought forward many interesting results, but until now, the parts of linear logic we have seen are devoid of meaning. Finding a semantics for linear logic and its fragments has proved to be a non-trivial endeavor due to its more complex calculus. We will give a short overview of different approaches to finding a semantics for linear logic. This also shows the variety in which linear logic can be used to formally describe mathematical objects. In particular, we focus on the approaches of phase semantics, which is well suited as a simple introduction, then show a more modular approach in the case of Kripke semantics, and finally present the most common and researched approach to semantics for linear logic: categorical semantics. We finish by giving a short outlook of game semantics, and the Geometry of Interaction, which is a -- as the name suggests -- geometric approach to give semantics to a logic that has its roots in linear logic. It has to be said that there are many other approaches as well, such as coherent semantics, which was developed by Girard in the process of a retrospect at his System F~\cite{GIRARD1986159} or finiteness semantics, which exhibit connections to model checking~\cite{grellois2015finitary}.

\subsection{Phase Semantics}\label{phasesem}

One of the first and simplest semantics of linear logic is given in the original article by~\textcite{Gir_87} and provides a physics flavored interpretation. The main idea behind this semantics is that we construct a phase space, whose underlying monoidal structure can be used to interpret the multiplicative part of linear logic. We then define a subset of phases, called facts, which suffice certain criteria and represent the true formulas. The presentation we give follows~\textcite{L_99}.

Given a multiplicative monoid $M$ and $X, Y \subseteq M$, we write $XY = \{\,xy\mid x\in X \text{ and } y\in Y\,\}$, which we will use to represent the $\otimes$-connective of linear logic in the model. To model linear implication, we define $X\multimap Y = \{\,z\in M\mid xz\in Y\text{ for all }x\in X\,\}$. The structure we work with is called the phase space.

\begin{definition}[Phase space]\label{def:phs}
    A \define{phase space} is a pair $(M, \bot^M)$, where $M$ is a commutative multiplicative monoid and $\bot^M\subseteq M$. If $X\subseteq M$, we write $X^\bot$ for $X\multimap \bot^M$.
\end{definition}

With this interpretation, we can prove various properties we would expect from linear logic, such as $X\subseteq Y^\bot \text{ if and only if } XY\subseteq \bot^M$, $XX^\bot \subseteq \bot^M$ or $(X^{\bot\bot}Y)^\bot = X\multimap Y^\bot$. Notice furthermore that $X\subseteq X^{\bot\bot}$. By also requiring the other direction of the subset relation, we now define facts, which are sets of all the phases for which a formula is true. With this definition, we can represent $\bot$ as the set of orthogonal phases, and $X^\bot$ will be an involutive operation representing linear negation.

\begin{definition}[Fact]\label{def:fact}
    A \define{fact} is a $X\subseteq M$ with $X = X^{\bot\bot}$.
\end{definition}

Of special interest are the facts $\bot^M = \{1\}^\bot$, $\mathbf{1}^M = \{1\}^{\bot\bot}$, $\top^M = \emptyset^\bot$ and $\mathbf{0}^M = \emptyset^{\bot\bot}$, since they provide the semantics of the constants of linear logic. Observe that with this definition, given an $X\subseteq M$, $X^{\bot\bot}$ is the smallest fact containing $X$.

Using facts, we can define the behaviour of the connectives of linear logic. If $\{y\}^\bot \subseteq \{x\}^\bot$, we write $x\sqsubseteq y$, and if $x\sqsubseteq y$ and $y\sqsubseteq x$, we write $x\equiv y$, reminiscent of the semantical congruence of formulas in classical logic. We can now define the following operations on facts:
\begin{center}
    \begin{minipage}[b]{0.3\textwidth}
        \begin{align*}
            X\parr Y &\coloneq (X^\bot Y^\bot)^\bot,\\
            X\with Y &\coloneq (X^\bot \cup Y^\bot)^\bot,\\
            ?X &\coloneq (X^\bot \cap I^M)^\bot,
        \end{align*}
    \end{minipage}
    \begin{minipage}[b]{0.3\textwidth}
        \begin{align*}
            X\otimes Y &\coloneq (XY)^{\bot\bot},\\
            X\oplus Y &\coloneq (X\cup Y)^{\bot\bot},\\
            !X &\coloneq (X\cap I^M)^{\bot\bot},
        \end{align*}
    \end{minipage}
\end{center}
where $I^M = \left\{\,x\in \mathbf{1}^M\mid x = x^2\,\right\}$. We can replace the $I^M$ in the last two definitions by any submonoid $K^M$ of $J^M = \left\{\,x\in \mathbf{1}^M \mid x\equiv x^2\,\right\}$. We call the $K^M$ an \emph{exponential structure}.

\begin{definition}[Phase model]\label{def:phm}
    Given a set of propositional variables $V$, a \define{phase model} is a phase space together with a fact $a^M$ for each $a\in V$. The interpretation is inductively defined as:\\[-4.5em]
    \begin{center}
        \begin{minipage}[t]{0.3\textwidth}
            \begin{align*}
                (a^\bot)^M &\coloneq (a^M)^\bot,\\
                (A\parr B)^M &\coloneq A^M \parr B^M,\\
                (A\with B)^M &\coloneq A^M \with B^M,
            \end{align*}
        \end{minipage}
        \begin{minipage}[t]{0.3\textwidth}
            \begin{align*}
                (A\otimes B)^M &\coloneq A^M \otimes B^M,\\
                (A\oplus B)^M &\coloneq A^M \oplus B^M,\\
                (?A)^M &\coloneq\ ?A^M,\\
                (!A)^M &\coloneq\ !A^M.
            \end{align*}
        \end{minipage}
    \end{center}
\end{definition}

A fact $A$ \emph{holds} in the model if $1\in A^M$, that is, if the fact contains the unit phase. More generally, the sequent $\vdash \Gamma$ holds in the model if $1\in \Gamma^M$, or equivalently, $(A_1^M)^\bot\cdots (A_n^M)^\bot \subseteq \bot^M$. The soundness of this semantics can easily be proved by induction on proofs.

Observe that we can also regard the monoid $M$ as a discrete poset, which makes $\mathcal{P}(M)$ a complete lattice. Since the poset of facts can then be thought of as a reflective sub-poset of $\mathcal{P}(M)$, it is also a complete lattice. The lattice-theoretic lens on the semantics for linear logic is further used in the next section, where we examine a relational semantics.

\subsection{Kripke Semantics}

We follow the paper by~\textcite{AD_93}, which gives a modular semantics up to \fragment{MALL}, and the ideas of which are simple enough to be explained in this thesis. \textcite{COUMANS201450} give a relational semantics for full linear logic, but their approach is much more involved. The paper of Allwein and Dunn considers also relevance logic, which for weak enough fragments splits the implication into two operators, implication and coimplication. Since those fragments are out of the scope of this thesis, we omit the details regarding these weaker fragments. The approach depends heavily on a representation theorem on residuated lattices, which are extensively used for the analysis of substructural logics~\cite{GJKO07}.

\begin{definition}[Girard monoid]\label{def:girmon}
    A \define[Girard monoid]{Girard monoid} is a tuple $\mathcal{D} = (A, \with, \oplus, \cdot^\bot, \otimes, \multimap, 1)$ such that
    \begin{enumerate}
        \item $(A, \with, \oplus, \cdot^\bot, 0, \top)$ is a lattice and $\cdot^\bot$ is a de Morgan negation on that lattice.
        \item $(A, \otimes, 1)$ is an commutative monoid.
        \item The monoid is ordered by the lattice with the relation $\preccurlyeq$.
        \item $a\otimes b\preccurlyeq c$ if and only if $a\otimes c^\bot\preccurlyeq b^\bot$ (antilogism).
        \item $a\otimes b\preccurlyeq c$ if and only if $a\preccurlyeq b\multimap c$ (residuation).\qedhere
    \end{enumerate}
\end{definition}

To provide a model that fits linear logic, we have to axiomatize the behavior of the connectives of the monoid. The relevant axioms are detailed in Table~\ref{gm-axioms}. Not listed are laws of associativity and commutativity, which are given in the obvious way. It is also possible to consider nonresiduated lattices, but this would make the following representations much more difficult. Since we want a duality between disjunction and conjunction in our lattice, we will define a dual for linear implication, $A \multimapdot B\coloneq (A\multimapinv B)^\bot$, which we will call the coimplication operator\footnote{To continue the example from the introduction, we could use this operator to formalize the sentence ``a comathematician is a device that turns cotheorems into ffee''.}. 

\begin{table}[ht!]
    \centering
    \caption{Axioms of the Girard monoid.}\label{gm-axioms}
    \begin{tabular}{l l}
        \toprule
        \textbf{Name} & \textbf{Axiom}\\ \hline
        $\otimes$ identity & $1\otimes a = a = a\otimes 1$\\
        $\otimes \oplus$ distribution & $a\otimes (b\oplus c) = (a\otimes b)\oplus(a\otimes c)$\\
        $\oplus\otimes$ distribution & $(a\oplus b)\otimes c = (a\otimes c)\oplus (b\otimes c)$\\ \hline
        $\otimes \multimapinv$ residuation & $a\otimes b\preccurlyeq c$ if and only if $a\preccurlyeq c\multimapinv b$\\
        $\otimes \multimap$ residuation & $a\otimes b\preccurlyeq c$ if and only if $b\preccurlyeq a\multimap c$\\ \hline
        $\multimap$ identity & $1\multimap a = a = a\multimapinv 1$\\
        $\multimap \with$ distribution & $a\multimap (b\with c) = (a\multimap b)\with (a\multimap c)$\\
        $\multimap \oplus$ distribution & $(a\oplus b)\multimap c = (a\multimap c) \with (b\multimap c)$\\
        $\multimap$ LR-permutation & $a\preccurlyeq c\multimapinv b$ if and only if $b\preccurlyeq a\multimap c$ \\\hline
        $\parr$ identity & $a\parr \bot = a = \bot \parr a$\\
        $\parr \with$ distribution & $a\parr (b\with c) = (a\parr b)\with (a\parr c)$\\
        $\with \parr$ distribution & $(a\with b)\parr c = (a\parr c)\with (b\parr c)$\\\hline
        $\parr \multimapdotinv$ residuation & $a\multimapdotinv c\preccurlyeq b$ if and only if $a\preccurlyeq b\parr c$\\
        $\parr \multimapdot$ residuation & $b\multimapdot a\preccurlyeq c$ if and only if $a\preccurlyeq b\parr c$\\\hline
        $\multimapdot$ identity & $a\multimapdotinv \bot = a = \bot \multimapdot a$\\
        $\multimapdotinv \with$ distribution & $a\multimapdotinv (b\with c) = (a\multimapdotinv b)\oplus (a\multimapdotinv c)$\\
        $\multimapdotinv \oplus$ distribution & $(a\oplus b)\multimapdotinv c = (a\multimapdotinv c)\oplus (b\multimapdotinv c)$\\
        $\multimapdot$ LR-permutation & $a\multimapdot b\preccurlyeq c$ if and only if $b\multimapdotinv c\preccurlyeq a$\\\hline
        Period 2 & $a^{\bot\bot}\preccurlyeq a$\\
        Order inversion & $a\preccurlyeq b^\bot$ implies $b\preccurlyeq a^\bot$\\
        \bottomrule
    \end{tabular}
\end{table}

To derive a linear Kripke frame from the Girard monoid, we will make use of a lattice representation theorem by~\textcite{U_76}. For a general lattice, we will define a doubly-ordered set which will be the basis for the Kripke frame. The theorem then gives us a representation of a general lattice by two lattices of sets, which we will call $\sharp$-lattice and $\flat$-lattice. We do this because lattices of sets are distributive regarding meet and join. We can thus interpret intersections as meets in the $\sharp$-lattice and as joins in the $\flat$-lattice. Furthermore, there is a Galois connection between the lattices. Thus, the representation theorem can be used to embed the lattice in a representation lattice, whose elements can be thought of as the worlds in which a statement is true.

\begin{definition}[Filter and ideal]\label{def:fm}
    Given a lattice $(P, \preccurlyeq)$, a subset $F$ of the lattice is a \define{filter}, if it is nonempty and for all $x, y\in F$, the meet $x\wedge y$ is also in $F$. A subset $I$ of the lattice is an \define{ideal}, if it is nonempty and for all $x, y\in I$, the join $x\vee y$ is also in $I$. A filter or ideal is called \emph{prime}, if its complement is a ideal or filter, respectively.
\end{definition}

With these two concepts, we can define the two orders we want. To do so, let $\mathcal{A} = (A, \wedge, \vee)$ be a nondistributive lattice. If $\nabla$ is a filter and $\Delta$ is an disjoint ideal, then we call $(\nabla, \Delta)$ a \emph{filter-ideal} pair. Now we can define a concept of relative maximality: $\nabla$ is $\Delta$-maximal if $\nabla$ is the maximal element of the set of filters disjoint from $\Delta$, and $\Delta$ is $\nabla$-maximal, if $\Delta$ is the maximal element in the set of ideals disjoint from $\nabla$. $(\nabla, \Delta)$ is maximal if $\nabla$ is $\Delta$-maximal and $\Delta$ is $\nabla$-maximal.

Now we can define the two order relations of our linear Kripke frame. For a lattice $\mathcal{A}$, let $X$ be the set of maximal filter-ideal pairs. For any $x\in X$ we write $x_1$ for the first and $x_2$ for the second element in $x$. We define $x\preccurlyeq_1 y$ if and only if $x_1\subseteq y_1$ and $x\preccurlyeq_2 y$ if and only if $x_2\subseteq y_2$. The representation lattice is the lattice of $\preccurlyeq_1$ increasing sets. We can now define our Kripke frame:

\begin{definition}[Linear Kripke frame]
    Given a lattice $\mathcal{A}$ with the set of all maximal filter-ideal pairs $X$ and orders $\preccurlyeq_1$ and $\preccurlyeq_2$ as defined above, the \define[linear Kripke frame]{linear Kripke frame} is defined as $\mathcal{K} = (X, \preccurlyeq_1, \preccurlyeq_2)$.
\end{definition}

Next, we will make the order correspondence between the two orders explicit. Let $X$ be a set with two quasiorders and $C\subseteq X$. We define the mappings $\sharp\colon \mathcal{P}(X)\rightarrow \mathcal{P}(X)$ and $\flat\colon \mathcal{P}(X)\rightarrow \mathcal{P}(X)$ by
\[\sharp C \coloneq \{\,x\mid x\preccurlyeq_1 y \text{ implies }y\not\in C\,\},\ \ \ \ \flat C \coloneq \{\,x\mid x\preccurlyeq_2 y\text{ implies }y\not\in C\,\}.\]

The mnemonic we employ with the function names is that when we view the doubly ordered set $X$ ordered by $\preccurlyeq_1$ and $C\subseteq X$, then $\sharp C$ is the region ``above'' every element of $C$, and $\flat C$ is the region ``below''. We can express the duality between the mappings by the following lemma.

\begin{lemma}
    The mappings $X\rightarrow \flat X$ and $X\rightarrow \sharp X$ define a Galois connection between the lattice of $\preccurlyeq_1$-increasing and the lattice of $\preccurlyeq_2$-increasing subsets of $X$.
\end{lemma}

This correspondence helps us in finding an embedding from a lattice into the representation lattice. The characterization of the elements of the representation lattice which can represent an element of the original lattice is done via stable sets: A set $C$ is \emph{$\sharp$-stable} if $\sharp\flat C = C$ and \emph{$\flat$-stable} if $\flat\sharp C = C$.

Urquhart's representation theorem gives us a mapping of elements of a lattice into the set of maximal filter-ideal pairs such that the element is a member of the filter of the pair. In symbols
\begin{align*}
    \beta\colon \mathcal{A} &\rightarrow \mathcal{P}(A)\\
    a &\mapsto \beta(a) = \{\,x\mid a\in x_1\,\}.
\end{align*}

It can be shown that this mapping gives us stable sets:
\begin{lemma}
    Let $\mathcal{A} = (A, \wedge, \vee)$ be a lattice, $a\in A$ and $X$ the set of maximal pairs of $\mathcal{A}$. Then
    \begin{enumerate}
        \item $\flat\beta(a) = \left\{\,x\mid x\in X \text{ and } a\in x_2\,\right\}$,
        \item $\beta(a)$ is an $\sharp$-stable set in $\beta(\mathcal{A})$,
    \end{enumerate}
    where $\beta(\mathcal{A})$ is the representation lattice extracted from the frame $\mathcal{K}$.
\end{lemma}

To better distinguish the operators from their representation, the representations will in the following be overset with a circle, e.\,g. $\overset{\circ}{\with}$. The representation lattice operations $\overset{\circ}{\with}$ and $\overset{\circ}{\oplus}$ for $\sharp$-stable sets $C$ and $D$ are defined by
\[C\overset{\circ}{\with} D \coloneq C\cap D,\text{ and } C\overset{\circ}{\oplus} D \coloneq \sharp(\flat C\cap \flat D).\]
Observe that the $\overset{\circ}{\with}$ operator has the function of the meet in the representation lattice (with a more lattice theory oriented notation, we could denote it as $\overset{\circ}{\wedge}$), and the $\overset{\circ}{\oplus}$ operator has the function of the join operation. The top $\top$ and bottom $0$ of the lattice are given by
\[\beta(\top) \coloneq \{\,x\mid\top\in x_1\,\},\text{ and }\beta(0)\coloneq \{\,x\mid 0\in x_1\,\}.\]
Given a linear Kripke frame $\mathcal{K}$ with a set of worlds $X$, $X$ will be the top of the lattice extracted from $\mathcal{K}$ and $\emptyset$ will be the bottom. The next task we have is to provide representations of all operations defined above in the representation lattice. As this is a very technical procedure, we will only convey the concept with the representation of implication, tensor, and negation, and refer to~\textcite{AD_93} for the rest.

The worlds in our Kripke semantics are the prime filters of our algebra, and we generate the relation $R$ via the operators $\otimes, \multimap$ as follows:
\begin{align*}
    Rxyz &\text{\phantom{xxx}iff\phantom{xxx}} \forall a, b (a\in x \text{ and }a\multimap b\in y\text{ implies }b\in z)\\
    Rxyz &\text{\phantom{xxx}iff\phantom{xxx}} \forall a, b (a\in x\text{ and }b\in y\text{ implies } a\otimes b\in z)
\end{align*}

These two definitions are equivalent, which can be proven using residuation. We need several more relations for the representation of the other connectives. They are denoted with $S$, $Q$, $\Theta$, $\Omega$, and $\Upsilon$ and their definition can be found in the paper by Allwein and Dunn. We furthermore demand monotonicity of $R$. Taking $z\preccurlyeq z'$ if $z\subseteq z'$, we get
\[Rxyz \text{\phantom{xxx}and\phantom{xxx}} z\preccurlyeq z' \text{\phantom{xxx}implies\phantom{xxx}} Rxyz'.\]
We can now define the operator $\overset{\circ}{\multimap}$ of the representation lattice with the relation $R$ as
\[C\overset{\circ}{\multimap}D \coloneq \{\,y \mid \forall x, z(Rxyz\text{ and }x\in C\text{ implies } z\in D)\,\}.\]
We omit the proof that this definition gives us the desired property
\[\beta(a\multimap b) = \beta(a)\overset{\circ}{\multimap} \beta(b).\]
The tensor is handled in a similar way. We define an operator $\overset{+}{\otimes}$ with $R$ as
\[C\overset{+}{\otimes}D \coloneq \{\,z\mid \forall x, y (Rxyz\text{ and } y\in D\text{ implies }x\in rC)\,\}.\]
From this, we can define the operator $\overset{\circ}{\otimes}$ as a $\sharp$-stable set via
\[C\overset{\circ}{\otimes}D \coloneq \sharp(C\overset{+}{\otimes} D).\]
Again, we omit the proof that this operator represents the tensor:
\[\beta(a\otimes b) = \beta(a)\overset{\circ}{\otimes}\beta(b).\]
For the negation, we make use of the \emph{generalized Routley-Meyer star operator}, which is a function $\cdot^\star\colon\mathcal{K}\rightarrow\mathcal{K}$ satisfying
\begin{align*}
    x\preccurlyeq_1 y&\text{ implies } x^\star\preccurlyeq_2 y^\star\\
    x\preccurlyeq_2 y&\text{ implies } x^\star\preccurlyeq_1 y^\star\\
    x^{\star\star} &= x.
\end{align*}
We define the $\cdot^\star$ operator as $(x_1, x_2)^\star \coloneq (x_2^\bot, x_1^\bot)$. We now describe how the lattice representation leads to the valuation semantics of the Kripke semantics. A point in the lattice, which represents a proposition, may either lie in the filter of a pair, the ideal of a pair, or neither of them. This leads to a three-valued semantics with the values \emph{true}, \emph{false} and \emph{indifferent}.

Given an $\sharp$-stable set $A$, we interpret $x\in A$ as $x\models_T A$ and $x\in \flat A$ as $x\models_F A$, where $T$ stands for truth, which we denote as $\top$ in linear logic, and $F$ stands for falsity, which we denote as $0$ in linear logic. We write $x\models_I A$ if and only if $x\not\models_T A$ and $x\not\models_F A$. Double subscripts indicate that either of the subscripts holds. First, we need a valuation function for the atomic variables.

\begin{definition}[Atomic valuation]
    Let $v$ be a function which maps atomic variables and worlds to the set $\{\, T, F, I\,\}$ and for an atomic variable $p$, let $P_1 \coloneq \{\,x\mid v(p, x) = T\,\}$ and $P_2 \coloneq \{\,x\mid v(p, x) = F\,\}$. Such a $v$ is a \define{valuation} just when
    \[P_1 = \sharp P_2\text{\phantom{xxx}and\phantom{xxx}} P_2 = \flat P_1.\qedhere\]
\end{definition}

The valuation of connectives can be directly derived from their representations. For example, consider $A\overset{\circ}{\oplus} B = \sharp(\flat A \cap \flat B)$. Then, from the definition of the representation it follows that
\[x\in \sharp(\flat A\cap \flat B)\text{\phantom{xxx}iff\phantom{xxx}}\forall y(x\preccurlyeq_1y\text{ implies } y\not\in \flat A \text{ and }y\not\in \flat B).\]
This gives us the interpretation
\[x\models_T A \oplus B\text{\phantom{xxx}iff\phantom{xxx}}\forall y(x\preccurlyeq_1 y\text{ implies } y\models_{TI}A\text{ or }x\models_{TI} B).\]

To construct the interpretation for \fragment{MALL}, we have to consider all connectives, and each time give a $T$- and $F$-interpretation. This culminates to the following rules.

\begin{itemize}
    \item Atomic variables
    \begin{itemize}
        \item $x\models_T p$ iff $v(p, x) = T$
        \item $x\models_F p$ iff $v(p, x) = F$
        \item $x\models_I p$ iff $v(p, x) = I$
    \end{itemize}
    \item With
    \begin{itemize}
        \item $x\models_T A \with B$ iff $x\models_T A$ and $x\models_T B$
        \item $x\models_F A\with B$ iff $\forall y(x\preccurlyeq_2 y\text{ implies }(y\models_{FI} A\text{ or }y\models_{FI}B))$
    \end{itemize}
    \item Multiplicative units
    \begin{itemize}
        \item $\forall x(x\models_F 0)$
        \item $\forall x(x\models_T \top)$
    \end{itemize}
    \item Additive units
    \begin{itemize}
        \item $\forall x(x\in \overset{\circ}{1} \text{ implies } x\models_T 1)$
        \item $\forall x(x\in \overset{\circ}{\bot}\text{ implies }x\models_F \bot)$
        \item $x\models_F 1$ iff $\forall y(x\preccurlyeq_2 y\text{ implies } y\not\in\overset{\circ}{1})$
        \item $x\models_T\bot$ iff $\forall y(x\preccurlyeq_1 y\text{ implies } y\not\in\overset{\circ}{\bot})$
    \end{itemize}
    \item Plus
    \begin{itemize}
        \item $x\models_T A\oplus B$ iff $\forall y(x\preccurlyeq_1 y\text{ implies } y\models_{TI} A\text{ or }y\models_{TI} B)$
        \item $x\models_F A\oplus B$ iff $x\models_F A$ and $x\models_F B$
    \end{itemize}
    \item Implication
    \begin{itemize}
        \item $x\models_T B \multimapinv A$ iff $\forall y,z((Rxyz\text{ and }y\models_T A)\text{ implies }z\models_T B)$
        \item $x\models_F B \multimapinv A$ iff $\forall x'\exists y,z(x\preccurlyeq_2 x'\text{ implies } (Rx'yz\text{ and }y\models_T A\text{ and }z\models_{FI} B))$
    \end{itemize}
    \item Tensor
    \begin{itemize}
        \item $z\models_T A\otimes B$ iff $\forall z'\exists x, y(z\preccurlyeq_1 z'\text{ implies }(Sxyz'\text{ and }y\models_T B\text{ and }x\models_{TI}A))$
        \item $z\models_F A\otimes B$ iff $\forall x, y((Sxyz\text{ and }y\models_T B)\text{ implies } x\models_F A)$
    \end{itemize}
    \item Par
    \begin{itemize}
        \item $z\models_T A\parr B$ iff $\forall x, y((\Omega xyz\text{ and }x\models_F A)\text{ implies }y\models_T B)$,
        \item $z\models_F A\parr B$ iff $\forall z'\exists x, y(z\preccurlyeq_2 z'\text{ implies }(\Omega xyz'\text{ and }x\models_F A\text{ and }y\models_{FI}B))$
    \end{itemize}
    \item Coimplication
    \begin{itemize}
        \item $y\models_T B\multimapdot A$ iff $\forall y'\exists x,z(y\preccurlyeq_2 y'\text{ implies }(\Theta xy'z\text{ and }x\models_F B\text{ and }z\models_{TI}A))$
        \item $y\models_F B\multimapdot A$ iff $\forall x,z((\Theta xyz\text{ and }x\models_F B)\text{ implies } y\models_F A)$
    \end{itemize}
    \item Negation
    \begin{itemize}
        \item $x\models_T A^\bot$ iff $x^\star\models_F A$
        \item $x\models_F A^\bot$ iff $\forall x'(x\preccurlyeq_2 x'\text{ implies } (x'\models_I A\text{ or }x'\models_T A))$
    \end{itemize}
\end{itemize}

We also have the following hereditary conditions for the operators on the model:
\begin{itemize}
    \item $\forall x(x\in \overset{\circ}{1}\text{ and } v(p, y) = T\text{ and } Rxyz \text{ implies }v(p, z) = T)$,
    \item $\forall x(x\in \overset{\circ}{1}\text{ and } v(p, x) = T\text{ and } Ryxz \text{ implies }v(p, z) = T)$,
    \item $\forall y\exists x(x\in \overset{\circ}{1}\text{ and }Sxyy)$, $\forall y\exists x(x\in \overset{\circ}{1}\text{ and }Qyxy)$,
    \item $\forall y(y\in\overset{\circ}{\bot}\text{ and }v(p,x) = F\text{ and }\Theta xyz \text{ implies }v(p,z) = F)$,
    \item $\forall y(y\in\overset{\circ}{\bot}\text{ and }v(p,x) = F\text{ and }\Theta yxz \text{ implies }v(p,z) = F)$,
    \item $\forall x\exists y(y\in \overset{\circ}{\bot}\text{ and }\Omega xyz)$, $\forall x\exists y(y\in\overset{\circ}{\bot}\text{ and }\Upsilon yxx)$.
\end{itemize}

\begin{definition}[Interpretation]
    An interpretation, $\models$, in a structure $\mathcal{K} = (X, \preccurlyeq_1, \preccurlyeq_2)$ is a function from well-formed formulas to the set $\{\,T, F, I\,\}$ such that it suffices the conditions above.
\end{definition}

\subsection{Category Theoretic Semantics}\label{sec:catlog}

We will now show how we can give a category-theoretic account of a semantics for linear logic. The main idea is that we give categories enough additional structure so that they are models of linear logic. There are several approaches to this, for an overview see~\textcite{dP_14}. A deep treatment of categorical semantics was given by~\textcite{Mell_09}, who reviews a number of approaches in detail. The approach we present here is due to~\textcite{S_89}.

\begin{definition}[Monoidal category]\label{def:monoidal_cat}
    A \define[category!monoidal]{monoidal category} is a category $\mathcal{C}$ with
    \begin{itemize}
        \item a functor $\otimes\colon \mathcal{C} \times\mathcal{C}\rightarrow \mathcal{C}$, called the tensor product,
        \item an object $1_\mathcal{C}\in \mathcal{C}$, called the unit object,
        \item a natural isomorphism
        \[\alpha\colon ((-)\otimes(-)) \otimes (-) \overset{\simeq}{\rightarrow} (-) \otimes ((-)\otimes (-))\]
        with components of the form
        \[\alpha_{A, B, C}\colon (A\otimes B)\otimes C \rightarrow A\otimes (B\otimes C),\]
        called the \define{associator},
        \item a natural isomorphism $\lambda\colon (1_\mathcal{C}\otimes (-)) \overset{\simeq}{\rightarrow} (-)$ with components of the form \\$\lambda_A\colon 1_\mathcal{C}\otimes A\rightarrow A$, called the \define[unitor!left]{left unitor}, and
        \item a natural isomorphism $\rho\colon ((-)\otimes 1_\mathcal{C}) \overset{\simeq}{\rightarrow} (-)$ with components of the form \\$\rho_A\colon A\otimes 1_\mathcal{C}\rightarrow A$, called the \define[unitor!right]{right unitor},
    \end{itemize}
    such that the following diagrams commute:
    \begin{enumerate}
        \item triangle identity
        \begin{center}
            \begin{tikzcd}
                (A \otimes 1_\mathcal{C})\otimes B \ar[r, "\alpha_{A, 1_\mathcal{C}, B}"] \ar[rd, "\rho_A \otimes 1_B", swap] & A\otimes (1_\mathcal{C}\otimes B) \ar[d, "1_A\otimes \lambda_B"]\\
                & A\otimes B\\
            \end{tikzcd}
        \end{center}
        \item pentagon identity
        \begin{displaymath}
            \begin{tikzcd}
                {} & (A\otimes B) \otimes (C\otimes D) \ar[rd, "\alpha_{A, B, C\otimes D}"] & {}\\
                ((A\otimes B)\otimes C)\otimes D \ar[ru, "\alpha_{A\otimes B, C, D}"] \ar[d, "\alpha_{A, B, C} \otimes 1_D", swap] & {} & (A\otimes (B\otimes (C \otimes D)))\\
                (A\otimes (B\otimes C)) \otimes D \ar[rr, "\alpha_{A,B\otimes C, D}"] & {} & A\otimes ((B\otimes C)\otimes D). \ar[u, "1_A \otimes \alpha_{B, C, D}", swap]
            \end{tikzcd}
        \end{displaymath}
    \end{enumerate}
    The two diagrams may seem complicated at first, but note that they simply enforce the behavior we would expect from a mapping that has a notion of unit and associativity, respectively.
\end{definition}

The symbolic similarities between the tensor product and unit object of monoidal categories and their counterparts in linear logic are no accident: observe that the definitions in the monoidal category and the ones in the sequent calculus cause them to behave in the same way. The tensor connector in linear logic is also commutative, which we represent in the categorical model as a braiding.

\begin{definition}[Braided monoidal category]\label{def:bmc}
    A \define[category!braided monoidal]{braided monoidal category} is a monoidal category $\mathcal{C}$ equipped with a natural isomorphism
    \[\tau_{A, B}\colon A\otimes B \overset{\simeq}{\rightarrow} B\otimes A\]
    called the \emph{braiding}, such that the following diagrams (called the hexagon identities) commute:
    \begin{center}
        \begin{tikzcd}
            (A\otimes B)\otimes C\ar[r, "\alpha_{A,B,C}"] \ar[d, "\tau_{A, B} \otimes 1_\mathcal{C}", swap] & A\otimes (B\otimes C) \ar[r, "\tau_{A, B\otimes C}"] & (B\otimes C)\otimes A \ar[d, "\alpha_{B, C, A}"]\\
            (B\otimes A)\otimes C \ar[r, "\alpha_{B,A,C}"] & B\otimes (A\otimes C) \ar[r, "1\otimes \tau_{A,C}"] & B\otimes (C\otimes A)\\
        \end{tikzcd}
        \begin{tikzcd}
            A\otimes (B\otimes C) \ar[r, "\alpha^{-1}_{A,B,C}"] \ar[d, "1\otimes \tau_{B, C}", swap] & (A\otimes B)\otimes C \ar[r, "\tau_{A\otimes B, C}"] & C\otimes (A \otimes B) \ar[d, "\alpha^{-1}_{C, A, B}"]\\
            A\otimes (C\otimes B) \ar[r, "\alpha^{-1}_{A,C,B}"] & (A\otimes C)\otimes B \ar[r, "\tau_{A,C}\otimes 1_\mathcal{C}"] & (C\otimes A)\otimes B\\
        \end{tikzcd}
    \end{center}
    Again, the diagrams look complicated on first sight, but they simply enforce the behavior we expect from a mapping that suffices a notion of commutativity. 
\end{definition}

Also, observe that when we commute the tensor in linear logic twice, we essentially do nothing. This leads us to the definition of symmetric monoidal categories.

\begin{definition}[Symmetric monoidal category]\label{def:smc}
    A \define[category!symmetric monoidal]{symmetric monoidal category} is a braided monoidal category for which the braiding
    \[\tau_{A,B}\colon A\otimes B\overset{\simeq}{\rightarrow} B\otimes A\]
    satisfies the condition
    \[\tau_{B,A}\circ \tau_{A,B} = 1_{A\otimes B}\]
    for all objects $A, B$.
\end{definition}

We can now define a mapping on these symmetric monoidal categories which represents linear implication. In category-theoretic terms, this is done via a special kind of morphism, the internal hom.

\begin{definition}[Internal hom]\label{def:inthom}
    Let $(\mathcal{C}, \otimes, 1)$ be a symmetric monoidal category. An \define{internal hom} in $\mathcal{C}$ is a functor
    \[-\multimap -\colon \mathcal{C}^\text{op}\times \mathcal{C}\rightarrow\mathcal{C}\]
    such that for every object $A\in\mathcal{C}$ we have a pair of functors
    \[((-)\otimes A), (A \multimap -)\colon \mathcal{C} \rightarrow\mathcal{C}\]
    which are adjoint
    \[((-)\otimes A)\dashv (A \multimap -).\]
    If this exists, $(\mathcal{C}, \otimes, 1, \multimap)$ is called a \define[category!symmetric closed monoidal]{symmetric closed monoidal category}.
\end{definition}

\begin{remark}
    We can always embed a \reftodef[category!symmetric monoidal]{symmetric monoidal category} $\mathcal{C}$ into some symmetric monoidal closed category $\vcenter{\hbox{\includegraphics[width=1.1em]{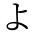}}}(\mathcal{C})$ via the Yoneda embedding. The Yoneda lemma further says that this embedding is full and faithful. The Yoneda lemma is one of the most used theorems in category theory.
\end{remark}

For a symmetric closed monoidal category, the evaluation map
\[\text{ev}_{A, B}\colon (A\multimap B) \otimes A \rightarrow B\]
is the $((-)\otimes A\dashv (A \multimap -))$-adjunct of the identity $1_{A \multimap B}\colon (A\multimap B)\rightarrow (A\multimap B)$. While this evaluation map seems to be an abstract definition at first, notice that in terms of logic, it simply describes the modus ponens.

\begin{definition}[$*$-autonomous category]\label{def:star_auto_cat}
    A \define[category!$*$-autonomous]{$*$-autonomous category} is a symmetric closed monoidal category $(\mathcal{C}, \otimes, 1, \multimap)$ with an object $\bot$ such that the canonical morphism $d_A\colon A\rightarrow (A\multimap \bot) \multimap \bot$, which is the transpose of the evaluation map $\text{ev}_{A, \bot}\colon (A\multimap \bot) \otimes A \rightarrow \bot$ is an isomorphism for all $A$. This object is called the \define[object!global dualizing]{global dualizing object}.
\end{definition}

The global dualizing object gives us furthermore an involution $(-)^*$. When we require that the category has finite products, we get the notion of linear categories that Seely describes in his paper.

\begin{definition}[Linear category]\label{lcat}
    A \define[category!linear]{linear category} is a $*$-autonomous category with finite products.
\end{definition}

Finally, to model the exponential, we present the definition given by Seely, with an additional criterion later provided by Bierman which ensures soundness of the model (cf.~\cite{dP_14}). They are as such called new-Seely categories. Note that this is one of several inequivalent ways to model the exponential. The other approaches all agree on the fact that the $!$-modality should be represented by a \reftodef{comonad}, while the $?$-modality should be a \reftodef{monad}. See~\textcite{S_89} or~\textcite{Mell_09} for further details.

\begin{definition}[New-Seely category]
    A \define[category!new-Seely]{new-Seely category}, $\mathcal{C}$, consists of
    \begin{itemize}
        \item A linear category $\mathcal{C}$, together with
        \item A comonad $(!, \varepsilon, \delta)$, and
        \item Two natural isomorphisms, $n\colon !A\otimes{} !B \overset{\cong}{\rightarrow}\ !(A\with B)$ and $p\colon 1 \overset{\cong}{\rightarrow}\ !\top$,
    \end{itemize}
    such that the adjunction between $\mathcal{C}$ and its co-Kleisli category is a monoidal adjunction, that is, an adjunction between two monoidal categories that respects the monoidal structure.
\end{definition}

\begin{proposition}\label{seely}
    New-Seely categories provide a semantics for linear logic. In particular, we have the following results.
    \begin{itemize}
        \item Given a linear logic $\mathcal{L}$, a new-Seely category $\mathcal{G}(\mathcal{L})$ may be constructed. The objects are formulas and the morphisms are equivalence classes of derivations of sequents.
        \item Given any new-Seely category $\mathcal{G}$, a linear logic $\mathcal{L}(\mathcal{G})$ may be constructed. The constants are the objects of $\mathcal{G}$ and the axioms are the morphisms of $\mathcal{G}$.
        \item $\mathcal{G} \simeq \mathcal{G}(\mathcal{L}(\mathcal{G}))$ and $\mathcal{L}$ is, in a suitable sense, equivalent to $\mathcal{L}(\mathcal{G}(\mathcal{L}))$.
    \end{itemize}
\end{proposition}

We will not provide a formal proof of the proposition, but rather describe intuitively how the connectives of linear logic are interpreted in the categorical model. The tensor connective~($\otimes$) is simply the tensor product of the category theory, negation~($\cdot^\bot$) is modeled by the global dualizing object's involution. As described before, linear implication~($\multimap$) is modeled by the \reftodef{internal hom}, and can with negation be defined as $(A\otimes B^\bot)^\bot$. The dual connective of the tensor, par~($\parr$), can be defined by $A\parr B \coloneq (A^\bot\otimes B^\bot)^\bot$. The with connective~($\with$) is represented by \reftodef[product]{products}, and the plus connective~($\oplus$) is represented by \reftodef[coproduct]{coproducts}, whose existence is guaranteed by the existence of products and the \reftodef[category!$*$-autonomous]{$*$-autonomy}. For the exponential modalities, the $!$-modality is modeled by a comonad, and the dual $?$-modality by a monad.

\subsection{Geometry of Interaction}

While the topic of Geometry of Interaction is not central to this thesis, we nevertheless feel the need to devote a short section to the explanation of the central concepts, because it sparked a plethora of research from when Girard presented it in the early nineties to this day. Girard himself examined Geometry of Interaction in various articles over the years~\cite{GoI1,GoI2,GoI3,GoI4,GoI5,GoI0}.

What makes Geometry of Interaction such an interesting research topic is that it allows us to give an algebraic characterization of proofs, namely through operator algebras: formulas are interpreted as Hilbert spaces and proofs then correspond to partial isometries. This is already reflected in the phase semantics from Section~\ref{phasesem}, but we will not introduce the mathematical machinery to give a formal account of transitioning from phase semantics to Geometry of Interaction, rather we will describe it in an informal manner: We can assign to each proof net a permutation matrix $\sigma$, which encodes the cut links of the proof net, as well as a proper orthogonal matrix $M$ which describes certain expressions built from a dynamic algebra, which in turn describe the possible paths inside the proof net. We can then derive the so-called \define{execution fomula} for the proof net
\[\text{Ex}(\sigma, M) = (1 - \sigma^2)\left(\sum_i M(\sigma M)\right)(1-\sigma^2),\]
which serves as a complete description of the proof net. This formula is also an invariant of the normalization process in \fragment{MLL}. Generally, Geometry of Interaction is viewed as a mathematical model capturing the dynamics of cut elimination.

This approach is often compared to the categorical semantics given above from a programming language perspective: while the categorical semantics corresponds to a form of \emph{denotational semantics} for linear logic, that is, describing the behavior of linear logic terms via mathematical objects, Geometry of Interaction can be perceived as a form of \emph{operational semantics}, where the meaning of the terms is described regarding some sort of execution.

Nevertheless, Geometry of Interaction can also be described in category-theoretic terms, using traced \reftodef[category!symmetric monoidal]{symmetric monoidal categories}. A survey is given by~\textcite{GoIsurvey}. In \textcite{GoI5}, he reformulated the approach from the ground up, now using von Neumann algebras which also account for light versions of linear logic. These logics play a big part in the study of implicit complexity, as we will describe in Section~\ref{llapplications}.

\subsection{Game Semantics}

The last way to define a semantics for linear logic that we present is a game semantics. This approach is due to~\textcite{BLASS1992183}, who proposed it in 1992. We describe the main ideas of the approach without proving soundness or correctness. For the game, let $\mathcal{A}$ and $\mathcal{B}$ be two players. We have four possible states: it is $\mathcal{A}$'s turn, it is $\mathcal{B}$'s turn, $\mathcal{A}$ has won, and $\mathcal{B}$ has won. The last two states loop indefinitely. When it is $\mathcal{A}$'s turn, $\mathcal{B}$ is winning and vice versa. This leads to two winning conditions: either the game is over or there is an infinite loop of $\mathcal{A}$'s or $\mathcal{B}$'s turn. We can now encode these four states with the constants of linear logic as in Table~\ref{gamestates}.

\begin{table}[ht!]
    \centering
    \caption{States of the game.}\label{gamestates}
    \begin{tabular}{l l}
        \toprule
        \textbf{Constant} & \textbf{Game state}\\ \hline
        $\top$ & It is $\mathcal{B}$'s turn, but they have no moves, the game loops indefinitely, $\mathcal{A}$ wins.\\
        $0$ & It is $\mathcal{A}$'s turn, but they have no moves, the game loops indefinitely, $\mathcal{B}$ wins.\\
        $1$ & $\mathcal{A}$ wins.\\
        $\bot$ & $\mathcal{B}$ wins.\\
        \bottomrule
    \end{tabular}
\end{table}

The next step is the simulation of connectives. They describe the action that the parties take in each step of the game.

\begin{description}
    \item[$A\with B$:] it is $\mathcal{B}$'s turn, they choose $A$ or $B$.
    \item[$A\oplus B$:] it is $\mathcal{A}$'s turn, they choose $A$ or $B$.
    \item[$A\otimes B$:] the games continue with $A$ and $B$ in parallel. If it is $\mathcal{A}$'s turn in either $A$ or $B$, it is $\mathcal{A}$'s turn. $\mathcal{A}$ wins if they win both games.
    \item[$A\parr B$:] the games continue with $A$ and $B$ in parallel. If it is $\mathcal{B}$'s turn in either $A$ or $B$, it is $\mathcal{B}$'s turn. $\mathcal{B}$ wins if they win both games.
    \item[$A^\bot$:] the roles of $\mathcal{A}$ and $\mathcal{B}$ are swapped and the game continues on $A$.
\end{description}

A game on $A$ is \emph{valid} if $\mathcal{A}$ has a winning strategy. This is sound and complete to the statement that $\vdash A$ is derivable in linear logic. Again, the connection between linear logic, categories, and games can be made explicit. \textcite{DBLP:journals/tcs/HylandS03} give some examples of how categories of games are models for linear logic.

\section{Other Models for Linear Logic}

Until now, we constructed semantics to precisely fit linear logic. Of course, with every treatment, different models for linear logic arose. We want to further convey the usefulness of linear logic to a variety of domains, and dedicate this section to giving some concrete examples of models, which are used in different mathematical areas. For the first two examples, we follow~\textcite{DBLP:journals/tcs/HylandS03} very closely.

\subsection{Sup Lattices}

A \define{sup lattice} is a poset that has joins of arbitrary subsets. It is a complete lattice, but the morphisms need only be $\vee$-preserving. It is widely used in topos theory. The category $\vee$\textbf{-Lat} of sup lattices is a model of classical linear logic with the following structure:

\emph{Multiplicative:} The tensor product $A\otimes B$ classifies the maps $A\times B\rightarrow C$ which are suprema preserving in every component. The linear function space $B\multimap C$ is the lattice of all maps from $B$ to $C$ which are $\vee$-preserving with the pointwise order.

\emph{Additive:} The additive structure of $\vee$\textbf{-Lat} is degenerate since it admits (infinite) biproducts, which means that a product is also a \reftodef{coproduct} and vice versa.

\emph{Exponential:} Since we have infinite biproducts for the additive structure, and especially $\oplus$, we can construct the free $\parr$-monoid by $\bot \oplus A\oplus (A\parr A)\oplus \cdots$.

\subsection{Vector Spaces}

Vector spaces are the central object of examination in modern linear algebra. The category \textbf{Vec}$_k$ of vector spaces over a field $k$ is a model for intuitionistic linear logic with the following structure.

\emph{Multiplicative:} The standard tensor product provides the $\otimes$-connective and the linear function space provides the $\multimap$-connective. Thus, we have given the multiplicative structure of the model.

\emph{Additive:} We again have a degenerate additive structure via the direct sum of vector spaces, which is a biproduct.

\emph{Exponential:} The exponential structure of \textbf{Vec}$_k$ constitutes in a free commutative coalgebra $!V$ on the vector space $V$. A coalgebra is defined as follows.
\begin{definition}[Coalgebra]
    A \define{coalgebra} over a field $k$ is a vector space $V$ over $k$ together with $k$-linear maps $\Delta\colon V\rightarrow V\otimes V$ and $\varepsilon\colon V\rightarrow k$ such that
    \begin{enumerate}
        \item $(1_V \otimes \Delta)\circ \Delta = (\Delta\otimes 1_V)\circ \Delta$
        \item $(1_V \otimes \varepsilon)\circ \Delta = 1_V = (\varepsilon \otimes 1_V)\circ \Delta$
    \end{enumerate}
    where $\otimes$ is the tensor product and $1_V$ the identity function in $V$.
\end{definition}
It is easy to see that $\Delta$ is the dual of multiplication in an algebra, while $\varepsilon$ is the dual of the unit. The maps are thus named \define[comultiplication!in a coalgebra]{comultiplication} and \define[counit!in a coalgebra]{counit}, respectively.

\subsection{Coherence Spaces}

Coherence spaces play a central role in stable domain theory and the examination of programming language semantics. They provide a model which gives a more graph-theoretic perspective on linear logic. It was actually during the investigation of these spaces by Girard, that linear logic was conceived. They appear in the original paper, and, more fleshed out, in~\textcite{10.5555/64805}.

\begin{definition}[Coherence space]
    A \define{coherence space} $X$ is a set where
    \begin{enumerate}
        \item $a\in X \wedge b\subseteq a \implies b\in X$ and
        \item For \emph{compatible} $a,b\in X$, that is $a\cup b\in X$, if $A\subseteq X$ is formed from pairwise compatible elements, then $\bigcup A\in X$.\qedhere
    \end{enumerate}
\end{definition}

We then define the web $W(X)$ of a coherence space to be a reflexive, undirected graph $(V, \frownsmile)$ with $V \coloneq \{\,z\mid \{\,z\,\}\in X\,\}$ and $x\frownsmile y$ if and only if $\{\,x,y\,\}\in X$. If $x\frownsmile y$, we say that $x$ and $y$ are coherent. Of further interest for the examination of coherence spaces are the relations of \emph{strict coherence} $x\strictfrown y$, where $x\frownsmile y$ and $x\neq y$, \emph{strict incoherence} $x\strictsmile y$ , where $\neg (x\frownsmile y)$, and \emph{incoherence} $x\smilefrown y$, where $\neg(x\strictfrown y)$. By abuse of notation, we also call the web a coherence space.

We can define the linear negation $X^\bot$ of $X$ as $x\frownsmile y$ in $X^\bot$ if and only if $x\smilefrown y$ in $X$. As in phase semantics, we have $X^{\bot\bot} = X$, so negation is involutive.

\emph{Multiplicative:} Multiplicatives are variations of the cartesian product, for example the set $V$ for $X\otimes Y$ would be $X\times Y$ and we have $(x, y) \frownsmile (x', y')$ in $X\otimes Y$ if and only if we have $x\frownsmile x'$ in $X$ and $y\frownsmile y'$ in $Y$. For $X\parr Y$, we have that $(x, y)\strictfrown (x', y')$ if and only if we have $x\strictfrown x'$ in $X$ or $y\strictfrown y'$ in $Y$.

\emph{Additive:} Additives are variations of the direct sum, for example the set $V$ for $X\with Y$ would be $\{\,0\,\}\times X \cup \{\,1\,\}\times Y$ and we have $(0, x)\frownsmile (0, x')$ in $X\with Y$ if and only if we have $x\frownsmile x'$ in $X$. The case for $y$ runs analogously, and the same holds for the $X\oplus Y$ case. They differ in the regard that we have $(0, x)\strictfrown (1, y)$ in $X\with Y$ for all $x\in X$ and $y\in Y$, and $(0, x)\strictsmile(1, y)$ in $X\oplus Y$ for all $x\in X$ and $y\in Y$.

We can easily check that the de Morgan equalities, associativity, commutativity, and distributivity laws we expect hold. The constants $1$ and $\bot$ are represented by the coherent space with exactly one element, which is unique up to isomorphism. The constants $0$ and $\top$ are represented by the coherent space with the empty web.

\emph{Exponential:} The exponential modality $!X$ is defined by the set $V \coloneq \{\,a\mid a\in X\text{ and } a\text{ is finite}\,\}$, and we have $a\frownsmile b$ in $!X$ if and only if $a\cup b\in X$. For $?X$, we have $V \coloneq \{\,a\mid a\in X^\bot\text{ and } a\text{ is finite}\,\}$, and $a\strictfrown b$ in $?X$ if and only if $a\cup b\not\in X^\bot$. Again, de Morgan equalities and distributivities are easily checked.

The category \textbf{Coh} of coherence spaces is a model for linear logic. Especially, since negation is involutive, it is a model of classical linear logic.

\section{Useful Properties of Linear Logic}

In this section, we investigate linear logic further and state some basic results which will be useful to us in its complexity-theoretic analysis. This constitutes in an embedding of classical and intuitionistic propositional logic into linear logic, cut-elimination for linear logic and a subtlety which arises with having two forms of conjunction and disjunction, and finally, we show which fragments admit the finite model property, giving us some first insight into the recursion-theoretic complexity of the fragments.

\subsection{Relation of Linear Logic to \fragmenttxt{LK} and \fragmenttxt{LJ}}

We will relate linear logic to classical and intuitionistic logic by giving an embedding of the two into linear logic. This also serves as an example of how we recover the expressibility of the two logics via the exponential modalities. The method we show follows~\textcite[Section 4]{Be_13}. We translate formulas from implicative-conjunctive propositional logic into linear logic as follows.

\begin{definition}
    Let $a$ be an atom and $A$ and $B$ formulas of implicative-conjunctive logic. We define the translation $\cdot^*$ into linear logic inductively as
    \[(a)^* \coloneq{} a\phantom{xxxxx}(A\rightarrow B)^*\coloneq{} !A^*\multimap B^*\phantom{xxxxx}(A\wedge B)^* \coloneq{} A^*\with B^*.\]
    Sequents are translated as $(A_1,...,A_n\vdash B)^* \coloneq{} \vdash{} ?(A^*_1)^\bot,\dots,?(A^*_n)^\bot, B^*$.
\end{definition}

We can extend this translation to also cover proofs. While the introduction rule for $\rightarrow$ can be directly translated into the introduction rule for $\parr$, an equivalent rule to the classical cut rule given below (ccut) is translated as
\[\prftree[r]{(ccut)}{\Gamma \vdash A\rightarrow B}{\Delta \vdash B}{\Gamma, \Delta \vdash B}\phantom{xx}\overset{\rightsquigarrow}{\phantom{O}}\phantom{xx}\prftree[r]{(cut)}{\vdash{}?(\Gamma^*)^\bot, ?A^\bot\parr B}{\prftree[r]{$(\otimes)$}{\prftree[r]{(of course)}{\vdash{}?(\Delta^*)^\bot, A}{\vdash{}?(\Delta^*)^\bot, !A}}{\prftree[r]{(id)}{}{\vdash B^\bot, B}}{\vdash{} ?(\Delta^*)^\bot, !A\otimes B^\bot, B}}{\vdash{}?(\Gamma^*)^\bot, ?(\Delta^*)^\bot, B}\]

In particular, this translation preserves provability between intuitionistic natural deduction and linear logic. Furthermore, we remember that we can translate classical logic to intuitionistic logic via the double negation translation, so consequently, we can also embed classical logic into linear logic. One way would be to compose both translations, but the process can be simplified. A translation with proof of correctness can be found in Troelstra~\cite{T_92}, Section~5.12.

The linear version of implication bears a major difference to its classical counterpart: a very central result for classical logic is the deduction theorem, by which we have that $\Gamma \cup \{\,\phi\,\} \vdash \psi$ implies $\Gamma \vdash \phi \rightarrow \psi$, thus relating the entailment relation with a logical connective. But notice how in the translation, we need an exponential modality for $A$, thus, this relation does not hold for linear implication, for which we would have to prepend a $!$-modality before the antecedent: We have that $\Gamma \cup \{\,A\,\}\vdash B$ is derivable in linear logic if and only if $\Gamma\vdash{} !A\multimap B$ is.

\subsection{Cut-elimination in Linear Logic and Additive Implication}

We now shift our focus to cut-elimination. This is, as we have seen, a very important property that logics can have. Luckily, linear logic admits cut-elimination as well, which we will also exploit for some of the complexity classifications of its fragments.

\begin{proposition}[Cut elimination for \fragment{LL}]\label{proposition:cut-ll}
    If a sequent is provable in \fragment{LL}, then it is provable in \fragment{LL} without a cut.
\end{proposition}

\begin{proof}
    See~\textcite[Theorem 9.3.4]{troelstra_schwichtenberg_2000}.
\end{proof}

As with classical logic, the proof is constructive, giving rise to an explicit procedure for eliminating cut rules. The next question is how complex this procedure is. The answer to this question for various fragments can be found in the paper by~\textcite{MT_03}. We have listed the results in Table~\ref{complcut}. They give an overview of how the complexity of cut elimination increases with regard to the expressivity of the fragment.

\begin{table}[ht!]
    \centering
    \caption{Complexity of cut-elimination in linear logic~\cite{MT_03}. Multiplicative soft linear logic (\fragment{MSLL}) and multiplicative light linear logic (\fragment{MLLL}) are presented in Section~\ref{llapplications}.}\label{complcut}
    \begin{tabular}{l l l l l}
        \toprule
        \fragment{MLL} & \fragment{MALL} & \fragment{MSLL} & \fragment{MLLL} & \fragment{MELL}\\ \hline
        \class{P}-complete & \class{coNP}-complete & \class{EXP}-complete & \class{2EXP}-complete & non-elementary\\
        \bottomrule
    \end{tabular}
\end{table}

There is, however, a subtlety arising with having the additive and multiplicative fragment present in linear logic. If we consider the \emph{additive cut rule}, which is not part of the sequent calculus of linear logic, we see that it is also not admissible.

\begin{proposition}
    The additive cut rule
    \[\prftree[r]{\textup{(cut add)}}{\vdash A, \Gamma}{\vdash A^\bot, \Gamma}{\vdash\Gamma}\]
    is not admissible in linear logic.
\end{proposition}

\begin{proof}
    While the formula $A\oplus A^\bot$ is not provable in linear logic, it is derivable via the additive cut rule:
    \[\prftree[r]
        {(cut add)}
        {\prftree[r]{($\oplus_1$)}{\prftree[r]{(id)}{}{\vdash A, A^\bot}}{\vdash A, A \oplus A^\bot}}
        {\prftree[r]{($\oplus_2$)}{\prftree[r]{(id)}{}{\vdash A, A^\bot}}{\vdash A \oplus A^\bot, A^\bot}}
        {\vdash A \oplus A^\bot}
    \]\vspace{-1em}
\end{proof}

While our presentation of linear logic is quite symmetric, we left out one connective until now: additive implication. This is quite sensible, as we will convey. First, the obvious definition of additive linear implication runs as follows.

\begin{definition}[Additive linear implication]\label{addimpl}
    $A\rightharpoonup B \coloneq A^\bot \oplus B$.
\end{definition}

But note that with this definition, neither reflexivity $A\rightharpoonup A$, nor the rule of modus ponens $A, A\rightharpoonup B \vdash B$ hold in linear logic, so the connective does not behave at all like what we would expect from an implication connective. For this reason, additive implication is usually omitted from presentations of linear logic, although it can be defined as above or, in a fragment with access to the multiplicatives, equivalently via multiplicative implication as $(A\multimap 0) \oplus B$.

\subsection{Finite Model Property in Linear Logic Fragments}

A further property we can establish for the ``standard'' linear logic we have introduced in this thesis is which fragments admit the finite model property. It is an interesting property to have because as we have established, it implies decidability. Note, however, that the converse direction need not necessarily hold. When it comes to linear logic, we have the following situation, established by~\textcite{10.2307/2275637}.

\begin{proposition}
    \fragment{MLL} and \fragment{MALL} admit the finite model property, \fragment{MELL} and \fragment{LL} do not.
\end{proposition}

Thus, we already can infer that \fragment{MLL} and \fragment{MALL} are decidable. We have however not gained any knowledge about decidability of \fragment{MELL} and \fragment{LL}.

\section{Further Variants of Linear Logic \& Applications}\label{llapplications}

This section is meant as an outlook on the many ways we can adjust linear logic to fit certain applications. We first present a short overview of the main variants of linear logic that enjoy active research, and then name some applications from various areas of mathematics and computer science. 

\subsection{Further Variants}

We give a short overview of the different ways we can adjust linear logic to our needs, and, if fitting, note some remarking properties of the logics that emerge. Of course, this list is very general and incomplete, since the field is an area of active research.

\begin{description}
    \item[Intuitionistic linear logic] is constructed in an analogous way to its classical counterpart, by restricting the right-hand side of the sequent to contain only a single formula. It is also used as a basis for many of the following logics.
    \item[Affine linear logic] is constructed when we reintroduce the weakening rule into linear logic. It is often treated together with the pure fragment in complexity-theoretic treatments since it often exhibits similar behavior and is easier to reason with.
    \item[Non-commutative linear logic] is constructed by also omitting the exchange rule of the calculus. The order of formulas in the sequents matter and they are viewed as lists. A treatment of them can be found in~\textcite{DBLP:journals/lmcs/Slavnov19}.
    \item[Elementary linear logic] is, like the next two logics we present, a variant of linear logic designed to reason about implicit complexity. Introduced together with light linear logic in~\textcite{GIRARD1998175}, it is one of the first and simplest such logics and is a sound and complete representation of the elementary functions. We do this by adjusting the rules of \fragment{ILL}$_1$ for the exponentials to be
    \[\prftree[r]{(!mf)}{\Gamma\vdash A}{!\Gamma\vdash{}!A}\phantom{xxx}\prftree[r]{(!cL)}{\Gamma, !A, !A \vdash C}{\Gamma, !A\vdash C}\phantom{xxx}\prftree[r]{(!wL)}{\Gamma\vdash C}{\Gamma, !A\vdash C}.\] 
    \item[Light linear logic] is again used for implicit complexity, but uses a new modality, §. The rules for exponentials from \fragment{ILL}$_1$ are now adapted to be
    \[\prftree[r]{(!f)}{\Gamma\vdash A}{!\Gamma\vdash{} !A}\phantom{xxx}\prftree[r]{(§)}{\Gamma, \Delta\vdash A}{!\Gamma, \text{§}\Delta\vdash \text{§}A}\phantom{xxx}\prftree[r]{(!wL)}{\Gamma\vdash C}{\Gamma, !A\vdash C}.\]
    The class of functions on binary lists representable in \fragment{LLL} is exactly \class{FP}.
    \item[Soft linear logic] also classifies the class \class{FP}, but with other rules for the exponentials.
    \[\prftree[r]{(!mf)}{\Gamma\vdash A}{!\Gamma\vdash{}!A}\phantom{xxx}\prftree[r]{(mplex)}{\Gamma, A^n\vdash C}{\Gamma, !A\vdash C}\] 
    \item[Differential linear logic] is the result of viewing the vector space model for linear logic as given above, and wondering if this can be further enriched. Differential linear logic is treated, for example, by~\textcite{DBLP:journals/corr/CliftM17} as well as~\textcite{DBLP:journals/mscs/Ehrhard18}. It extends the exponential rules and has as models finiteness spaces and linear and continuous functions.
    \item[Hybrid linear logic] is an extension of \fragment{ILL} by the notions of worlds and hybrid connectives. It is, for example, treated in~\textcite{DBLP:journals/mscs/ChaudhuriDOP19}.
\end{description}

\subsection{Applications}

We close this chapter with a short outlook on how these variants of linear logic can be applied in a variety of scenarios. Again, we only highlight a small portion of the many research directions that have made use of linear logic.

\subsubsection{Quantum Logic}

\textcite{Gir_87} already had the idea that linear logic would be a prime candidate for quantum logics. In his introduction to phase semantics, he wrote:

\vspace{0.7em}
``One of the wild hopes that this suggests is the possibility of a direct connection with quantum mechanics\dots but let’s not dream too much!'' \cite[Section II.5]{Gir_87}.
\vspace{0.7em}

The idea of this was then fleshed out by~\textcite{615518}, who showed how linear logic can be used as a dynamic logic to describe quantum mechanics with an application to VLSI design. Indeed, especially the fact that information cannot simply be copied or destroyed can be viewed as baking the no-cloning theorem of quantum mechanics directly into the logical apparatus of linear logic.

Further treatments of this idea can be found in work by~\textcite{DBLP:conf/lics/AbramskyC04, DBLP:journals/mscs/AbramskyD06}. He laid the focus especially on categorical logics for quantum mechanics, a viewpoint that works well with linear logic, as we have seen.

Further connections between physics, topology, logic, and computation are presented by~\textcite{Baez_2010}, again using closed symmetric monoidal categories as a unifiying notion.

The idea of reasoning about quantum systems with linear logic is still an area of active research, with continuations found in formal systems for quantum programming languages. We describe this further in one of the following sections.

\subsubsection{Parallel Computation}

Another interpretation of linear logic atoms, other than resources, is that of concurrently acting and communicating agents. We can differentiate between the independent acting and the synched acting agents as we have seen above, via the notions of context-sharing and context-free connectives. \textcite{DBLP:journals/pacmpl/AschieriG20} use this notion together with the strong connection of linear logic to the $\lambda$-calculus to develop a $\lambda$-calculus with parallelism and communicating primitives, which can serve as a formal foundation of concurrent functional programs.

Another calculus that is widely used for the verification of concurrent processes is the $\pi$-calculus. Typing in this calculus is often done via so-called sessions types. They describe the input/output behavior of processes and provide vital guarantees such as deadlock freedom and fidelity. Based on \fragment{ILL},~\textcite{DBLP:journals/mscs/CairesPT16} give a type system that incorporates the key features of sessions types in the $\pi$-calculus.

\subsubsection{Cryptographic Protocols}

Linear logic also sees usage for the specification of cryptographic protocols. Again using linear logic to model parallel acting agents, an approach based on logic programming with linear Horn formulas by~\textcite{DBLP:conf/icalp/ComptonD99} describes cryptographic protocols and their attack vectors.

In~\textcite{CERVESATO20018} we see how linear logic can be incorporated as a foundation for the multiset rewriting model, which is a specification language for cryptographic protocols. The model was also used to prove undecidability results for cryptographic protocols.

\subsubsection{Implicit Computational Complexity}

In this thesis, we use methods of computational complexity to examine linear logic. But since the objects of study in linear logic are resources, we can turn this process around and use linear logic to encode complexity properties by restricting the calculus in such a way that only functions of a certain complexity can be constructed. This study is known as implicit computational complexity since contrary to classical approaches, we do not consider a model of computation that gives explicit complexity bounds like Turing machines or Boolean circuits.

This approach bears some similarity to the study of descriptive complexity. But while descriptive complexity examines the expressivity and complexity of logics by which problems they can encode, implicit complexity tries to provide calculi in which exactly the \emph{functions} of a certain complexity class are representable, without giving a complexity bound explicitly. This approach has its roots with the paper of~\textcite{DBLP:journals/cc/BellantoniC92}, who first gave a characterization of the polytime functions in this way. We have already seen this in the discussion on elementary, light, and soft linear logic, where the general approach is to limit the expressibility of the exponential fragment in such a way that we can only construct formulas to a specific degree.

\textcite{Baillot11, BAILLOT20153} gives an identification of elementary linear logic with the polynomial time class and the exponential time hierarchy by considering variants of the logic with fixed points and weakening. In another approach,~\textcite{DBLP:journals/tcs/BaillotG20} use \fragment{ElemLL} in conjunction with a $\lambda$-calculus with size types, which amounts to be a variation on Gödel's System T. An untyped and non-affine version of these results was developed by~\textcite{DBLP:journals/tcs/Laurent20}.

Combining the approach to use linear logic as a specification logic for quantum systems and for calculi for implicit complexity,~\textcite{Lag_10} give a quantum $\lambda$-calculus based on soft linear logic which is able to capture the quantum complexity classes \class{EQP}, \class{BQP}, and \class{ZQP}. The calculus they describe has quantum data be manipulated by classical control.

\subsubsection{Type Theory and Programming Languages}\label{sec:pl}

Another field where linear logic is applied with great success is the construction of resource-aware type systems. We have already seen that linear logic is often combined with type systems, a challenge in this is however to combine linear logic with another state-of-the-art foundation for modern type systems, \emph{dependent types}. They are the basis for dependent type theory, on which grounds many modern theorem provers like Agda, Coq, or Idris are built. A dependent type system is very expressive, which enables the language designer to move the detection of many errors, for example, array bound checking, from the runtime to the type checking phase. Efforts to combine linear and dependent types were made by~\textcite{DBLP:conf/birthday/McBride16}, which were later revised and extended to quantitative type theory by~\textcite{DBLP:conf/lics/Atkey18}.

\textcite{DBLP:journals/corr/abs-2004-13472} present a different approach, using fibrations on \reftodef[category!monoidal]{monoidal categories} to combine the two type theories. The result is a functional quantum programming language, which enforces the no-cloning theorem on the language level, treats quantum circuits as first-class citizens, and uses dependent types to index families of quantum circuits over classical parameters. A similar approach can be found in~\textcite{DBLP:journals/lmcs/RennelaS19}; they use enriched categories as a basis for the semantics.

Another example we have touched on before is the System F by Girard and Reynolds (cf.~\cite[Chapter 11]{10.5555/64805}), also called the polymorphic $\lambda$-calculus, which in contrast to the simply typed $\lambda$-calculus, also allows universal quantification over types and is as such a second-order calculus. It is a central formalism for proof theory and programming language theory, often in variations such as System F$_\omega$, the higher-order polymorphic $\lambda$-calculus, or System F$_{<:}$, where the system is extended by subtyping.

An example of a more mainstream programming language that aims to incorporate a linear typing system is Haskell~\cite{Bernardy_2018}. Haskell is a general-purpose programming language and one of the most widely used modern functional programming languages. The paper shows that a linear type system can be used efficiently in practice, leading to streamlined code which uses linear types. It furthermore inspired other strategies for programming language type systems, such as ownership typing in Rust.

\subsubsection{Proof Theory, Algebra, and Program Synthesis}

Of course, linear logic is also widely used in proof theory. Some recent advances in this field are the proof of the undecidability of the logic of action lattices~\cite{DBLP:conf/lics/Kuznetsov19}, or a categorical treatment of the proof theory of co-intuitionistic linear logic by~\textcite{DBLP:journals/corr/Bellin14}. He also builds models in \reftodef[category!monoidal]{monoidal categories} with additional structure.

\textcite{murfet2017logic} uses the connections between linear logic and \reftodef[category!symmetric monoidal]{symmetric closed monoidal categories} to give a correspondence between proofs in linear logic and algorithms for constructing morphisms in said categories. He starts from \fragment{ILL}, and later discusses \fragment{ILL}$_2$, in which every recursive function which is provably total in second-order Peano arithmetic can be encoded. This also has direct connections to linear algebra.

Another interesting direction of research is program synthesis. First, \textcite{Clift_2020} give an encoding of Turing machines into intuitionistic differential linear logic and then use this encoding to analyze the derivatives of programs~\cite{clift2019derivatives}. For a program $P$, the derivative $\partial P$ describes how the output of a program changes with an infinitesimal change of its input. They then use this for a machine learning approach to program synthesis via gradient descent. An even more elaborate approach to this is given in~\textcite{clift2021geometry}, where they associate programs to singularities of analytic functions and approximate weights using Markov chain Monte Carlo methods.
\chapter{Known Complexity Properties of Linear Logic}\label{ch:classic-comp}

In this chapter, we present the current state of research regarding the complexity of deciding provability in linear logic and its fragments. It will become apparent that there is no uniform way to classify the complexity of the fragments, each proof employs a different strategy. We also have some blank spots in the lattice: there is no known complexity characterization of \fragment{ELL} and \fragment{AELL} yet.

\section{Full Linear Logic is $\Sigma_1^0$-complete}\label{sec:ll-comp}

We start with the complexity of full propositional linear logic (\fragment{LL}) because it will show us that further complexity-theoretic examination of the full fragment is unnecessary: it is undecidable.

\begin{defproblem}[\fragment{LL}-\textsf{Provability}]\label{ll_prove2}
    \begin{description}
        \item[Input:] An \fragment{LL} sequent $\Gamma$.
        \item[Output:] Is $\Gamma$ provable in \fragment{LL}?
    \end{description}
\end{defproblem}

It is easy to see that the problem is recursively enumerable and that it is as such contained in $\Sigma^0_1$. Now we show that the problem is undecidable. To keep this thesis concise, we will not be performing a fully rigorous proof. Nevertheless, we will summarize the main ideas of the proof following the work of~\textcite{LMSS_92}. The proof consists of the following major steps:
\begin{enumerate}
    \item We define \emph{linear logic with theories} and prove a cut-standardization theorem for it.
    \item Using the cut-standardization, we show that pure linear logic is sound and complete in encoding linear logic with theories.
    \item We define \emph{and-branching two counter machines} and show that the halting problem for two counter machines with zero tests, which is known to be undecidable, can be reduced to their halting problem.
    \item We show that there is a sound and complete encoding of and-branching counter machines into linear logic with theories.
\end{enumerate}

\subsection*{Linear Logic with Theories}

\begin{definition}[Linear logic with theories]\label{llwt}
    \define[linear logic!with theories]{Linear logic with theories} is an extension of linear logic by theories, which are finite sets of axioms of the form
    \[\vdash C, p_{i_1}^\bot, p_{i_2}^\bot, \dots, p_{i_n}^\bot,\]
    where $C$ is a \fragment{MALL}-formula, and $p_{i_j}^\bot$ are negative literals, for $i,j,n\in\mathbb{N}$.

    For any theory $\mathbb{T}$, a sequent $\vdash \Gamma$ is provable in $\mathbb{T}$ if and only if we are able to derive $\vdash \Gamma$ using the standard set of linear logic proof rules, in combination with axioms from $\mathbb{T}$.
\end{definition}

\begin{remark}
    The notion of ``theory'' in this context differs slightly from the established meaning in classical logic, that is, a set of closed formulas closed under deduction. We employ the terminology employed by~\textcite{LMSS_92}. Note that the definition we give here is trivial in classical logic, since every classical formula can be reproduced arbitrarily often.
\end{remark}

Intuitively, axioms can be seen as reusable sequents which can occur as a leaf of a proof tree. We now go on to construct a pendant of the cut-elimination theorem of \reftodef[\textsf{LK}]{\fragment{LK}} in linear logic with theories. In this setting, instead of talking about cut-free proofs, we talk about \emph{directed proofs}, proofs in which all cuts have at least one premise which is an axiom. This leads to the following lemma:

\begin{lemma}[Cut-standardization in linear logic with theories]\label{cs-llwt}
    If there is a proof of $\vdash\Gamma$ in theory $\mathbb{T}$, then there is a directed proof of $\vdash\Gamma$ in theory $\mathbb{T}$.
\end{lemma}

To prove this lemma, the proof of cut-elimination of pure linear logic can be augmented to handle the cases where there are axioms in the premises. This proof is in turn performed mainly in the same way the cut-elimination for \fragment{LK} is. We eliminate cut rules inductively based on the degree of the formula. We have to note, however, that the induction is quite more involved than the proof for \fragment{LK} because of the additional information that a proof in linear logic carries with it. The linear logic counterpart of the mix rule is
\begin{displaymath}
    \prftree[r]{(cut!)}
        {\vdash \Sigma, (?A)^n}
        {\vdash \Delta, !A^\bot}
        {\vdash \Sigma, \Delta},
\end{displaymath}
where $n \ge 1$ and $(?A)^n$ denotes a multiset of formulas.

\subsection*{Embedding Linear Logic with Theories in Pure Linear Logic}

Now, the task is to give a sound and complete embedding of linear logic with theories in pure linear logic. Since the axioms we introduced consist only of \fragment{MALL} formulas, we actually extended just the \fragment{MALL} fragment. Its decision problem is, in its pure form, \class{PSPACE}-complete, which we will show in Section~\ref{sec:mall-comp}, but with axioms, it becomes undecidable, implying undecidability for \fragment{LL} as well.

The translation $\ulcorner\mathbb{T}\urcorner$ of a theory $\mathbb{T}$ with $k$ axioms is defined as a multiset of pure linear logic formulas by
\[\ulcorner\{\,t_1, t_2, \dots, t_k\,\}\urcorner = ?\ulcorner t_1\urcorner, ?\ulcorner t_2\urcorner, \dots, ?\ulcorner t_k\urcorner,\]
where $\ulcorner t_i\urcorner, 1\leq i\leq k$ is defined as
\[\ulcorner\vdash C, p_a^\bot, p_b^\bot, \dots, p_z^\bot\urcorner \coloneqq (C \parr p_a^\bot \parr p_b^\bot \parr \cdots \parr p_z^\bot) = (C^\bot \otimes p_a \otimes p_b \otimes \cdots \otimes p_z).\]

Note how this is an example of how the modalities provide enough expressibility for linear logic to be $\Sigma^0_1$-complete. We omit the proof that the translation is sound and complete. Next, we present the machine model which provides an intermediate problem for our reduction.

\subsection*{And-branching Two Counter Machines}

The halting problem for \define[two counter machine!with zero tests]{two counter machines with zero tests} is known to be undecidable, but it is difficult to encode the zero test instruction into linear logic. So instead, we define a very similar machine model, which is strong enough to simulate the zero test by being able to branch at certain steps in the computation.

\begin{definition}[And-branching two counter machine]\label{ab2cm}
    An \define[two counter machine!and-branching]{and-branching two counter machine} is a quadruple $M = (Q, \delta, Q_I, Q_F)$, where
    \begin{itemize}
        \item $Q$ is a finite set of states,
        \item $Q_I\in Q$ is the initial state,
        \item $Q_F\in Q$ is the final state,
        \item $\delta$ is the finite set of transitions of the form
        \begin{center}
            \begin{tabular}{ll}
                \textbf{Transition} & \textbf{Action}\\
                $(Q_i, \texttt{incr}, A, Q_j)$ & $(Q_i, A, B) \mapsto (Q_j, A + 1, B)$\\
                $(Q_i, \texttt{incr}, B, Q_j)$ & $(Q_i, A, B) \mapsto (Q_j, A, B + 1)$\\
                $(Q_i, \texttt{decr}, A, Q_j)$ & $(Q_i, A, B) \mapsto (Q_j, A - 1, B)$\\
                $(Q_i, \texttt{decr}, B, Q_j)$ & $(Q_i, A, B) \mapsto (Q_j, A, B - 1)$\\
                $(Q_i, \texttt{fork}, Q_j, Q_k)$ & $(Q_i, A, B) \mapsto \left\{\,(Q_j, A, B), (Q_k, A, B)\,\right\}$\\
            \end{tabular}
        \end{center}
        where $Q_i, Q_j, Q_k\in Q$ and $A, B$ represent counters. The $\texttt{decr}$ instructions do not apply if the corresponding counter is zero.
    \end{itemize}
    An \emph{instantaneous description} (ID) is a finite list of triples $(Q_i, A, B)$ with $Q_i\in Q$ and $A, B\in \mathbb{N}$. We define the accepting triple as $(Q_F, 0 , 0)$, and an ID is accepting if every element of the ID is an accepting triple.
\end{definition}

It is easy to see that to simulate two counter machines with zero tests, we can substitute the zero test by branching. We can therefore use and-branching machines instead of machines with zero tests in the following.

\subsection*{Encoding And-branching Two Counter Machines into Linear Logic with Theories}

Given an and-branching two counter machine $M = (Q, \delta, Q_I, Q_F)$, we define a set of propositions:
\[\left\{\,q_i \mid Q_i\in Q\,\right\} \cup \left\{\,q_i^\bot \mid Q_i\in Q\,\right\} \cup \left\{\,a, a^\bot, b, b^\bot\,\right\}.\]

We then translate the relation $\delta$ into axioms of a linear logic theory as follows:
\begin{align*}
    (Q_i, \texttt{incr}, A, Q_j) \phantom{xx}\rightsquigarrow &\phantom{xx}\vdash q_i^\bot, (q_j \otimes a),\\
    (Q_i, \texttt{incr}, B, Q_j) \phantom{xx}\rightsquigarrow &\phantom{xx}\vdash q_i^\bot, (q_j \otimes b),\\
    (Q_i, \texttt{decr}, A, Q_j) \phantom{xx}\rightsquigarrow &\phantom{xx}\vdash q_i^\bot, a^\bot, q_j,\\
    (Q_i, \texttt{decr}, B, Q_j) \phantom{xx}\rightsquigarrow &\phantom{xx}\vdash q_i^\bot, b^\bot, q_j,\\
    (Q_i, \texttt{fork}, Q_j, Q_k) \phantom{xx}\rightsquigarrow &\phantom{xx}\vdash q_i^\bot, (q_j \otimes q_k).\\
\end{align*}

The intuition behind this translation becomes clear when we denote the axioms with the use of linear implication. For example, the translation of $(Q_i, \texttt{incr}, A, Q_j)$ would be denoted as $\vdash q_i \multimap (q_j \otimes a)$. We see now clearly that we switch from the state $q_i$ to the state $q_j$ (or ``using'' the state $q_i$ to get the state $q_j$), gaining an $a$ in the process.

Now, we can define the translation of a triple $(Q_i, x, y)$ as
\[\theta\left((Q_i, x, y)\right) \coloneq\ \ \vdash q_i^\bot, a^{\bot^x}, b^{\bot^y}, q_F,\]
where
\[C^n \coloneqq \underbrace{C, C, \dots, C}_n.\]

The translation of an ID is simply the set of translations of the elements of the ID:
\[\theta\left(\{\,E_1, E_2, \dots, E_m\,\}\right) \coloneq \left\{\,\theta(E_1), \theta(E_2), \dots, \theta(E_m)\,\right\}.\]

It remains to be shown that with this translation, an and-branching two counter machine $M$ accepts an input $s$ if and only if every sequent in $\theta(s)$ is provable in the theory derived from $M$. Since the proof requires a technical induction on the height of the deduction trees, we do not perform it here and instead, refer to the proofs of Lemmas 3.5 and 3.6 in~\textcite{LMSS_92}.

We have seen how we can reduce the decision problem for two counter machines with zero tests to \fragment{LL}\textsf{-Provability}, from which the desired result immediately follows.

\begin{theorem}[Complexity of \fragmenttxt{LL}]\label{LL_undec}
    \fragment{LL}-$\mathsf{Provability}$ is $\Sigma_1^0$-complete.
\end{theorem}


\section{Multiplicative Exponential Linear Logic is \fragmenttxt{TOWER}-hard}\label{sec:mell-comp}

With the multiplicative-exponential fragment, we arrive at the most pressing complexity-theoretic question regarding linear logic: it is not yet known whether provability in \fragment{MELL} is decidable or not. Either way, we already know of lower bounds that would rule out the usage of the fragment in a practical setting. The provability problem of \fragment{MELL} is known to be \class{TOWER}-hard.

\begin{defproblem}[\fragment{MELL}-\textsf{Provability}]\label{mell_prove}
    \begin{description}
        \item[Input:] An \fragment{MELL} sequent $\Gamma$.
        \item[Output:] Is $\Gamma$ provable in \fragment{MELL}?
    \end{description}
\end{defproblem}

\textcite{dG_04} showed the inter-reducibility of \fragment{MELL} and a certain decision problem on so-called branching vector addition systems with states, an automaton model akin to the and-branching two counter machines we met earlier. \textcite{LS_15} then showed that this decision problem is \class{TOWER}-hard. The proof we present follows Lazi\'c and Schmitz and consists of two major steps:
\begin{enumerate}
    \item We show that the problem of reachability in BVASS is \class{TOWER}-hard.
    \item We show that reachability in BVASS can be reduced to \fragment{MELL}-\textsf{Provability}.
\end{enumerate}

\subsection*{Reachability in BVASS is \fragmenttxt{TOWER}-hard}

\begin{definition}[Branching vector addition system with states]\label{bvass}
    A \define{branching vector addition system with states} (BVASS) is a tuple $M = (Q, d, \delta_u, \delta_s)$, where
    \begin{itemize}
        \item $Q$ is a finite set of states,
        \item $d\in \mathbb{N}$ is the dimension,
        \item $\delta_u \subseteq Q\times \mathbb{Z}^d\times Q$ is a finite set of unary rules,
        \item $\delta_s \subseteq Q^3$ is a finite set of split rules.
    \end{itemize}
    We denote $(q, \bar{u}, q_1)\in \delta_u$ by $q\overset{\bar{u}}{\longrightarrow} q_1$ and $(q, q_1, q_2)\in \delta_s$ by $q\longrightarrow q_1 + q_2$.
\end{definition}

\begin{remark}
    In their paper, \textcite{LS_15} actually work with a generalization of both and-branching two counter machines, which they call alternating VASS, and BVASS. The model they use is called \emph{alternating branching VASS with full zero tests} ($\text{ABVASS}_{\,\bar{0}}$). They show that this model is inter-reducible to full propositional linear logic.
\end{remark}

\begin{definition}[Deduction semantics of BVASS]\label{sembvass}
    A \emph{configuration of a} BVASS is a pair $(q, \overline{v})\in Q\times \N^d$. To each rule type of the BVASS, we associate a deduction rule as follows:
    \[\prftree[r]{(unitary)}{q, \overline{v}}{q_1, \overline{v} + \overline{u}}\phantom{xxxx}\prftree[r]{(split)}{q, \overline{v}_1 + \overline{v}_2}{q_1, \overline{v}_1\ \ \ \ q_2, \overline{v}_2}\qedhere\]
\end{definition}

Given a BVASS $B$ and a finite set of states $Q_\ell$, we denote the fact that there is a deduction tree $D$ in $B$ with the root label $(q, \overline{v})$ and leaf labels in $Q_\ell \times \left\{\,\overline{0}\,\right\}$ by the \define{root judgement} $B, Q_\ell \triangleright q_r, \overline{0}$. The following rules hold for the root judgements:
\[\prftree[r]{if $q_\ell\in Q_\ell$}{}{B, Q_\ell \triangleright q_\ell, \overline{0}}\phantom{xxxx}\prftree[r]{if $q\overset{\bar{u}}{\longrightarrow} q_1$}{B, Q_\ell \triangleright q_1, \overline{v} + \overline{u}}{B, Q_\ell \triangleright q, \overline{v}}\]
\[\prftree[r]{if $q\longrightarrow q_1 + q_2$}{B, Q_\ell \triangleright q_1, \overline{v}_1}{B, Q_\ell \triangleright q_2, \overline{v}_2}{B, Q_\ell \triangleright q, \overline{v}_1 + \overline{v}_2}\]

From the root judgment, a decision problem naturally arises. This problem provides a bridge between a decision problem for Minsky machines, which is known to be \class{TOWER}-hard, and \fragment{MELL}. We will reduce this problem to \fragment{MELL}-\textsf{Provability} once we establish its hardness.

\begin{defproblem}[\textsf{BVASS}-\textsf{Reachability}]\label{bvass-reach}
    \begin{description}
        \item[Input:] A BVASS $B$, a finite set of states $Q_\ell$, a state $q_r$.
        \item[Output:] $B, Q_\ell \triangleright q_r, \overline{0}$?
    \end{description}
\end{defproblem}

To establish the lower complexity bound, we will make use of a well-known type of counter machine, the Minsky machine. They are named after Marvin Minsky, who formalized the model in 1961~\cite{10.2307/1970290} and have many established complexity properties.

\begin{definition}[Minsky machine]\label{minmach}
    A \define[Minsky machine]{Minsky machine} is a tuple $M = (Q, C, \delta_\uparrow, \delta_\downarrow, \delta_z)$, where
    \begin{itemize}
        \item $Q$ is a finite set of states,
        \item $C$ is a finite set of counters,
        \item $\delta_\uparrow \subseteq Q \times C\times Q$ is a finite set of increment rules,
        \item $\delta_\downarrow \subseteq Q\times C\times Q$ is a finite set of decrement rules,
        \item $\delta_z \subseteq Q\times C\times Q$ is a finite set of zero test rules.
    \end{itemize}
    We denote $(q, c, q_1)\in \delta_\uparrow$ by $q\overset{\incr c}{\longrightarrow} q_1$, $(q, c, q_1)\in \delta_\downarrow$ by $q\overset{\decr c}{\longrightarrow} q_1$, and $(q, c, q_1)\in \delta_z$ by $q\overset{c=0}{\longrightarrow} q_1$.
\end{definition}

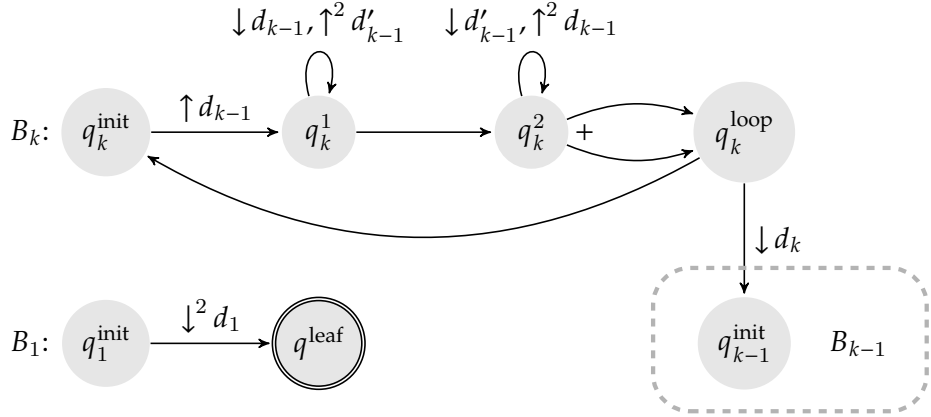
\begin{figure}[h]
    \centering
    \begin{tikzpicture}[->,>=stealth',shorten >=1pt,auto,node distance=2.8cm, semithick]
    \tikzstyle{every state}=[fill=gray!20!white,draw=none]
    
    \node[state]          (A)                    {$q_k^\text{init}$};
    \node[state]          (B) [right of=A]       {$q_k^1$};
    \node[state]          (C) [right of=B]       {$q_k^2$};
    \node[state]          (D) [right of=C]       {$q_k^\text{loop}$};
    \node[state]          (E) [below of=D]       {$q_{k-1}^\text{init}$};
    \node (Bk) [left of=A,node distance=1cm] {$B_k$:};
    \node (Plus) [right of=C,node distance=0.7cm] {$+$};

    \node[state]          (F) [below of=A]       {$q_{1}^\text{init}$};
    \node[state, accepting, draw]  (G) [right of=F]       {$q^\text{leaf}$};
    \node (B1) [left of=F,node distance=1cm] {$B_1$:};

    \path (A) edge node {$\incr d_{k-1}$} (B);
    \path (B) edge [loop above] node {$\decr d_{k-1}, \incr^2 d_{k-1}'$} (B);
    \path (B) edge node {} (C);
    \path (C) edge [loop above] node {$\decr d_{k-1}', \incr^2 d_{k-1}$} (C);
    \path (C) edge [bend left=20] node {} (D);
    \path (C) edge [bend right=20] node {} (D);
    \path (D) edge [bend left] node {} (A);
    \path (D) edge node {$\decr d_{k}$} (E);
    
    \path (F) edge node {$\decr^2 d_{1}$} (G);

    \draw  [ultra thick, dashed, black!30, rounded corners=14pt] (7.2, -3.7) -- (10.8, -3.7) -- (10.8, -1.8) -- (7.2, -1.8) -- cycle;

    \node  (lower) [right of=E, node distance=1.5cm]    {$B_{k-1}$};

    \end{tikzpicture}
    \caption{A hierarchy of BVASS. The symbol $\uparrow^n$ denotes $n$-fold increment, $\downarrow^n$ denotes $n$-fold decrement, and the two arrows with the $+$ a split rule.}
    \label{exbvass}
\end{figure}

Bounded Minsky machine halting problems provide natural hardness results for non-elementary complexity classes. This is covered thoroughly in~\textcite{Schmitz_2016}.

\begin{defproblem}[\textsf{F}$_3$-\textsf{MM}]\label{f3-mm}
    \begin{description}
        \item[Input:] A Minsky machine $M$, two states $q_0, q_H$.
        \item[Output:] Does $M$ have a computation starting from $q_0$ with all counters at zero, and ending in $q_H$ such that all counter values are at most tower$(|M|)$?
    \end{description}
\end{defproblem}

The name of the problem comes from the fact that this problem in Minsky machines is the trivially complete problem for the third level of the hierarchy that~\textcite{Schmitz_2016} defines. The following lemma immediately follows.

\begin{lemma}
    $\mathbf{\mathsf{F}}_3$-$\mathsf{MM}$ is \class{TOWER}-hard.
\end{lemma}

In the proof, we will make use of a hierarchy of BVASS given in Figure~\ref{exbvass}. The last lemma we need ensures that for each level in the hierarchy $k$ the existence of a deduction tree is bounded from below by tower$(k)$.

\begin{lemma}\label{exdedtree}
    For every $k\geq 1$ and vector of naturals $\overline{v}_0$ such that $\overline{v}_0(d_i) = \overline{v}_0(d_i') = 0$ for all $i < k$, we have that $B_k$ has a $(q^\textup{init}_k, \overline{v}_0)$-rooted $\left\{\,q^\textup{leaf}\,\right\}$-leaf-covering deduction tree if and only if $\overline{v}_0(d_k) \geq \textup{tower}(k)$.
\end{lemma}

\begin{proof}
    See~\textcite[Lemma 20]{LS_15}.
\end{proof}

With this work done, we can now show the \class{TOWER}-hardness of \fragment{BVASS}-\textsf{Reachability}.

\begin{figure}[t]
    \centering
    \begin{tikzpicture}[->,>=stealth',shorten >=1pt,auto,node distance=2.8cm, semithick]
    \tikzstyle{every state}=[fill=gray!20!white,draw=none]
    
    \node[state]          (A)                    {};
    \node[state]          (B) [right of=A]       {};

    \node[state]          (C) [right of=B]       {};
    \node[state]          (D) [right of=C]       {};
    
    \node[state]          (E) [below of=A]       {};
    \node[state]          (F) [right of=E]       {};
    \node[state]          (G) [right of=F]       {};
    \node[state]          (H) [right of=G]       {};
    \node[state]          (I) [below of=H]       {$q_K^\text{init}$};

    \node (B1) [left of=A,node distance=1cm] {$\incr c$:};
    \node (B2) [left of=C,node distance=1cm] {$\decr c$:};
    \node (B3) [left of=E,node distance=1cm] {$c = 0$:};

    \path (A) edge node {$\incr c, \decr \hat{c}$} (B);

    \path (C) edge node {$\decr c, \incr \hat{c}$} (D);

    \path (E) edge node {} (F);
    \path (F) edge node {} (G);
    
    \path (G) edge node {} (H);
    \path (G) edge node {} (I);
    \node (plus) [below right=-0.3cm and 0.1cm of G] {$+$};
    \path (F) edge [loop below] node {$\decr \hat{c}, \incr d_{K}, \incr c'$} (F);
    \path (G) edge [loop below] node {$\decr c', \incr \hat{c}$} (G);

    \draw  [ultra thick, dashed, black!30, rounded corners=14pt] (6.1, -6.6) -- (9.7, -6.6) -- (9.7, -4.7) -- (6.1, -4.7) -- cycle;

    \node  (lower) [left of=I, node distance=1.5cm]    {$B_{k}$};

    \end{tikzpicture}
    \caption{Simulating Minsky machine operations.}
    \label{mmsim}
\end{figure}
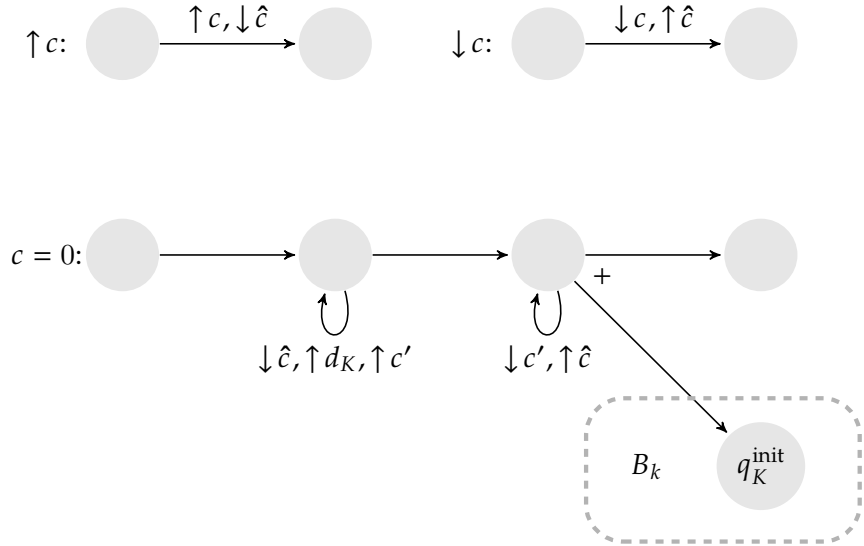

\begin{lemma}
    $\mathsf{F}_3$-$\mathsf{MM} \leq_m^\textup{log} \mathsf{BVASS}$-$\mathsf{Reachability}$.
\end{lemma}

\begin{proof}
    We show that, given a Minsky machine $M$ of size $K$ and two states $q_0, q_H$, we can construct in logarithmic space a BVASS $B(M)$, a state $q_r$ and a finite set $Q_\ell \coloneq \left\{\,q_H, q^\text{leaf}\,\right\}$, such that $M\in \textsf{\textbf{F}$_3$}\textsf{-MM}$ if and only if $B(M)\in \textsf{BVASS}\textsf{-Reachability}$.

    We represent each counter $c$ of $M$ with three counters in $B(M)$ which we denote $c$, $\hat{c}$, and $c'$. The initial part of $B(M)$ consists of a constant BVASS with a start state, input counter, final state, and output counter. It produces on input $m$ non-deterministically a value $\leq \text{tower}(m)$. We use this to initialize every $\hat{c}$ with a value $\leq \text{tower}(K)$.

    We then simulate the computation of $M$ on start state $q_0$ with the operations given in Figure~\ref{mmsim}. Observe that, since we apply the opposite operation of $c$ to $\hat{c}$ in the simulation of increments and decrements, $c + \hat{c}$ stays constant. For the zero test, we copy the value of $\hat{c}$ to $d_K$ using $c'$ and employ a hierarchy of BVASS defined in Figure~\ref{exbvass}.

    We can now show the correctness of the reduction, that is, $(M, q_0, q_H)\in\textsf{\textbf{F}$_3$}\textsf{-MM}$ if and only if $(B(M), Q_\ell, q_r)\in \textsf{BVASS}\textsf{-Reachability}$:
    \begin{description}
        \item[``$\implies$'':] We simulate a $0$-initialized tower$(K)$-bound computation of $M$ from $q_0$ to $q_H$ with $B(M)$ as follows:
        \begin{itemize}
            \item each $\hat{c}$ is initialized to tower$(K)$,
            \item we simulate a zero test by setting $c = 0, \hat{c} = \text{tower}(K), c' = 0, d_K = 0$ before the loops, and $c = 0, \hat{c} = \text{tower}(K), c' = 0, d_K = \text{tower}(K)$ before the split
            \item at every start of $B_K$, $d_K = \text{tower}(K)$ and all other counters are $0$.
        \end{itemize}
        By Lemma~\ref{exdedtree} we obtain a $(q_r, \overline{0})$-rooted $Q_\ell$-leaf-covering deduction tree of $B(M)$.
        \item[``$\impliedby$'':] Given a BVASS $B(M)$ with a $(q_r, \overline{0})$-rooted $Q_\ell$-leaf-covering deduction tree $D$, we obtain a $0$-initialized tower$(K)$-bounded computation of $M$ from $q_0$ to $q_H$ by observing that, by construction, $D$ consists of a path $\pi$ which consists of the simulation of increments, decrements and zero tests as shown in Figure~\ref{mmsim}. The computation is furthermore subject to the following properties:
        \begin{itemize}
            \item After $\hat{c}$ is initialized in $D$, we have $c + \hat{c} + c' \leq \text{tower}(K)$.
            \item For each simulation of a zero test of $c$, Lemma~\ref{exdedtree} ensures that the value of $d_K$ is tower$(K)$ before the split, $0$ after the split on $\pi$, and we have $c = 0, \hat{c} = \text{tower}(K)$, and $c' = 0$ before the loops.
            \item Any erroneous decrements of $c$ due to branchings can only occur after the last zero test of $c$, since such a decrement gives us $c + \hat{c} + c' < \text{tower}(K)$. Thus, such an erroneous decrement can not lead to an erroneous simulation.
            \item Similarly, only the last transfer of $c'$ to $\hat{c}$ may be incomplete.
        \end{itemize} 
        Thus, the computation of $B(M)$ has the required properties.\qedhere
    \end{description}
\end{proof}

\subsection*{Reachability in BVASS is Reducible to \fragmenttxt{MELL}-\textsf{Provability}}

\begin{lemma}
    $\mathsf{BVASS}$-$\mathsf{Reachability} \leq_m^\textup{log} \fragment{MELL}$-$\mathsf{Provability}$.
\end{lemma}

\begin{proof}
    We make again use of linear logic with theories from Definition~\ref{llwt} and their translation to pure linear logic. We encode a configuration $(q, \overline{v})\in Q\times \mathbb{N}^d$ of a BVASS $B = (Q, d, \delta_u, \delta_s)$ as
    \[\theta(q, \overline{v}) \coloneq \,\vdash q^\bot, \left(e_1^\bot\right)^{\overline{v}(1)},\dots, \left(e_d^\bot\right)^{\overline{v}(d)},\]
    where $Q\uplus\left\{\,e_i \mid i = 1,\dots,d\,\right\}$ is included in the set of atomic propositions. Notice the similarity to the translation of and-branching two counter machines. The definition of the axioms is also handled similarly:
    \begin{align*}
        q\overset{\overline{e}_i}{\longrightarrow} q_1 \phantom{xx}\rightsquigarrow &\phantom{xx}\vdash q^\bot, q_1 \otimes e_i,\\
        q\overset{-\overline{e}_i}{\longrightarrow} q_1 \phantom{xx}\rightsquigarrow &\phantom{xx}\vdash q^\bot, e_i^\bot, q_1,\\
        q \longrightarrow q_1 + q_2 \phantom{xx}\rightsquigarrow &\phantom{xx}\vdash q^\bot, q_1 \parr q_2.\\
    \end{align*}

    We thus have to show that for all $(q, \overline{v}) \in Q\times \mathbb{N}^d$, we have $B, Q_\ell \triangleright q, \overline{v}$ if and only if $\vdash \theta(q, \overline{v}), ?Q_\ell$ in linear logic with theories.

    We prove this again by induction on the height of the deduction tree. The first two rules are covered by~\textcite{LMSS_92}, only the split rule remains.

    We perform a direct proof by assuming that we have $B, Q_\ell \triangleright q, \overline{v}$ as a direct result of a split rule $q \longrightarrow q_1 + q_2$ with $\overline{v} = \overline{v}_1 + \overline{v}_2$, $B, Q_\ell \triangleright q_1, \overline{v}_1$ and $B, Q_\ell \triangleright q_2, \overline{v}_2$. We have by induction hypothesis $\vdash \theta(q_1, \overline{v}_1), ?Q_\ell$ and $\vdash \theta(q_2, \overline{v}_2)$, so we can prove with $(\otimes)$ that
    \[\vdash q_1^\bot \otimes q_2^\bot, (c_1^\bot)^{\overline{v}_1(1)+\overline{v}_2(1)},\dots,(c_d^\bot)^{\overline{v}_1(d)+\overline{v}_2(d)}, ?Q_\ell, ?Q_\ell.\]
    After $|Q_\ell|$ contractions, we can apply a directed cut with $\vdash q^\bot, q_1\parr q_2$, and obtain $\vdash \theta(q, \overline{v}), ?Q_\ell$ as desired.

    For the other direction, observe that the only rules that allow the application of a directed cut as given is a $(\otimes)$-rule followed by a series of weakening rules. Thus, we get $\vdash \theta(q_1, \overline{v}_1), ?Q_1$ and $\vdash \theta(q_2, \overline{v}_2), ?Q_2$ with $\overline{v} = \overline{v}_1 + \overline{v}_2$ and $Q_1 \cup Q_2\subseteq Q_\ell$. This gives us, by induction hypothesis, $B, Q_1 \triangleright q_1, \overline{v}_1$ and $B, Q_2 \triangleright q_2, \overline{v}_2$. Because $Q_1\subseteq Q_\ell$ and $Q_2\subseteq Q_\ell$, this entails $B, Q_\ell \triangleright q_1, \overline{v}_1$ and $B, Q_\ell \triangleright q_2, \overline{v}_2$ from which we can derive $B, Q_\ell \triangleright q, \overline{v}$ via a split rule.
\end{proof}

This concludes the reduction from the problem of BVASS reachability to the problem of \fragment{MELL} provability. We arrive at the following result:

\begin{theorem}[Complexity of \fragmenttxt{MELL}]\label{MELL_tower}
    \fragment{MELL}-$\mathsf{Provability}$ is \class{TOWER}-hard.
\end{theorem}

\section{Multiplicative Additive Linear Logic is \fragmenttxt{PSPACE}-complete}\label{sec:mall-comp}

Multiplicative additive linear logic, along with \fragment{MLL} counts to the most used and studied fragments of linear logic. The decision problem for provability of the fragment falls also in one of the most studied classes of complexity theory: the problem is \class{PSPACE}-complete. In this section, we again follow the work of~\textcite{LMSS_92}.

\begin{defproblem}[\fragment{MALL}-\textsf{Provability}]\label{mall_prove}
    \begin{description}
        \item[Input:] An \fragment{MALL} sequent $\Gamma$.
        \item[Output:] Is $\Gamma$ provable in \fragment{MALL}?
    \end{description}
\end{defproblem}

Both directions of the proof use the fact that the \fragment{MALL}-fragment enjoys cut-elimination. Luckily, this result is easily established since the proof of the cut-elimination can be adapted from the proof for the full fragment.

\begin{proposition}\label{cutelim-mall}
    Any sequent provable in \fragment{MALL} is provable without the cut rule.
\end{proposition}

\begin{proof}
    Since \fragment{MALL} is a fragment of \fragment{LL}, the procedure for Proposition~\ref{proposition:cut-ll} applies. Furthermore, since \fragment{MALL} has the \reftodef{subformula property} (cf. Definition~\ref{subform}), a cut-free proof of a \fragment{MALL} sequent contains only \fragment{MALL} formulas. Because all rules which apply to \fragment{MALL} formulas are already in \fragment{MALL}, it follows that the proof is a \fragment{MALL} proof.
\end{proof}

As is standard, the proof for \class{PSPACE}-completeness consists of two main parts: showing that the problem is contained in the class and that it is hard for the class. The first direction is an easy deduction from the fact that \fragment{MALL} admits cut-elimination, the second direction uses an encoding of the canonical \class{PSPACE}-complete problem, the evaluation of quantified Boolean formulas.

\subsubsection{Membership in \fragmenttxt{PSPACE}}
We start with the easier direction, the membership in \class{PSPACE}. Here, we use the cut-elimination in conjunction with the results about alternating time complexity we established above.

\begin{proposition}\label{inpspace-mall}
    \fragment{MALL}-$\mathsf{Provability} \in \class{PSPACE}$.
\end{proposition}

\begin{proof}
    By Proposition~\ref{cutelim-mall}, any provable \fragment{MALL} sequent has a cut-free \fragment{MALL} proof. In such a proof are at most two premises per rule and each premise is strictly smaller than the consequent. Thus, the depth of a cut-free \fragment{MALL} proof is bounded linearly w.\,r.\,t. the length of the final sequent of the proof.

    An alternating Turing machine can therefore decide in linear time if a cut-free proof is correct. For this procedure, it can use $\exists$-branching to guess a reduction in the cut-free proof, and $\forall$-branching to generate and check the proofs of both premises of a rule in parallel. By Proposition~\ref{atime-pspace}, we obtain \class{PSPACE}-membership.
\end{proof}

Alternatively, we could also prove \class{PSPACE}-membership by analyzing the space requirements of a nondeterministic Turing machine which generates a cut-free proof in a depth-first manner: the depth of the proof tree is bounded linearly w.\,r.\,t. the length of the final sequent of the proof. The extra memory required is bounded by the linear number of sequents, each again being linearly bounded in length. This gives us a quadratic space upper bound.

\subsubsection{\fragmenttxt{PSPACE}-Hardness}
\begin{proposition}\label{pspacehard-mall}
    \fragment{MALL}-$\mathsf{Provability}$ is \class{PSPACE}-hard.
\end{proposition}

We prove this by reduction from \problem{QBF} in two major steps:
\begin{itemize}
    \item We show that the evaluation of quantifier-free Boolean functions can be simulated by cut-free proof search in \fragment{MALL}.
    \item We then show that the evaluation of a QBF can be encoded into a \fragment{MALL}-sequent by simulating the Boolean quantifiers $\exists$ and $\forall$ with $\oplus$ and $\with$.
\end{itemize}

We denote with $\sigma(G)$ the \fragment{MALL}-sequent which represents the QBF $G$. A nice intuition for the following encodings becomes apparent when viewing the QBF as a Boolean circuit: variables are modeled as input signals, and connectives are modeled as gates with various input signals and an output signal. The sequent, therefore, consists of the encoding of the QBF $\llbracket G\rrbracket_g$, where $g$ is the output signal. We define $\llbracket G\rrbracket_g$ inductively on the length of the quantifier prefix.

For a quantifier-free Boolean function $M$, $[M]_g$ denotes the \fragment{MALL}-translation of $M$ which is defined inductively on the structure of $M$. Keep in mind that we use one-sided sequents. To avoid confusion, we use lower case letters to denote linear logic variables, upper case letters to denote classical variables, and upper case letters with bars to denote sets of variables when necessary.

In the definition, we will use the following auxiliary formulas:
\begin{align*}
    \textsc{not}(x, y) &\coloneq (x\otimes y)\oplus(x^\bot \otimes y^\bot)\\
    \textsc{and}(x, y, b) &\coloneq (x\otimes y\otimes b^\bot) \oplus (x^\bot \otimes y^\bot \otimes b) \oplus (x\otimes y^\bot \otimes b) \oplus (x^\bot \otimes y \otimes b)\\
    \textsc{copy}(x) &\coloneq (x \otimes (x^\bot \parr x^\bot)) \oplus (x^\bot \otimes (x\parr x))\\
    \textsc{copyAll}(\overline{X}) &\coloneqq \parr_{X_i\in\overline{X}}\textsc{copy}(x_i).
\end{align*}

Observe that \textsc{not} and \textsc{and} are simply encoding the truth table for their corresponding classical connectives and that \textsc{copy} and \textsc{copyAll} duplicate their inputs. This is needed to simulate the multiple usages of classical variables. With these formulas, we can now define the encoding of QBF formulas into sequents of \fragment{MALL}. Since $\{\,\neg, \wedge\,\}$ is a complete basis of Boolean logic, we only have to deal with these connectives in the propositional part of the QBF.

\begin{definition}\label{mall-encoding-qbf}
    Given a QBF $G$, the representing \fragment{MALL}-sequent $\sigma(G)$ is defined as
    \begin{align*}
        \sigma(G) &\coloneqq\ \,\vdash q_n, \llbracket G\rrbracket_g, g & q_n, g \text{ new,}\\
        \left\llbracket \left(\forall X_{i+1} G\right)\right\rrbracket_g &\coloneqq (q_{i+1}^\bot \otimes ((x_{i+1}\parr q_i)\with (x_{i+1}^\bot \parr q_i))), \llbracket G\rrbracket_g & q_{i+1} \text{ new,}\\
        \left\llbracket \left(\exists X_{i+1} G\right)\right\rrbracket_g &\coloneqq (q_{i+1}^\bot \otimes ((x_{i+1}\parr q_i)\oplus (x_{i+1}^\bot \parr q_i))), \llbracket G\rrbracket_g & q_{i+1} \text{ new,}\\
        \llbracket M\rrbracket_g &\coloneqq (q_0^\bot \otimes [M]_g) & q_0 \text{ new,}\\
        [X]_g &\coloneqq (x^\bot \otimes g) \oplus (x \otimes g^\bot) & \\
        [\neg N]_g &\coloneqq \textsc{not}(a, g) \parr [N]_a & a\text{ new,}\\
        [N\wedge P]_g &\coloneqq \begin{cases}
            \textsc{and}(a, b, g) \parr [N]_a \parr [P]_b \parr& \text{Var}(N) \cap \text{Var}(P) \neq \emptyset,\\
            \phantom{xxxx}\textsc{copyAll}(\text{Var}(N) \cap \text{Var}(P)), &\\
            \textsc{and}(a, b, g) \parr [N]_a \parr [P]_b, & \text{Var}(N) \cap \text{Var}(P) = \emptyset,
        \end{cases} & a,b\text{ new.}
    \end{align*}
    The annotation `$a$ new' indicates that $a$ must not occur anywhere else in the encoding.
\end{definition}

We look closer at the complexity of this encoding. First, we note that the encoding rule for the quantifiers adds only a constant length to the formula. The rule for literals demands logarithmic extra space in the size of $G$. The cost of \textsc{not} and \textsc{and} formulas is fixed w.\,r.\,t. the representation of the literals and the intersection can be modeled via bit-vectors describing the sets of variables. Since by this definition, we have to copy sets of variables at each conjunction, we have to use more than logarithmic space, but this can be remedied by determining the number of occurrences of each variable at the beginning of the algorithm and constructing the corresponding number of \textsc{copy} formulas. Thus, we see that we can implement the encoding in logarithmic space.

We now proceed to show that this encoding is correct with the two steps as described above: first, we show the correctness of the quantifier-free part, then we show how we can simulate the evaluation of the QBF by simulating the quantifiers.

Let $\mathfrak{I}$ be an assignment for a set of Boolean variables $\overline{Y}$, and $\overline{X}\subseteq \overline{Y}$. Then $\mathfrak{I}/\overline{X}$ is the assignment $\mathfrak{I}$ restricted to $\overline{X}$. By abuse of notation, we write $\mathfrak{I}/M$ for $\mathfrak{I}/\text{Var}(M)$ for a Boolean formula $M$. We also write $\langle\mathfrak{I}\rangle$ for the encoding of assignments into linear logic, which we do in the following manner:
\begin{align*}
    \langle\mathfrak{I}\rangle &\coloneq \langle X_1\rangle_\mathfrak{I},\dots,\langle X_n\rangle_\mathfrak{I}\\
    \langle X_i\rangle_\mathfrak{I} &\coloneq \begin{cases}
        x_i^\bot, & \mathfrak{I}(X_i) = 1,\\
        x_i, & \text{otherwise.}
    \end{cases}
\end{align*}

We use the linear negation of the variables because we work in the one-sided variant of the sequent calculus. Thus, the assignment $X_1, \neg X_2$ is encoded by $x^\bot_1, x_2$.

\begin{lemma}\label{lem24}
    Given sets of variables $\overline{X}$ and $\overline{Y}$, and an assignment $\mathfrak{I}$ for $\overline{X}\cup \overline{Y}$, there is a deduction of the sequent $\vdash \langle\mathfrak{I}\rangle, \textup{\textsc{copyAll}}(\overline{X} \cap \overline{Y}), \Gamma$ from the sequent $\vdash \langle\mathfrak{I}/\overline{X}\rangle, \langle\mathfrak{I}/\overline{Y}\rangle, \Gamma$. 
\end{lemma}

\begin{proof}
    The derivation is straightforward. Intuitively, instead of evaluating $\overline{X}$ and $\overline{Y}$ separately, we evaluate both, duplicating all variables that appear in $\overline{X}$ as well as $\overline{Y}$.
\end{proof}

\begin{lemma}\label{lem25}
    Let $M$ be a Boolean formula and $\mathfrak{I}$ an assignment for the variables in $M$, then
    \begin{enumerate}
        \item if $\mathfrak{I} \models M$, then $\vdash \langle\mathfrak{I}\rangle, [M]_g, g,$
        \item if $\mathfrak{I} \not\models M$, then $\vdash \langle\mathfrak{I}\rangle, [M]_g, g^\bot$.
    \end{enumerate}
\end{lemma}

\begin{proof}
    By induction on the structure of $M$.

    \begin{description}
        \item[Base case: $M \equiv X$.] Suppose $\mathfrak{I}(X) = 1$, then $\mathfrak{I} \models M$ and $\langle \mathfrak{I}\rangle = x^\bot$. By expansion of the definition of $[M]_g$, we get the following proof:
        \begin{displaymath}
            \prftree[r]{(R$\oplus$)}
            {\prftree[r]{($\otimes$)}{\prftree[r]{(id)}{}{\vdash x^\bot, x}}{\prftree[r]{(id)}{}{\vdash g^\bot, g}}{\vdash x^\bot, (x\otimes g^\bot), g}}
            {\vdash x^\bot, (x^\bot \otimes g)\oplus (x\otimes g^\bot), g}
        \end{displaymath}
        The proof for the case when $\mathfrak{I}(X) = 0$ is executed analogously.
        \item[Induction step.] We distinguish by cases according to the possible definitions of $[M]_g$. We will only perform the proof of one case, since the other cases can be handled in the same way.
          
        Let $M\equiv N\wedge P\textrm{ and Var}(N) \cap \textrm{Var}(P) \neq \emptyset$. Furthermore, suppose that $\mathfrak{I}/N\models N$ and $\mathfrak{I}/P \not\models P$, so that $\mathfrak{I}\not\models N\wedge P$. By expanding $[M]_g$, $\textsc{and}(a, b, g)$ and using Lemma~\ref{lem24}, we deduce:
        \begin{displaymath}
            \prftree[r]{($\parr$)}{
                \prftree[r]{}{
                    \prftree[r]{}{
                        \prftree[r]{($\otimes$)}{
                            \prftree[r]{($\otimes$)}{
                                \prftree[r]{(id)}{}{\vdash g, g^\bot}
                            }
                            {
                                \prftree[r]{($\otimes$)}{
                                    \overset{\vdots}{\vdash \langle\mathfrak{I}/N\rangle, [N]_a, a}
                                }
                                {
                                    \overset{\vdots}{\vdash \langle\mathfrak{I}/P\rangle, [P]_b, b^\bot}
                                }{\vdash \langle\mathfrak{I}/N\rangle, \langle\mathfrak{I}/P\rangle, (a\otimes b^\bot), [N]_a, [P]_b}
                            }{\vdash \langle\mathfrak{I}/N\rangle, \langle\mathfrak{I}/P\rangle, (a\otimes b^\bot, g), [N]_a, [P]_b, g^\bot}
                        }{\vdash \langle\mathfrak{I}/N\rangle, \langle\mathfrak{I}/P\rangle, \textsc{and}(a, b, g), [N]_a, [P]_b, g^\bot}
                    }{\vdots}
                }{\vdash \langle\mathfrak{I}\rangle, \textsc{and}(a, b, g), \textsc{copyAll}(\text{Var}(N)\cap \text{Var}(P)), [N]_a, [P]_b, g^\bot}
            }{\vdash \langle\mathfrak{I}\rangle, \textsc{and}(a, b, g)\parr \textsc{copyAll}(\text{Var}(N)\cap \text{Var}(P))\parr [N]_a\parr [P]_b, g^\bot}
        \end{displaymath}

        By applying the induction hypothesis to $\mathfrak{I}/N, N$ and $a$, and, respectively to $\mathfrak{I}/P, P$ and $b$, we can prove the remaining subgoals. We omit the other cases of $N\wedge P$ and of the structure of $M$ as they are handled similarly.\qedhere
    \end{description}
\end{proof}

\begin{lemma}\label{lem26}
    If $\vdash \Gamma$ is a provable \fragment{MALL}-sequent, then for any assignment of truth values to the atoms in $\Gamma$, there exists a formula $A$ in the sequence $\Gamma$ such that $A$ is true under the classical interpretation.
\end{lemma}

\begin{proof}
    By induction on the structure of cut-free \fragment{MALL}-proofs.
\end{proof}

The following lemma shows that we can simulate the evaluation of Boolean formulas regarding an assignment by assigning truth values to our linear logic encoding.

\begin{lemma}\label{lem27}
    Let $M$ be a Boolean formula and $\mathfrak{I}$ be an assignment for the variables in $M$. There exists an assignment $\mathfrak{K}$ of truth values to the atoms in $\langle\mathfrak{I}\rangle$ and $[M]_g$ such that for every formula $A$ in the sequence $\langle\mathfrak{I}\rangle$, $[M]_g$, assignment $\mathfrak{K}$ falsifies $A$ under the classical interpretation, and $\mathfrak{K}(g) = 1$ if and only if $\mathfrak{I}\models M$.
\end{lemma}

\begin{proof}
    By induction on the construction of $[M]_g$.
\end{proof}

\begin{lemma}\label{lem28}
    If $\mathfrak{I}$ is an assignment for the variables in a given Boolean formula $M$, then
    \begin{enumerate}
        \item if $\vdash \langle\mathfrak{I}\rangle, [M]_g, g$ is provable, then $\mathfrak{I}\models M$,
        \item if $\vdash \langle\mathfrak{I}\rangle, [M]_g, g^\bot$ is provable, then $\mathfrak{I}\not\models M$.
    \end{enumerate}
\end{lemma}

\begin{proof}
    Immediate from Lemmas~\ref{lem26} and~\ref{lem27}.
\end{proof}

\begin{lemma}\label{lem29}
    $\vdash \langle\mathfrak{I}\rangle, [M]_g, g$ is provable if and only if $\mathfrak{I}\models M$.
\end{lemma}

\begin{proof}
    Follows from Lemmas~\ref{lem25} and~\ref{lem28}.
\end{proof}

This concludes the correctness proof for the encoding of the quantifier-free part of the QBF. Next, we will deal with the encoding of the quantifiers.

\begin{lemma}
    If $q$ is a positive or negative literal and the sequent $\vdash q, \Gamma$ contains no constants, then $\vdash q, \Gamma$ is provable only if either $\Gamma \equiv q^\bot$ or $\Gamma$ contains at least one occurrence of a subformula either of the form $q^\bot\,\clubsuit\, A$ or the form $A\,\clubsuit\, q^\bot$, where $\clubsuit$ may be either $\oplus, \with$ or $\otimes$.
\end{lemma}

\begin{proof}
    By induction on cut-free \fragment{MALL}-proofs of $\vdash q, \Gamma$.
\end{proof}

\begin{lemma}\label{lem211}
    Let $M$ be a Boolean formula in the variables $X_1, \dots, X_n$, then for any $m$, $0\leq m \leq n$, and assignment $\mathfrak{I}$ for $X_{m+1}, \dots, X_n$, the relation $\mathfrak{I}\models Q_m X_m \cdots Q_1X_1 M$ holds if and only if the sequent $\vdash q_m, \langle\mathfrak{I}\rangle, \left\llbracket Q_m X_m \cdots Q_1 X_1 M\right\rrbracket, g$ is provable in \fragment{MALL}.
\end{lemma}

\begin{proof}
    By induction on $m$.
\end{proof}

This completes the correctness proof for the encoding of QBFs in \fragment{MALL} we have given. We can now prove the main theorem of this section.

\begin{theorem}[Complexity of \fragmenttxt{MALL}]\label{MALL_pspace}
    \fragment{MALL}-$\mathsf{Provability}$ is \class{PSPACE}-complete.
\end{theorem}

\begin{proof}
    Taking $m = n$ in Lemma~\ref{lem211}, it follows that a closed QBF $G$ is valid if and only if $\sigma(G)$ is provable in \fragment{MALL}. Furthermore, the encoding requires logarithmic space, as we have seen.
\end{proof}

\section{Focussed \fragmenttxt{MALL} Proofs and \fragmenttxt{PH}}

Now that we have established that \fragment{MALL} is \class{PSPACE}-complete, a natural question arises: can we systematically limit the expressivity of \fragment{MALL} to find corresponding fragments for classes in the polynomial hierarchy? The answer is yes, and~\textcite{DBLP:journals/jar/Das20} found such a correspondence by using multi-focussed proof search. The result is quite recent, and we will convey the main ideas in the construction, omitting the proofs. One main difference to other approaches to focussed proof search is that we have a third, \emph{deterministic} phase, accounting for invertible rules that do not branch. This phase is needed because otherwise, the estimation on upper bound of the complexity would be too coarse to fit in the desired complexity classes. In the presentation of the calculus, we use the metavariables for formulas restricted to some connectors as indicated in Table~\ref{metaform}. We use as delimiters $\Downarrow$ and $\Uparrow$ to distinguish between the parts of the sequent that are in different phases. The first arrow stands for the nondeterministic phase, the second for the co-nondeterministic phase.

\begin{table}[ht!]
    \centering
    \caption{Metavariables for multisets of formulas.}\label{metaform}
    \begin{tabular}{l l l}
        \toprule
        \textbf{Variable} & \textbf{Description} & \textbf{Connective}\\ \hline
        $M$ & negative and not deterministic & $\with$\\
        $N$ & negative & $\with, \parr$\\
        $O$ & deterministic & $\otimes, \parr$\\
        $P$ & positive & $\otimes, \oplus$\\
        $Q$ & positive and not deterministic & $\oplus$\\
        \bottomrule
    \end{tabular}
\end{table}

In his paper, Das first proves the correspondence between the levels of the hierarchy of quantified Boolean formulas and the focussing hierarchy for \fragment{MALL} with weakening by defining encodings in both directions which preserve quantifier alternations and phase alternations, respectively. He then adapts these encodings for \fragment{MALL} without weakening. Since we do not convey the details of the construction, we directly establish the correspondence between the QBF hierarchy and the focussing hierarchy for \fragment{MALL} without weakening.

The sequent calculus for the multi-focussed \fragment{MALL} fragment bears a strong similarity to the sequent calculus restricted to \fragment{MALL}. We introduce new rules to introduce and eliminate the division of the sequents into the parts of different phases. The rules are named $(D)$ and $(D^\bot)$ for ``decide'', and $(R)$ and $(R^\bot)$ for ``release'', respectively. We can then assign to the rules their respective phases. Notice that the unit rules are clearly deterministic, as is the $(\parr)$-rule. The rules for $\oplus$, $\otimes$, and $(R)$ are nondeterministic, and the rules for $\with$ and $(R^\bot)$ are co-nondeterministic. We employ the convention that we denote with $a, b,$ etc. atomic formulas, and with \textbf{a}, \textbf{b}, etc. and \textbf{A}, \textbf{B}, etc. multisets of (atomic) formulas.

\begin{definition}[Multi-focussed \fragmenttxt{MALL}]
    We consider the following sequent calculus, called \fragment{FMALL}.

    \textbf{Deterministic phase:}
    \begin{displaymath}
        \prftree[r]{(id)}{}{\vdash A, A^\bot}\hspace{2em} \prftree[r]{($\bot$)}{\vdash \Gamma}{\vdash \Gamma, \bot}\hspace{2em}\prftree[r]{(1)}{}{\vdash 1}\hspace{2em}\prftree[r]{($\top$)}{}{\vdash \Gamma, \top}\hspace{2em} \prftree[r]{($\parr$)}{\vdash \Gamma, A, B}{\vdash \Gamma, A\parr B}
    \end{displaymath}
    \[\prftree[r]{(D)}{\vdash \textbf{a}, \textbf{P}\Downarrow \textbf{P}'}{\vdash \textbf{a}, \textbf{P}, \textbf{P}'}\hspace{2em} \prftree[r]{(D$^\bot$)}{\vdash \textbf{a}, \textbf{P}\Uparrow \textbf{M}}{\vdash \textbf{a}, \textbf{P}, \textbf{M}}\]

    \textbf{Nondeterministic phase:}
    \begin{displaymath}
        \prftree[r]{($\oplus$)}{\vdash \Gamma \Downarrow \Delta, A_i}{\vdash \Gamma \Downarrow \Delta, A_0 \oplus A_1}\hspace{2em}\prftree[r]{($\otimes$)}{\vdash \Gamma \Downarrow\Sigma, A}{\vdash \Delta\Downarrow\Pi, B}{\vdash \Gamma, \Delta \Downarrow \Sigma, \Pi, A\otimes B}\hspace{2em}\prftree[r]{(R)}{\vdash \Gamma, \textbf{a}, \textbf{N}}{\vdash \Gamma \Downarrow \textbf{a}, \textbf{N}}
    \end{displaymath}
    
    \textbf{Co-nondeterministic phase:}
    \begin{displaymath}
        \prftree[r]{($\with$)}{\vdash \Gamma \Uparrow \Delta, A}{\vdash \Gamma \Uparrow \Delta, B}{\vdash \Gamma\Uparrow \Delta, A\with B}\hspace{2em}\prftree[r]{(R$^\bot$)}{\vdash \Gamma, \textbf{P}, \textbf{O}}{\vdash \Gamma \Uparrow \textbf{P}, \textbf{O}}
    \end{displaymath}
    Where \textbf{P}' and \textbf{M} are nonempty and $i\in \{\,0, 1\,\}$.
\end{definition}

To establish a connection to \fragment{MALL}, we need the notion of a bi-focussed subsystem of \fragment{FMALL}. That the full fragment is then also complete for \fragment{MALL} is a simple observation.

\begin{definition}
    A \fragment{FMALL} proof is \define[focussed proof]{focussed} if \textbf{P}' in $(D)$ is always a singleton. It is \define[focussed proof!co-focussed]{co-focussed} if \textbf{M} in $(D^\bot)$ is always a singleton. A proof that is both focussed and co-focussed is called \define[focussed proof!bi-focussed]{bi-focussed}.
\end{definition}

\begin{proposition}
    The class of bi-focussed \fragment{FMALL}-proofs is complete for \fragment{MALL}.
\end{proposition}

On the grounds of the sequent calculus we have just defined, we can build a hierarchy of sets of formulas. The intuition behind this hierarchy is similar to the polynomial hierarchy, but instead of counting the number of alternations of quantifiers, we count the number of alternations between the nondeterministic and co-nondeterministic phases.

\begin{definition}[Focussing hierarchy]
    A \reftodef{cedent} $\Gamma$ of \fragment{MALL} is
    \begin{itemize}
        \item $\Sigma^f_0$-provable (and also $\Pi^f_0$-provable) if $\vdash \Gamma$ is provable by using only deterministic rules.
        \item $\Sigma^f_{k+1}$-provable if there is a derivation of $\vdash \Gamma$, using only deterministic and non-determini\-stic rules, from sequents $\vdash \Gamma_i$ which are $\Pi^f_{k}$-provable.
        \item $\Pi^f_{k+1}$-provable if every maximal path from $\vdash \Gamma$, bottom-up, through deterministic and co-nondeterministic rules ends at a $\Sigma^f_k$-provable sequent.\qedhere
    \end{itemize}
\end{definition}

The idea is then to construct a complexity measure on \fragment{FMALL} formulas as follows. This corresponds directly to the hierarchy given above but has the advantage that we can calculate it directly.

\begin{definition}[Complexity measures for \fragmenttxt{FMALL}]
    Let $\Phi$ be a \fragment{FMALL} proof. We define
    \begin{itemize}
        \item The \emph{nondeterministic complexity} of $\Phi$, written $\sigma(\Phi)$, as the maximum number of alterations, bottom-up, between $(D)$ and $(D^\bot)$ steps in a branch through $\Phi$, setting $\sigma(\Phi) = 1$ if $\Phi$ has only $(D)$ steps.
        \item The \emph{co-nondeterministic complexity} of $\Phi$, written $\pi(\Phi)$, as the maximum number of alterations, bottom-up, between $(D)$ and $(D^\bot)$ steps in a branch through $\Phi$, setting $\pi(\Phi) = 1$ if $\Phi$ has only $(D^\bot)$ steps.
    \end{itemize}
    For a \reftodef{cedent} $\Gamma$, we further define the following:
    \begin{itemize}
        \item $\sigma(\Gamma)$ is the least $k\in \N$ s.\,t. there is a \fragment{FMALL} proof $\Phi$ of $\vdash \Gamma$ with $\sigma(\Phi) = k$.
        \item $\pi(\Gamma)$ is the least $k\in \N$ s.\,t. there is a \fragment{FMALL} proof $\Phi$ of $\vdash \Gamma$ with $\pi(\Phi) = k$.\qedhere
    \end{itemize}
\end{definition}

Unfortunately, while we can calculate these complexity measures, we can not \emph{efficiently} calculate them. To use them for an encoding, we thus have to give an efficiently calculable overapproximation for them. For this, we implicitly assume an order on the \fragment{FMALL}-formulas. It can be shown that the complexity measures are confluent for different orderings. In the definition, the metavariable \textbf{c} stands for formulas of the form
\[c \Coloneqq \bot \bnfsep c\oplus x \bnfsep x \oplus c \bnfsep x^\bot\oplus c\bnfsep c\oplus x^\bot \bnfsep c\oplus c.\]

\begin{definition}[Overapproximation for the complexity measures]
    We define the overapproximation for the complexity measures of \fragment{FMALL} sequents inductively as follows.
    \begin{align*}
        \lceil\sigma\rceil{}(\textbf{a}, \textbf{c})&\coloneq 1&\\
        \lceil\sigma\rceil{}(\Gamma, A\parr B)&\coloneq 
        \lceil\sigma\rceil{}(\Gamma, A, B)&\\
        \lceil\sigma\rceil{}(\textbf{a}, \textbf{c}, \textbf{P}, P)&\coloneq \lceil\sigma\rceil{}(\textbf{a}, \textbf{c}, \textbf{P} \Downarrow, P) & P \text{ is least in }\textbf{P},P\\
        \lceil\sigma\rceil{}(\textbf{a}, \textbf{c}, \textbf{P}, \textbf{M}, M)&\coloneq 1 + \lceil\pi\rceil (\textbf{a}, \textbf{c}, \textbf{P}, \textbf{M}, M)&\\
        &&\\
        \lceil\pi\rceil{}(\textbf{a}, \textbf{c})&\coloneq 1&\\
        \lceil\pi\rceil{}(\Gamma, A\parr B)&\coloneq 
        \lceil\pi\rceil{}(\Gamma, A, B)&\\
        \lceil\pi\rceil{}(\textbf{a}, \textbf{c}, \textbf{P}, P)&\coloneq 1 + \lceil\sigma\rceil{}(\textbf{a}, \textbf{c}, \textbf{P}) &\\
        \lceil\pi\rceil{}(\textbf{a}, \textbf{c}, \textbf{P}, \textbf{M}, M)&\coloneq \lceil\pi\rceil (\textbf{a}, \textbf{c}, \textbf{P}, \textbf{M} \Uparrow M)& M \text{ is least in }\textbf{M}, M\\
        &&\\
        \lceil\sigma\rceil{}(\Gamma \Downarrow A\oplus B)&\coloneq \begin{cases}
            \lceil\sigma\rceil{}(\Gamma, A)&\lceil\sigma\rceil{}(A) \geq \lceil\sigma\rceil{}(B)\\
            \lceil\sigma\rceil{}(\Gamma, B)&\text{otherwise}
        \end{cases}\\
        \lceil\sigma\rceil{}(\Gamma \Downarrow A\otimes B)&\coloneq \begin{cases}
            \lceil\sigma\rceil{}(\Gamma, A)&\lceil\sigma\rceil{}(A) \geq \lceil\sigma\rceil{}(B)\\
            \lceil\sigma\rceil{}(\Gamma, B)&\text{otherwise}
        \end{cases}\\
        \lceil\sigma\rceil{}(\Gamma \Downarrow X)&\coloneq \lceil\sigma\rceil (\Gamma, X)& X\text{ is } a \text{ or } c \text{ or }N.\\
        &&\\
        \lceil\pi\rceil{}(\Gamma \Uparrow A\with B)&\coloneq \begin{cases}
            \lceil\pi\rceil{}(\Gamma, A)&\lceil\pi\rceil{}(A) \geq \lceil\pi\rceil{}(B)\\
            \lceil\pi\rceil{}(\Gamma, B)&\text{otherwise}
        \end{cases}\\
        \lceil\pi\rceil{}(\Gamma \Downarrow X)&\coloneq \lceil\pi\rceil (\Gamma, X)& X\text{ is } O \text{ or } P.
    \end{align*}
    Where $\lceil\sigma\rceil{}$ is the overapproximation for the nondeterministic complexity and $\lceil\pi\rceil{}$ is the overapproximation for the co-nondeterministic complexity, respectively.
\end{definition}

With this complexity measure at hand, we can define sets of formulas whose decision problem is complete for each level of the polynomial hierarchy. As stated above, the proofs that the encoding (and the converse encoding from Boolean formulas to linear logic formulas not covered here) is sound, complete and efficiently computable can be found in the paper by~\textcite{DBLP:journals/jar/Das20}.

\begin{theorem}[Focussed \fragmenttxt{MALL} and \fragmenttxt{PH}]
    We have for $k \geq 1$ that
    \begin{itemize}
        \item $\left\{\,A\mid \lceil\sigma\rceil (A)\leq k \text{ and \fragment{MALL} proves }A\,\right\}$ is $\Sigma^p_k$-complete and
        \item $\left\{\,A\mid \lceil\pi\rceil (A)\leq k \text{ and \fragment{MALL} proves }A\,\right\}$ is $\Pi^p_k$-complete.
    \end{itemize}
\end{theorem}


\section{Multiplicative Linear Logic is \fragmenttxt{NP}-complete}\label{sec:mll-comp}

When we restrict \fragment{MALL} to omit the additive fragment, the decision also becomes easier\footnote{That is, of course, assuming $\class{NP}\neq \class{PSPACE}$.}. The problem is \class{NP}-complete. The complexity characterization of the \fragment{MLL}-fragment can be done in a very concise way by using the hardness result from Horn fragments we will establish later, which constitute a subset of \fragment{MLL}. For membership, the knowledge we have gained about \fragment{MLL} proof nets will become useful.

\begin{defproblem}[\fragment{MLL}-\textsf{Provability}]\label{mll_prove}
    \begin{description}
        \item[Input:] An \fragment{MLL} sequent $\Gamma$.
        \item[Output:] Is $\Gamma$ provable in \fragment{MLL}?
    \end{description}
\end{defproblem}

\begin{theorem}[Complexity of \fragmenttxt{MLL}]\label{MLL_np}
    \fragment{MLL}-$\mathsf{Provability}$ is \class{NP}-complete.
\end{theorem}

\begin{proof}
    The hardness of the decision problem follows directly from Corollary~\ref{hornnphard}.

    For membership, guess the cut-free \fragment{MLL} proof, which, by the \reftodef{subformula property} is bound polynomially w.\,r.\,t. the sequent we get as an input. We can then construct the proof net which corresponds to the proof, and verify it in polynomial time with the graph parsing procedure we presented in Section~\ref{ssec:proofnets}.
\end{proof}

\begin{remark}
    An interesting property of \fragment{MLL} is that even the restriction to constant-only \fragment{MLL} (that is, the \fragment{MLL}-fragment without the axiom) is already \class{NP}-complete~\cite{DBLP:journals/tcs/LincolnW94}. The problem used in the reduction to show \class{NP}-hardness is \textsf{3-Partition}, the same as for the Horn fragment used in Corollary~\ref{hornnphard}.
\end{remark}

\section{Additive Linear Logic is in \fragmenttxt{P}}

\begin{defproblem}[\fragment{ALL}-\textsf{Provability}]\label{all_prove}
    \begin{description}
        \item[Input:] An \fragment{ALL} sequent $\Gamma$.
        \item[Output:] Is $\Gamma$ provable in \fragment{ALL}?
    \end{description}
\end{defproblem}

For the provability of \fragment{ALL}, \textcite{DBLP:conf/lics/HeijltjesH15} have found an efficient algorithm that makes use of the structure of additive proof nets and works in the spirit of the graph parsing algorithm discussed in Section~\ref{ssec:proofnets}. We use a simulation of \fragment{ALL}$^-$-derivations via Petri nets and extend this approach to full \fragment{ALL} later. We use two-sided sequents since they convey the ideas most clearly. The construction of the fitting inference rules for \fragment{ALL} ((L$\with$), (L$\oplus$), (R$\with$), and (R$\oplus$)) is obvious. We will focus on the idea of the encoding and leave out the correctness proofs. They can be found in the aforementioned paper.

\begin{definition}[Additive proof nets]
    Given a sequent $\Gamma \vdash \Delta$, a \emph{link} $A \linking{} B$ connects a source subformula in $A\in \Gamma$ to a target subformula $B\in \Delta$. An \emph{axiom link} is a link between occurrences of the same atom.

    A \emph{(axiom) linking} on a sequent $\Gamma\vdash \Delta$ is a set of (axiom) links on $\Gamma\vdash \Delta$. We write $\lambda : \Gamma\vdash \Delta$ for a linking $\lambda$ on $\Gamma\vdash \Delta$.

    A \emph{resolution} $r$ for an \fragment{ALL}$^-$-formula $A$ is a function choosing one child for each subformula that is a product. A linking is \emph{discrete} if every resolution for $\Gamma\vdash \Delta$ retains exactly one link in $\lambda$.

    An \define[proof net!additive]{additive proof net} is a discrete axiom linking.
\end{definition}

To gain some intuition for how these proof nets and our encoding to Petri nets work, we will accompany it with the proof for associativity of $\with$. The sequent calculus presentation for this proof is as follows.

\[\prftree[r]{(R$\with$)}
                {\prftree[r]{(R$\with$)}
                    {\prftree[r]{(L$\with$)}
                        {\prftree[r]{(id)}{}{P\vdash P}}
                        {P\with (Q\with R)\vdash P}
                    }
                    {\prftree[r]{(L$\with$)}
                        {\prftree[r]{(L$\with$)}
                            {\prftree[r]{(id)}{}{Q\vdash Q}}
                            {Q\with R\vdash Q}
                        }
                        {P\with (Q\with R)\vdash Q}
                    }
                    {P\with (Q\with R)\vdash P\with Q}
                }
                {\prftree[r]{(L$\with$)}
                    {\prftree[r]{(L$\with$)}
                        {\prftree[r]{(id)}{}{R\vdash R}}
                        {Q\with R\vdash R}
                    }
                    {P\with (Q\with R)\vdash R}
                }
                {P\with (Q\with R) \vdash (P\with Q)\with R}\]

This proof can be represented as a proof net with axiom links.

\begin{center}
    \begin{tikzpicture}
        \node at (0,0) {$P$};
        \node at (0.3,0) {$\with$};
        \node at (0.55,0) {$($};
        \node at (0.75,0) {$Q$};
        \node at (1.15,0) {$\with$};
        \node at (1.5,0) {$R$};
        \node at (1.7,0) {$)$};
        \node at (-0.2,-1) {$($};
        \node at (0,-1) {$P$};
        \node at (0.35,-1) {$\with$};
        \node at (0.75,-1) {$Q$};
        \node at (0.95,-1) {$)$};
        \node at (1.2,-1) {$\with$};
        \node at (1.5,-1) {$R$};
        \draw [very thick] (0, -0.2) -- (0, -0.8);
        \draw [very thick] (0.75, -0.2) -- (0.75, -0.8);
        \draw [very thick] (1.5, -0.2) -- (1.5, -0.8);
    \end{tikzpicture} 
\end{center}

\begin{definition}[Petri net]
    A transition on a set $P$ is a pair $(s, t)$ with $s, t\subseteq P$. A \define[Petri net]{Petri net} $N = (P, \petritrans{})$ consists of a set of nodes $P$ and a transition relation $\petritrans{}$. A \emph{marking} $M\subseteq P$ consists of \emph{tokens}.

    \emph{Firing} is the rewrite relation on markings defined by
    \[M\rightsquigarrow (M\setminus s)\cup t\]
    for $s\subseteq M$, $s\petritrans{} t$, and $t\cap M = \emptyset$.

    A node $r\in P$ is called \emph{root} if it is not in the source of any transition. A Petri net is called \emph{rooted} if it has a unique root.
\end{definition}

\begin{figure}[t]
    \centering
    \begin{minipage}[t]{0.4\textwidth}
        \centering
        \scalebox{0.5}{\begin{tikzpicture}[thick,
            node distance=4cm,
            on grid,
            every transition/.style={fill=black,minimum width=.3cm, minimum height=0.3cm},
            every place/.style={draw=black!75},
            every label/.style={black!75}]
        
            \node[place] (place1) {};
            
            \node[place,
                right=of place1] (place2) {};
            
            \node[place,
                right= of place2] (place3) {};
            
            \node[transition,
            right=2cm of place1] (T1) {};
            \node[transition,
            right=2cm of place2] (T2) {};
            
            \draw (place1.east) edge[ultra thick] (T1.west)
                (T1.east) edge[ultra thick, post] (place2.west);
            \draw (place3.west) edge[ultra thick] (T2.east)
                (T2.west) edge[ultra thick, post] (place2.east);

            \node [below=1cm of place1, black!50, scale=2.5] {$P^\bot$};
            \node [below=1cm of place2, black!50, scale=2.5] {$\oplus$};
            \node [below=1cm of place3, black!50, scale=2.5] {$Q^\bot$};
        \end{tikzpicture}}
    \end{minipage}
    \begin{minipage}[t]{0.4\textwidth}
        \centering
        \scalebox{0.5}{\begin{tikzpicture}[thick,
            node distance=4cm,
            on grid,
            every transition/.style={fill=black,minimum width=.3cm, minimum height=0.3cm},
            every place/.style={draw=black!75},
            every label/.style={black!75}]
        
            \node[place] (place1) {};
            
            \node[place,
                right=of place1] (place2) {};
            
            \node[place,
                right= of place2] (place3) {};
            
            \node[transition,
            above=1cm of place2] (T1) {};
            
            \draw (place1.45) edge[bend left=10, ultra thick] (T1.west)
                (T1.east) edge[bend left=10, ultra thick] (place3.135)
                (T1.south) edge[ultra thick, post] (place2.north);

            \node [below=1cm of place1, black!50, scale=2.5] {$P$};
            \node [below=1cm of place2, black!50, scale=2.5] {$\with$};
            \node [below=1cm of place3, black!50, scale=2.5] {$Q$};
        \end{tikzpicture}}
    \end{minipage}
    \caption{Petri net simulation of the formulas $P^\bot\oplus Q^\bot$ and $P\with Q$.}
    \label{all2pn}
\end{figure}

We can encode \fragment{ALL}$^-$-formulas as Petri nets as shown in Figure~\ref{all2pn}. Notice that both nets are rooted. We can thus simulate the inductive definition of \fragment{ALL}$^-$-formulas by cascading the Petri nets, where the roots are the output of the subformula. For this, we write $N(A)$ for the Petri net generated by the \fragment{ALL}$^-$ formula $A$.

Formally, for two nets $N_1 = (P_1, \petritrans_1)$ and $N_2 = (P_2, \petritrans_2)$, we do this in the following way, using connectives corresponding to the linear logic connectives.

\begin{align*}
    N_1 \with N_2 &= (P_1 \uplus P_2 \uplus \{\,r\,\}, \petritrans_1 \uplus \petritrans_2 \uplus \petritrans_\with)\\
    N_1 \oplus N_2 &= (P_1 \uplus P_2 \uplus \{\,r\,\}, \petritrans_1\uplus \petritrans_2 \uplus \petritrans_\oplus)\\
\end{align*}

where $r_1, r_2\petritrans_\with r$ and $r_1\petritrans_\oplus r$ and $r_2\petritrans_\oplus r$.

To simulate the additive proof nets via Petri nets, we place a token in the place of the Petri net which corresponds to the axiom linking in the proof net.

We then define the cartesian product of two Petri nets as
\[N_1 \times N_2 \coloneq (P_1\times P_2, \petritrans),\]
where the transition relation is defined as
\begin{align*}
    \{\,p_1\,\} \times s_2 \petritrans{} \{\,p_1\,\} \times t_2 \hspace{5em}& \text{for all } p_1\in P_1\text{ and }s_2\petritrans_2 t_2\\
    s_1\times \{\,p_2\,\} \petritrans{} t_1 \times \{\,p_2\,\} \hspace{5em}& \text{for all }p_2\in P_2 \text{ and } s_1\petritrans_1 t_1.\\
\end{align*}

We want to check provability of sequents of the form $\Gamma\vdash \Delta$. The simulation on these is done via cartesian products of Petri nets, where for a sequent $\Gamma\vdash \Delta$, we construct the net for the sequent as $N(\Gamma\vdash \Delta) \coloneq N(\Gamma^\bot) \times N(\Delta)$.

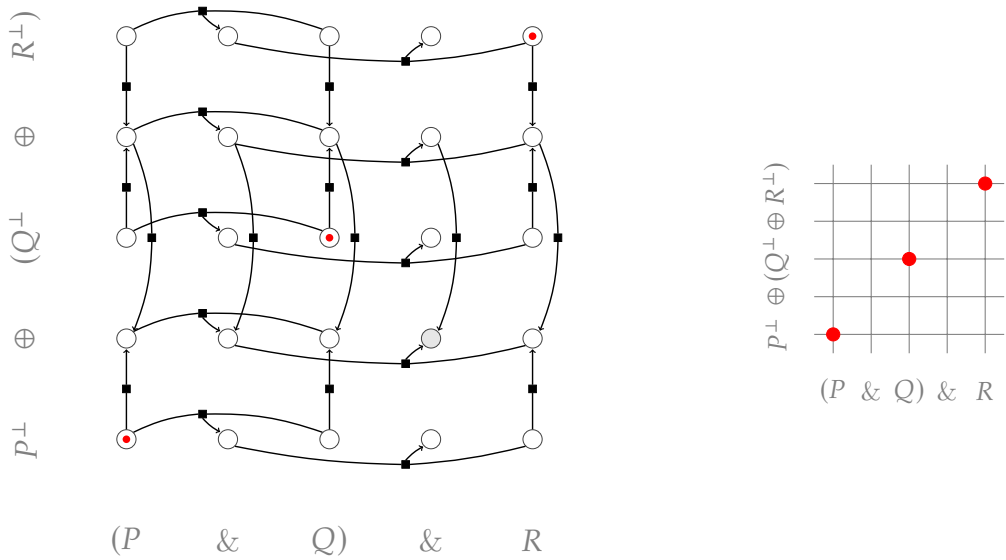
\begin{figure}[t]
    \centering
    \begin{minipage}{0.6\textwidth}
        \includegraphics[width=0.8\textwidth]{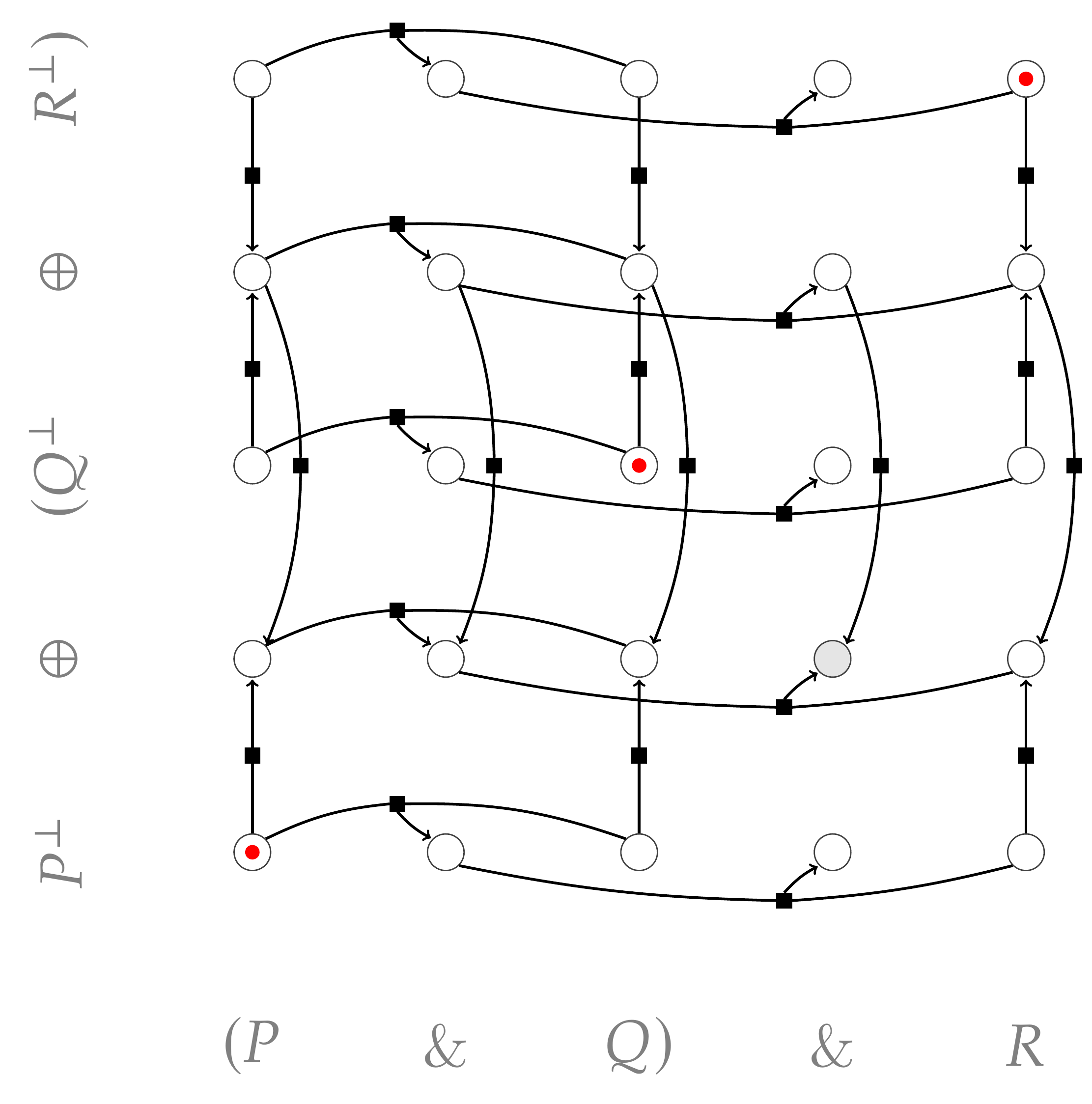}
    \end{minipage}
    \hspace{1em}
    \begin{minipage}{0.3\textwidth}
        \begin{tikzpicture}[scale=0.5]
            \draw[step=1.0,black!50,thin] (0.5,0.5) grid (5.5,5.5);
            \node [black!50, scale=0.9] at (1, -0.5) {$(P$};
            \node [black!50, scale=0.9] at (2, -0.5) {$\with$};
            \node [black!50, scale=0.9] at (3, -0.5) {$Q)$};
            \node [black!50, scale=0.9] at (4, -0.5) {$\with$};
            \node [black!50, scale=0.9] at (5, -0.5) {$R$};
            
            \node [black!50, rotate=90, scale=0.9] at (-0.5, 5) {$R^\bot)$};
            \node [black!50, rotate=90, scale=0.9] at (-0.5, 4) {$\oplus$};
            \node [black!50, rotate=90, scale=0.9] at (-0.5, 3) {$(Q^\bot$};
            \node [black!50, rotate=90, scale=0.9] at (-0.5, 2) {$\oplus$};
            \node [black!50, rotate=90, scale=0.9] at (-0.5, 1) {$P^\bot$};

            \draw[red,fill=red] (1,1) circle (1ex);
            \draw[red,fill=red] (3,3) circle (1ex);
            \draw[red,fill=red] (5,5) circle (1ex);
        \end{tikzpicture}
    \end{minipage}
    \caption{A Petri net for the proof of associativity along with its grid notation}
    \label{gridnot}
\end{figure}

Observe that the definition of the cartesian product of Petri nets leads to very regular nets. They can, however, be cumbersome to read and write, especially for larger sequents. To give a more concise notation for the cartesian product of a Petri net, we employ the grid notation, an example of which can be seen in Figure~\ref{gridnot}. 

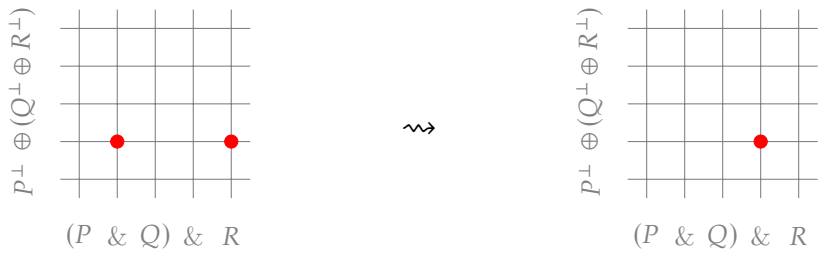
\begin{figure}[t]
    \centering
    \begin{minipage}{0.3\textwidth}
        \begin{tikzpicture}[scale=0.5]
            \draw[step=1.0,black!50,thin] (0.5,0.5) grid (5.5,5.5);
            \node [black!50, scale=0.9] at (1, -0.5) {$(P$};
            \node [black!50, scale=0.9] at (2, -0.5) {$\with$};
            \node [black!50, scale=0.9] at (3, -0.5) {$Q)$};
            \node [black!50, scale=0.9] at (4, -0.5) {$\with$};
            \node [black!50, scale=0.9] at (5, -0.5) {$R$};
            
            \node [black!50, rotate=90, scale=0.9] at (-0.5, 5) {$R^\bot)$};
            \node [black!50, rotate=90, scale=0.9] at (-0.5, 4) {$\oplus$};
            \node [black!50, rotate=90, scale=0.9] at (-0.5, 3) {$(Q^\bot$};
            \node [black!50, rotate=90, scale=0.9] at (-0.5, 2) {$\oplus$};
            \node [black!50, rotate=90, scale=0.9] at (-0.5, 1) {$P^\bot$};

            \draw[red,fill=red] (2,2) circle (1ex);
            \draw[red,fill=red] (5,2) circle (1ex);
        \end{tikzpicture}
    \end{minipage}
    \hspace{1em}
    $\rightsquigarrow$
    \hspace{4em}
    \begin{minipage}{0.3\textwidth}
        \begin{tikzpicture}[scale=0.5]
            \draw[step=1.0,black!50,thin] (0.5,0.5) grid (5.5,5.5);
            \node [black!50, scale=0.9] at (1, -0.5) {$(P$};
            \node [black!50, scale=0.9] at (2, -0.5) {$\with$};
            \node [black!50, scale=0.9] at (3, -0.5) {$Q)$};
            \node [black!50, scale=0.9] at (4, -0.5) {$\with$};
            \node [black!50, scale=0.9] at (5, -0.5) {$R$};
            
            \node [black!50, rotate=90, scale=0.9] at (-0.5, 5) {$R^\bot)$};
            \node [black!50, rotate=90, scale=0.9] at (-0.5, 4) {$\oplus$};
            \node [black!50, rotate=90, scale=0.9] at (-0.5, 3) {$(Q^\bot$};
            \node [black!50, rotate=90, scale=0.9] at (-0.5, 2) {$\oplus$};
            \node [black!50, rotate=90, scale=0.9] at (-0.5, 1) {$P^\bot$};

            \draw[red,fill=red] (4,2) circle (1ex);
        \end{tikzpicture}
    \end{minipage}
    \caption{An example of a legal firing in grid notation}
    \label{legalfir}
\end{figure}

Legal firings in the grid notation are, as in the case of Petri nets, determined by the parse trees of the two labeling formulas. An example of a legal firing can be seen in Figure~\ref{legalfir}.

Next, we will define alterations to the firing relations that will provide us with mechanisms for efficient proof search, and thereby an efficient algorithm for deciding provability. First, we will introduce a process called spawning. The intuitive idea behind this process is that we adapt the firing relation to keeping tokens in every state that we already visited. When we then reach the root link of the sequent with exhaustive spawning, the sequent we represent with the Petri net is provable.

\begin{definition}[Spawning]
    The \define{spawning rewrite relation} is generated by the following steps:
    \begin{itemize}
        \item given a link $A\linking B$ or $A\linking C$, add $A\linking B\oplus C$
        \item given two links $A\linking B$ and $A\linking C$, add $A\linking B\with C$
        \item given two links $A\linking C$ and $B\linking C$, add $A\oplus B\linking C$
        \item given a link $A\linking C$ or $B\linking C$, add $A\with B\linking C$.
    \end{itemize}
    The \emph{provability grid} of a sequent $A\vdash B$ is the result of exhaustive spawning on $\lambda_{A, B} A\vdash B$.
\end{definition}

\begin{proposition}
    A sequent $A\vdash B$ is provable in \fragment{ALL}$^-$ if and only if its provability grid contains the root link $A\linking B$.
\end{proposition}

With the help of the \reftodef{subformula relation} extended to a product order, that is we have $A\linking B\leq A'\linking B'$ if and only if $A$ is a subformula of $A'$ and $B$ is a subformula of $B'$, an algorithm for implementing the spawning relation has to only iterate once over the grid of a sequent $C\vdash D$. Because we construct a witness along with the calculation, we get the following result.

\begin{corollary}
    \fragment{ALL}$^-$-$\mathsf{Provability}\in \class{P}$.
\end{corollary}

We now extend the spawning rewrite relation to also include unit links, thus extending our result to the \fragment{ALL} fragment. Again, this rewrite relation gives rise to an efficient algorithm for proof search. We call this relation the saturation rewrite relation.

\begin{definition}[Saturation]
    The \define{saturation rewrite relation} is generated by the follwing steps:
    \begin{itemize}
        \item given $A\linking 1$ and $B\linking 1$, add $A\oplus B\linking 1$ and vice versa
        \item given $A\linking 1$ or $B\linking 1$, add $A\with B\linking 1$ and vice versa
        \item given $0\linking B$ or $0\linking C$, add $0\linking B\oplus C$ and vice versa
        \item given $0\linking B$ and $0\linking C$, add $0\linking B\with C$ and vice versa.\qedhere
    \end{itemize}
\end{definition}

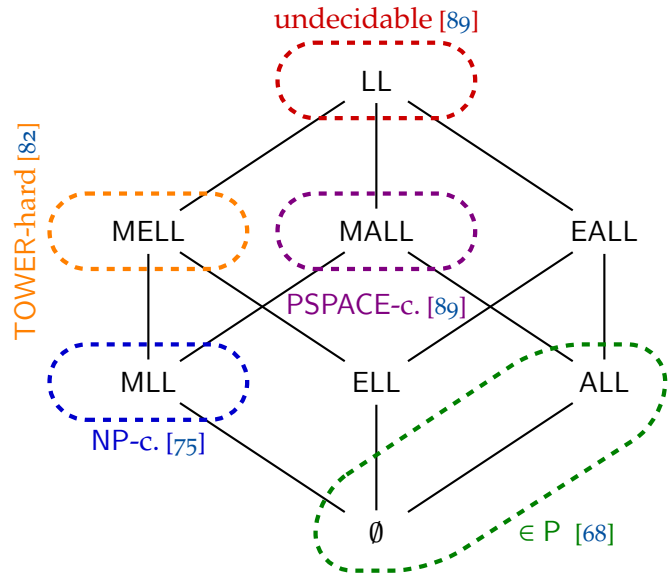
\begin{figure}[t]
    \centering
    \begin{tikzpicture}
        \node at (3,0) [] {$\emptyset$};
    
        \node at (0,2) [] {\fragment{MLL}};
        \node at (3,2) [] {\fragment{ELL}};
        \node at (6,2) [] {\fragment{ALL}};
    
        \node at (0,4) [] {\fragment{MELL}};
        \node at (3,4) [] {\fragment{MALL}};
        \node at (6,4) [] {\fragment{EALL}};

        \node at (3,6) [] {\fragment{LL}};

        \draw [thick, shorten >= 5mm, shorten <= 5mm] (3,0)  -- (0,2);
        \draw [thick, shorten >= 3mm, shorten <= 3mm] (3,0)  -- (3,2);
        \draw [thick, shorten >= 5mm, shorten <= 5mm] (3,0)  -- (6,2);
            
        \draw [thick, shorten >= 3mm, shorten <= 3mm] (0,2)  -- (0,4);
        \draw [thick, shorten >= 5mm, shorten <= 5mm] (0,2)  -- (3,4);
            
        \draw [thick, shorten >= 5mm, shorten <= 5mm] (3,2)  -- (0,4);
        \draw [thick, shorten >= 5mm, shorten <= 5mm] (3,2)  -- (6,4);
            
        \draw [thick, shorten >= 5mm, shorten <= 5mm] (6,2)  -- (3,4);
        \draw [thick, shorten >= 3mm, shorten <= 3mm] (6,2)  -- (6,4);

        \draw [thick, shorten >= 5mm, shorten <= 5mm] (3,6)  -- (0,4);
        \draw [thick, shorten >= 3mm, shorten <= 3mm] (3,6)  -- (3,4);
        \draw [thick, shorten >= 5mm, shorten <= 5mm] (3,6)  -- (6,4);

        \draw  [ultra thick, dashed, blue!80!black, rounded  corners =14pt] (-1.3, 1.5) -- (1.3, 1.5) -- (1.3, 2.5) -- (-1.3, 2.5) -- cycle;

        \draw  [ultra thick, dashed, orange, rounded  corners =14pt] (-1.3,4.5) -- (1.3,4.5) -- (1.3,3.5)  -- (-1.3,3.5) -- cycle;

        \draw  [ultra thick, dashed, red!80!black, rounded  corners =14pt] (1.7,6.5) -- (4.3,6.5) -- (4.3,5.5)  -- (1.7,5.5) -- cycle;

        \draw  [ultra thick, dashed, blue!50!red, rounded  corners =14pt] (1.7, 3.5) -- (4.3, 3.5) -- (4.3, 4.5) -- (1.7, 4.5) -- cycle;

        \draw  [ultra thick, dashed, green!50!black, rounded  corners =14pt] (2.2, 0.5) -- (5.2, 2.5) -- (6.8, 2.5) -- (6.8, 1.5) -- (3.7, -0.5) -- (2.2, -0.5) -- cycle;

        \node at (3,3) [blue!50!red] {\class{PSPACE}-c.\footnotesize{~\cite{LMSS_92}}};
        \node at (5.5,0) [green!50!black] {$\in$ \class{P}\footnotesize{~\cite{DBLP:conf/lics/HeijltjesH15}}};
        \node at (-1.6,4) [orange, rotate=90] {\class{TOWER}-hard\footnotesize{~\cite{LS_15}}};
        \node at (0,1.2) [blue!80!black] {\class{NP}-c.\footnotesize{~\cite{K_92}}};
        \node at (3,6.8) [red!80!black] {undecidable\footnotesize{~\cite{LMSS_92}}};
    \end{tikzpicture}
    \caption{Lattice of linear logic fragments with complexity classification.}
    \label{ll-fragcomp}
\end{figure}

\begin{theorem}[Complexity of \fragmenttxt{ALL}]
    \fragment{ALL}-$\mathsf{Provability}\in \class{P}$.
\end{theorem}
This concludes the current state of knowledge regarding complexity characterization of the main syntactic fragments of linear logic. An overview of what we have seen so far is presented in Figure~\ref{ll-fragcomp}. For simplicity, we omitted the focussing hierarchy, it would be placed right under \fragment{MALL}, which is the limit of the hierarchy. In classical propositional logic, the corresponding problem to the problem of provability we examine here is \class{TAUT}, the class of propositional tautologies. \class{TAUT} is \class{coNP}-complete for the whole propositional fragment, which emphasizes the vast increase in complexity we get for linear logic.

\section{The Complexity of Provability in Various Horn-Fragments}

We now move on to the complexity characterization of the Horn fragments of linear logic. The results that we present in this chapter are all due to the paper by~\textcite{K_92}. We consider decision problems of the following form, where $\spadesuit \in \mathcal{P}(\{\,\oplus, !, \with\,\})$, so we have a decision problem for each element in the lattice of Horn fragments.

\begin{defproblem}[$(\spadesuit)$\textsf{-Horn-Provability}]\label{hornprov}
    \begin{description}
        \item[Input:] Simple conjunctions $W, Z$, a multiset of $(\spadesuit)$-Horn clauses $\Gamma$.
        \item[Output:] Is $W, \Gamma \vdash Z$ provable in $(\spadesuit)$-Horn?
    \end{description}
\end{defproblem}

One nice fact about linear Horn fragments is that Kanovich found a computational model which corresponds to the fragments we will consider: branching Horn programs. 

\begin{definition}[Branching Horn program]\label{bhp}
    A \define[branching Horn program]{branching Horn program} is a finite binary tree such that for each edge of it, a Horn implication is associated with this edge.
\end{definition}

\begin{definition}[Semantics of branching Horn programs]\label{semhorn}
    Given a branching Horn program $P$ and a simple conjunction $W$, we use induction on $P$ to assign the simple conjunction $\val(W, v)$ to each vertex $v$ of $P$:
    \begin{itemize}
        \item For the root $v$, $\val(W, v) = W$.
        \item For a vertex $v$ and its child $v_1$, let $(X\multimap Y)$ be the linear Horn implication associated to the edge $(v, v_1)$.
        
        If $\val(W, v)$ is defined and $X\subseteq \val(W, v)$, then
        \[\val(W, v_1)\coloneq (Y\otimes (\val(W, v) - X)),\]
        where $A - B$ denotes the simple conjunctions that represents the difference of the associated multisets of $A$ and $B$. Otherwise, $\val(W, v_1)$ is undefined.
    \end{itemize}
    If for each leaf $v$ of $P$, $\val(W, v)$ is defined and $\val(W, v) = Z$, we say that
    \[P(W) = Z.\]

    \begin{itemize}
        \item To a non-branching non-terminal vertex $v$ and its child $v_1$, we assign the Horn implication $A$ to the edge $(v, v_1)$. We may associate a formula of the form $A$, $(A\with B)$ or $(B\with A)$ with this vertex $v$.
        \item For a branching vertex $v$ with its children $v_1$ and $v_2$, and the Horn implications $(X\multimap Y_1)$ and $(X\multimap Y_2)$ assigned to the edges $(v, v_1)$ and $(v, v_2)$, respectively, we associate with this vertex $v$ the $(\oplus)$-Horn implication $(X\multimap (Y_1 \oplus Y_2))$.\qedhere
    \end{itemize}
\end{definition}

\begin{example}[Branching Horn program]
    The following is a branching Horn program which transforms $p$ into $t$.
    \begin{center}
        \begin{tikzpicture}[>=stealth', shorten >= 1mm, shorten <= 1mm,
            squarednode/.style={rectangle, draw, very thick, minimum size=5mm},
            ]
            \node[squarednode] (v0) {$v_0$};

            \node[squarednode] (v1) [below left=of v0] {$v_1$};
            \node[squarednode] (v2) [below right=of v0] {$v_2$};

            \node[squarednode] (v3) [below left=of v1] {$v_3$};
            \node[squarednode] (v4) [below right=of v1] {$v_4$};
            \node[squarednode] (v5) [below=of v2] {$v_5$};

            \node[squarednode] (v6) [below=of v3] {$v_6$};
            \node[squarednode] (v7) [below=of v4] {$v_7$};
            \node[squarednode] (v8) [below=of v5] {$v_8$};
            
            \draw[->] (v0.south) -- (v1.north) node[midway, left] {$(p\multimap q)$};
            \draw[->] (v0.south) -- (v2.north) node[midway, right] {$(p\multimap r)$};
            
            \draw[->] (v1.south) -- (v3.north) node[midway, left] {$(q\multimap r)$};
            \draw[->] (v1.south) -- (v4.north) node[midway, right] {$(q\multimap q)$};
            \draw[->] (v2.south) -- (v5.north) node[midway, right] {$(r\multimap q)$};
            
            \draw[->] (v3.south) -- (v6.north) node[midway, left] {$(r\multimap t)$};
            \draw[->] (v4.south) -- (v7.north) node[midway, left] {$(q\multimap t)$};
            \draw[->] (v5.south) -- (v8.north) node[midway, right] {$(q\multimap t)$};
        \end{tikzpicture}
    \end{center}
    We can see that each branch of this program uses each formula from the multiset
    \[(p\multimap (q\oplus r)), (q\multimap (r\oplus q)), ((q\multimap t)\with (r\multimap t))\]
    exactly once. It is thus a model for the $(\oplus, \with)$-Horn fragment.
\end{example}

Next, we show that branching Horn programs stand in complete correspondence to generalized Horn implications. Since the proof of this result is very technical, we give just the main idea. The full proof can be found in~\textcite{DBLP:journals/apal/Kanovich94}.

\begin{proposition}[Soundness and completeness of Horn programs]\label{completenesshorn}
    For any $\Gamma$ and $\Delta$ consisting of generalized Horn implications, a sequent of the form
    \[W, \Gamma, !\Delta \vdash Z\]
    is derivable in linear logic if and only if we can construct a branching Horn program $P$ such that
    \begin{itemize}
        \item All formulas used in the program $P$ are from either $\Gamma$ or $\Delta$.
        \item For every branch $b$ of $P$, each formula from $\Gamma$ is used on this branch $b$ exactly once.
        \item For every branch $b$ of $P$, each formula from $\Delta$ may be used on this branch $b$ any number of times.
        \item $P(W) = Z$.
    \end{itemize}
\end{proposition}

\begin{proof}
    The idea is to construct an intermediate calculus for generalized Horn sequents. Then we show that arbitrary derivations in linear logic can be encoded in this calculus, the other direction is trivial.
    
    Then we show that derivations in the intermediate calculus can be transformed into branching Horn programs and vice versa. The intermediate calculus can be found in the paper by~\textcite{DBLP:journals/apal/Kanovich94}.
\end{proof}

\begin{remark}
    From this, results regarding space complexity immediately follow: in particular, all $!$-free linear Horn fragments are solvable in deterministic linear space.
\end{remark}

We will focus on the \class{NP}- and \class{PSPACE}-completeness results. For the sake of completeness, we will sketch the proof ideas for the other fragments.

\begin{theorem}[Complexity of $(!)$-\textsf{Horn} and $(!, \with)$-\textsf{Horn}]
    The decision problems $(!)$-$\mathsf{Horn}$-$\mathsf{Provability}$ and $(!, \with)$-$\mathsf{Horn}$-$\mathsf{Provability}$ are decidable.
\end{theorem}

\begin{proof}
    The general idea is to reduce the problem of $(!, \with)$\textsf{-Horn-Provability} to the problem of reachability in vector addition systems found in~\textcite{10.1145/800076.802477}. This problem for vector addition systems is known to be polynomially reducible to the problem of reachability in Petri nets, which is decidable.
\end{proof}

\begin{theorem}[Complexity of $(!, \oplus)$-\textsf{Horn}]
    $(!, \oplus)$-$\mathsf{Horn}$-$\mathsf{Provability}$ is undecidable.
\end{theorem}

\begin{proof}
    The proof of~\textcite{LMSS_92} also applies to the $(!, \oplus)$-Horn fragments.
\end{proof}

We now prove the complexity properties of the fragments which are of the main interest for this thesis. We will first show \NP-hardness of \textsf{Horn-Provability} from which the \NP-hardness for the other two fragments immediately follows. The following problem is \NP-complete according to~\textcite{10.5555/574848}.
\begin{defproblem}[\textsf{3-Partition}]\label{3Part}
    \begin{description}
        \item[Input:] $b\in \mathbb{N}, m, k\in \mathbb{Z}$, $s \in \mathbb{Z}_+^k$ such that $k = 3m$ and $\frac{b}{4} < s_i < \frac{b}{2}, s_i\in s, 1\leq i\leq k$.
        \item[Output:] Can $\{\,1, 2, \dots,k\,\}$ be partitioned into $m$ disjoint sets $S_1, S_2, \dots, S_j, \dots, S_m$ such that for each $1\leq j\leq m$
        \[\sum_{i\in S_j} s_i = b?\] 
    \end{description}
\end{defproblem}

To reduce \textsf{3-Partition} to \textsf{Horn-Provability}, we will encode instances of \textsf{3-Partition} as a Horn sequent. For this, let $\textit{PR}_s$ be a multiset consisting of the Horn implications
\[\left\{\,(p\multimap (q^{b-s_i} \otimes r^{s_i}))\,\right\}, 1\leq i\leq k,\]
and let $\textit{RP}_m$ be the multiset of $m$ copies of th Horn implication
\[((q^{2b}\otimes r^b) \multimap p^3).\]

We can now show that the Horn sequent
\[p^3, \textit{PR}_s, \textit{RP}_m \vdash p^3\]
is derivable in the Horn fragment of linear logic if and only if the corresponding instance is a member of \textsf{3-Partition}. The left to right implication is trivial, the right to left implication uses an $m$-fold application of the following lemma.

\begin{lemma}[Pulsing]
    If the Horn sequent
    \[p^3, \textit{PR}_s, \textit{RP}_m \vdash p^3\]
    is derivable in linear logic, then we can find different integers $i_1, i_2, i_3$ such that
    \begin{enumerate}
        \item $1\leq i_1, i_2, i_3 \leq k$,
        \item $s_{i_1} + s_{i_2} + s_{i_3} = b$,
        \item the Horn sequent
        \[p^3, \textit{PR}_{s'}, \textit{RP}_{m-1} \vdash p^3\]
        is also derivable in linear logic, where $s'$ is the following $(k-3)$-dimensional vector:
        \[(s_1, \dots, s_{i_1 - 1}, s_{i_1 + 1},\dots,s_{i_2 - 1}, s_{i_2 + 1},\dots,s_{i_3 - 1}, s_{i_3 + 1}, \dots,s_{k}).\]
    \end{enumerate}
\end{lemma}

\begin{proof}
    By Proposition~\ref{completenesshorn}, there exists a Horn program for the sequent $p^3, \textit{PR}_s, \textit{RP}_m \vdash p^3$, of which the first four steps have the following form. We label each vertex $v$ with $\val(p^3, v)$ on the left for convenience.
    \begin{center}
        \begin{tikzpicture}[>=stealth', shorten >= 1mm, shorten <= 1mm,
            squarednode/.style={rectangle, draw, very thick, minimum size=5mm},
            ]
            \node[squarednode] (v0) {$v_0$};
            \node[squarednode] (v1) [below=of v0] {$v_1$};
            \node[squarednode] (v2) [below=of v1] {$v_2$};
            \node[squarednode] (v3) [below=of v2] {$v_3$};
            \node[squarednode] (v4) [below=of v3] {$v_4$};

            \node (v0f) [left=0.1cm of v0] {$p^3$};
            \node (v1f) [left=0.1cm of v1] {$p^2\otimes q^{b - s_{i_1}} \otimes r^{s_{i_1}}$};
            \node (v2f) [left=0.1cm of v2] {$p \otimes q^{2b - s_{i_1} - s_{i_2}} \otimes r^{s_{i_1} + s_{i_2}}$};
            \node (v3f) [left=0.1cm of v3] {$q^{3b - s_{i_1} - s_{i_2} - s_{i_3}} \otimes r^{s_{i_1} + s_{i_2} + s_{i_3}}$};
            \node (v4f) [left=0.1cm of v4] {$p^3$};
            
            \draw[->] (v0.south) -- (v1.north) node[midway, right] {$(p\multimap (q^{b - s_{i_1}} \otimes r^{s_{i_1}}))$};
            \draw[->] (v1.south) -- (v2.north) node[midway, right] {$(p\multimap (q^{b - s_{i_2}} \otimes r^{s_{i_2}}))$};
            \draw[->] (v2.south) -- (v3.north) node[midway, right] {$(p\multimap (q^{b - s_{i_3}} \otimes r^{s_{i_3}}))$};
            \draw[->] (v3.south) -- (v4.north) node[midway, right] {$((q^{2b}\otimes r^b)\multimap p^3)$};
        \end{tikzpicture}
    \end{center}
    Observe that the Horn implications used in the first three vertices stem from $\textit{PR}_s$, while the last one stems from $\textit{RP}_m$. The last step is possible because by the definition of \textsf{3-Partition}, we have $s_{i_1} + s_{i_2} + s_{i_3} = b$. We can construct a program $P'$ by starting from vertex $v_4$. We then have $P'(p^3) = p^3$, but as a Horn sequent, as desired,
    \[p^3, \textit{PR}_{s'}, \textit{RP}_{m-1} \vdash p^3,\qedhere\]
\end{proof}

\begin{corollary}\label{hornnphard}
    $(\with)$-$\mathsf{Horn}$-$\mathsf{Provability}$, $(\oplus)$-$\mathsf{Horn}$-$\mathsf{Provability}$ and $\mathsf{Horn}$-$\mathsf{Provability}$ are \NP-hard.
\end{corollary}

This concludes the encoding of \textsf{3-Partition} into the Horn fragment of linear logic. We next show that the three problems are in \NP. We do this in two major steps: we first show that $(\with)$\textsf{-Horn-Provability} (and thus \textsf{Horn-Provability}) is in \NP. Then we reduce $(\oplus)$\textsf{-Horn-Provability} to \textsf{Horn-Provability}.

\begin{lemma}\label{horn-np}
    $(\with)$-$\mathsf{Horn}$-$\mathsf{Provability}\in \NP$.
\end{lemma}

\begin{proof}
    We search for a derivation by guessing a corresponding \reftodef[branching Horn program]{branching Horn program}, which has no branching vertices and is bounded in length by the length of the sequent. We can thus verify the solution in polynomial time.
\end{proof}

To reduce sequents of the $(\oplus)$-Horn fragment to ones of the Horn fragment, we make use of the following lemma.

\begin{lemma}[Inverse]
    Let $\Delta$ be a multiset consisting of only Horn implications and let both linear Horn sequents $Y_1, \Delta \vdash Z$ and $Y_2, \Delta \vdash Z$ be derivable. Then $Y_1 = Y_2$.
\end{lemma}

\begin{proof}
    This is due to the fact that linear Horn sequents are balanced with respect to occurrences of positive and negative literals.
\end{proof}

When we apply this lemma to Horn programs, we get for a vertex $v$ with children $v_1, v_2$ that $\val(W, v_1) = \val(W, v_2)$. Together with the applicability conditions for the $(\oplus)$-Horn implication $(X\multimap (Y_1 \oplus Y_2))$ that is used in $v$, we get $Y_1 = Y_2$. We can infer that for reverse computations from the leaves of a Horn program to the root, each $(\oplus)$-Horn implication $(X\multimap (Y_1 \oplus Y_2))$ is actually non-branching. We can thus replace all $(\oplus)$-Horn implications by Horn implications and arrive at the Horn fragment, which we have shown to be in \NP.

\begin{corollary}\label{oplushornnp}
    $(\oplus)$-$\mathsf{Horn}$-$\mathsf{Provability}\in \NP$.
\end{corollary}

Thus, $(\with)$\textsf{-Horn-Provability}, $(\oplus)$\textsf{-Horn-Provability} and \textsf{Horn-Provability} are in \NP. We summarize the results of Lemma~\ref{horn-np} and Corollaries~\ref{oplushornnp} and~\ref{hornnphard} in the following theorem. 

\begin{theorem}[Complexity of $(\with)$-\textsf{Horn}, $(\oplus)$-\textsf{Horn}, and \textsf{Horn-Provability}]
    The decision problems $(\with)$-$\mathsf{Horn}$-$\mathsf{Provability}$, $(\oplus)$-$\mathsf{Horn}$-$\mathsf{Provability}$ and $\mathsf{Horn}$-$\mathsf{Provability}$ are \NP-complete.
\end{theorem}

The last remaining fragment which we will consider is $(\oplus, \with)$-Horn. This fragment is \PSPACE-complete. While membership in \PSPACE follows directly from Proposition~\ref{completenesshorn}, we will show \PSPACE-hardness by embedding the pure implicative fragment of intuitionistic logic, whose \PSPACE-completeness was shown by~\textcite{STATMAN197967}, into the linear $(\oplus, \with)$-Horn fragment. The encoding consists of the following major steps:
\begin{enumerate}
    \item Interpret intuitionistic conjunctions as multiplicative conjunctions.
    \item Interpret intuitionistic Horn implications as linear Horn implications.
    \item Interpret embedded intuitionistic implications as linear $(\oplus)$-Horn implications.
\end{enumerate}

We will first define a special type of sequent, which can represent every intuitionistic implicative formula~\cite{10.1007/3-540-54415-1_67}, and has properties that help us embedding intuitionistic implicative formulas into the $(\oplus)$-Horn fragment.

\begin{definition}[Intuitionistic task sequent]\label{its}
    A sequent $W, \Gamma \vdash Z$ is an \define[sequent!intuitionistic task]{intuitionistic task sequent} if
    \begin{enumerate}
        \item $Z$ has no two different occurrences of one and the same literal.
        \item each formula of $\Gamma$ is either of the form
        \begin{enumerate}[label=(\alph*)]
            \item $(V \rightarrow Y)$, where $V$ is written without repetitions, or
            \item $((U\rightarrow V)\rightarrow Y)$, where $V$ is written without repetitions.\qedhere
        \end{enumerate}
    \end{enumerate}
\end{definition}

Restriction to sequents of the form defined above leads to a calculus with a quite manageable number of inference rules. Remember that in this context, for simple conjunctions $X$ and $Y$ representing multisets $L$ and $M$, $X \otimes Y$ is interpreted as the union of $L$ and $M$.

\begin{definition}[Calculus of intuitionistic task sequents without contraction]\label{taskseq}
    The calculus of intuitionistic task sequents without contraction consists of the following rules:

    \textbf{Axiom}
    \begin{displaymath}
        \prftree[r]{(id)}
            {}
            {X, \Gamma \vdash Z}
    \end{displaymath}
    \phantom{xxxxx}where $Z\subseteq X$.\\[0.5ex]

    \textbf{Logical rules}
        \begin{center}
            \begin{minipage}[b]{0.45\textwidth}
                \begin{displaymath}
                    \prftree[r]{(L$\otimes$)}
                        {X, \Gamma \vdash Z}
                        {Y, \Gamma \vdash Z}
                \end{displaymath}
            \end{minipage}
            \begin{minipage}[b]{0.45\textwidth}
                \begin{displaymath}
                    \prftree[r]{(L$\rightarrow$)}
                        {(X\otimes V\otimes Y), \Gamma \vdash Z}
                        {(X\otimes V), (V\rightarrow Y), \Gamma\vdash Z}
                \end{displaymath}
            \end{minipage}
        \end{center}
        \begin{displaymath}
            \prftree[r]{(L$\rightarrow\rightarrow$)}
                {(X\otimes U), \Gamma \vdash V}
                {(X\otimes Y), \Gamma\vdash Z}
                {X, ((U\rightarrow V)\rightarrow Y), \Gamma\vdash Z}
        \end{displaymath}
        \phantom{xxxxx}where in L$\otimes$, we have $X = Y$.
\end{definition}

\textcite{10.1007/3-540-54415-1_67} shows that a task sequent $W, \Gamma\vdash Z$ is valid in intuitionistic logic if and only if it is derivable in the calculus. For an arbitrary intuitionistic task sequent, we assume that the embedded implications in $\Gamma$ are enumerated from 1 to $k$. To give an embedding into $(\oplus)$-Horn with weakening which can be embedded into $(\oplus)$-Horn via Corollary~\ref{hornweak}, we introduce new literals $t, r_0, r_1,\dots, r_k$ in the following manner.

\begin{definition}
    For each $A\in \Gamma$, a multiset $A^\oplus$ is defined as
    \begin{enumerate}
        \item $(V\rightarrow Y)^\oplus$ is the multiset consisting of $k+1$ Horn implications of the form
        \[((r_i \otimes V)\multimap (r_i \otimes V\otimes Y)) \text{ for }0\leq i\leq k.\]
        \item $((U_j\rightarrow Z_j)\rightarrow Y_j)^\oplus$ is the multiset consisting of the Horn implication
        \[((r_j \otimes Z_j) \multimap (t\otimes Z))\]
        as well as $k+1$ $(\oplus)$-Horn implications of the form
        \[(r_i \multimap ((r_j \otimes U_j)\oplus (r_i\otimes Y_j)))\text{ for }0\leq i\leq k.\]
    \end{enumerate}
    We denote the result of replacing every formula $A$ in $\Gamma$ by $A^\oplus$ by $\Gamma^\oplus$.
\end{definition}

\begin{proposition}
    The task sequent $W, \Gamma \vdash Z$ is valid in intuitionistic logic if and only if the sequent
    \[(r_0\otimes W), ((r_0\otimes Z)\multimap (t\otimes Z_0)), \Gamma^\oplus \vdash (t\otimes Z)\]
    is derivable in the fragment of linear $(\oplus)$-Horn sequents.
\end{proposition}

\begin{proof}
    We leave the proof to the reader. The main idea is to simulate the only branching rule in the calculus of intuitionistic task sequents by a rule of the following form.
    \begin{displaymath}
        \prftree[r]{}
            {(r_j\otimes U_j), \Delta\vdash Z}
            {(r_i \otimes Y_j), \Delta\vdash Z}
            {r_i, (r_i\multimap ((r_j\otimes U_j)\oplus (r_i\otimes Y_j))), \Delta \vdash Z}
    \end{displaymath}
    The simulation of the other rules is straightforward.
\end{proof}

When we compose the result given in Corollary~\ref{hornweak} with this result, we have an embedding of the pure implicative fragment of intuitionistic logic into the linear $(\oplus, \with)$-Horn fragment. We, therefore, arrive at the following result.

\begin{theorem}[Complexity of $(\oplus, \with)$-\textsf{Horn}]
    $(\oplus, \with)$-$\mathsf{Horn}$-$\mathsf{Provability}$ is \PSPACE-complete.
\end{theorem}

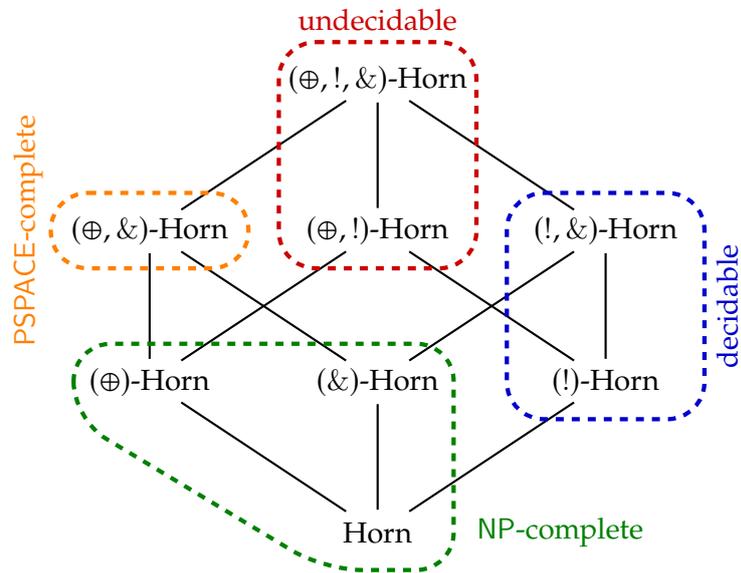
\begin{figure}[t]
    \centering
    \begin{tikzpicture}
        \node at (3,0) [] {Horn};
    
        \node at (0,2) [] {$(\oplus)$-Horn};
        \node at (3,2) [] {$(\with)$-Horn};
        \node at (6,2) [] {$(!)$-Horn};
    
        \node at (0,4) [] {$(\oplus, \with)$-Horn};
        \node at (3,4) [] {$(\oplus, !)$-Horn};
        \node at (6,4) [] {$(!, \with)$-Horn};

        \node at (3,6) [] {$(\oplus, !, \with)$-Horn};

        \draw [thick, shorten >= 5mm, shorten <= 5mm] (3,0)  -- (0,2);
        \draw [thick, shorten >= 3mm, shorten <= 3mm] (3,0)  -- (3,2);
        \draw [thick, shorten >= 5mm, shorten <= 5mm] (3,0)  -- (6,2);
            
        \draw [thick, shorten >= 3mm, shorten <= 3mm] (0,2)  -- (0,4);
        \draw [thick, shorten >= 5mm, shorten <= 5mm] (0,2)  -- (3,4);
            
        \draw [thick, shorten >= 5mm, shorten <= 5mm] (3,2)  -- (0,4);
        \draw [thick, shorten >= 5mm, shorten <= 5mm] (3,2)  -- (6,4);
            
        \draw [thick, shorten >= 5mm, shorten <= 5mm] (6,2)  -- (3,4);
        \draw [thick, shorten >= 3mm, shorten <= 3mm] (6,2)  -- (6,4);

        \draw [thick, shorten >= 5mm, shorten <= 5mm] (3,6)  -- (0,4);
        \draw [thick, shorten >= 3mm, shorten <= 3mm] (3,6)  -- (3,4);
        \draw [thick, shorten >= 5mm, shorten <= 5mm] (3,6)  -- (6,4);

        \draw  [ultra thick, dashed, green!50!black, rounded  corners =14pt] (-1, 1.5) -- (-1,2.5) -- (4,2.5) -- (4,-0.5) -- (2.3,-0.5) -- cycle;

        \draw  [ultra thick, dashed, orange, rounded  corners =14pt] (-1.3,4.5) -- (1.3,4.5) -- (1.3,3.5)  -- (-1.3,3.5) -- cycle;

        \draw  [ultra thick, dashed, red!80!black, rounded  corners =14pt] (1.7,6.5) -- (4.3,6.5) -- (4.3,3.5)  -- (1.7,3.5) -- cycle;

        \draw  [ultra thick, dashed, blue!80!black, rounded  corners =14pt] (4.7,4.5) -- (7.3,4.5) -- (7.3,1.5) -- (4.7,1.5) -- cycle;

        \node at (5.4,0) [green!50!black] {\class{NP}-complete};
        \node at (-1.6,4) [orange, rotate=90] {\class{PSPACE}-complete};
        \node at (3,6.8) [red!80!black] {undecidable};
        \node at (7.6,3) [blue!80!black, rotate=90] {decidable};
    \end{tikzpicture}
    \caption{Lattice of linear Horn fragments with complexity classification. All bounds are established in~\textcite{K_92}.}
    \label{horncomp}
\end{figure}

In conclusion, we get the complexity classification of linear Horn fragments shown in Figure~\ref{horncomp}. The examination of the complexity properties of Horn fragments has brought forward some interesting and counterintuitive results: contrary to classical or intuitionistic logic, where there exists a gap between the complexities of the full fragments and the Horn fragments, at least for \fragment{MLL}, the complexity of the whole fragment is the same as even the simplest Horn fragment. For some further complexity bounds of fragments close to those we examined in this thesis, see also Table~\ref{appendix:1}.

\chapter{Complexity of \fragmenttxt{ELL} and Ideas for a Structural Approach}\label{ch:new}

In this chapter, we develop the complexity-theoretic treatment of linear logic further. This is done in two ways. The first, more ``incremental'' result is that we give a first complexity-theoretic treatment for \fragment{ELL}, a more exotic fragment of linear logic. We will establish that provability can be decided efficiently in this fragment. The second, ``deeper'' treatment we give is that we propose ideas for a more structured approach to the complexity characterization of the fragments. We will discuss the current state, the feasibility of such an approach, and the challenges that present themselves.

\section{Exponential Linear Logic is in \fragmenttxt{P}}

By examining the complexity of provability of certain linear logic fragments, we saw that there are still some open questions. In this section, we will provide a proof that answers the question of whether the provability problem for \fragment{ELL} is efficiently decidable by providing an algorithm that decides the problem in quadratic time. For this, we will make use of a lattice structure that the exponential modalities exhibit.

\subsection{The Lattice of Exponential Modalities}

Since we have not yet looked very closely at the exponential modalities, we will now establish some notions that enable us to deal with them in a formal manner. First, we lay down what constitutes an exponential modality. That this definition behaves well with linear logic can be directly inferred by the rules of the sequent calculus. The construction of the lattice is mathematical folklore, and can be found in the~\textcite{llwiki}, for example.

\begin{definition}[Exponential modality]\label{expmod}
    An \define{exponential modality} $\mu$ is an arbitrary (possibly empty) sequence of the two exponentials $!$ and $?$. We simply write $\mu A$ for the application of a modality $\mu$ to the formula $A$. We denote the empty exponential modality by $\varepsilon$.
\end{definition}

Next, we exhibit some structure in the modalities which will lead to the definition of the lattice. The best way to do this is to define a preorder relation on the modalities.

\begin{definition}[Preorder of exponential modalities]\label{preomod}
    We define the \define{preorder of exponential modalities} as
    \[\mu\lesssim \nu \coloneq \mu A\vdash\nu A \text{ for all formulas } A.\]
    It induces an equivalence relation
    \[\mu \sim \nu \coloneq \mu \lesssim \nu \text{ and } \nu\lesssim\mu.\qedhere\]
\end{definition}

When establishing the lattice, we will make extensive use of the following lemma, the proof of which can be easily derived by the inference rules of the sequent calculus. When viewing the proof of Proposition~\ref{thismakesitwork} as inductive, this lemma would establish the base cases.

\begin{lemma}\label{useful}
    For any formula $A$, we have
    \begin{enumerate}
        \item[\ding{192}] $!A\vdash{} A$ and $A\vdash{} ?A$,
        \item[\ding{193}] $!A\vdash{} !!A$ and $??A\vdash{} ?A$,
        \item[\ding{194}] $!A\vdash{} !?!A$ and $?!?A\vdash{} ?A$.
    \end{enumerate}
\end{lemma}

Another powerful property we have for the exponential modalities is that of functoriality. This lemma enables the ``induction step'' in the proof of the following proposition.

\begin{lemma}[Functoriality]
    If $A$ and $B$ are two formulas with $A\vdash B$ then, for any exponential modality $\mu$, we have $\mu A\vdash \mu B$.
\end{lemma}

With these two lemmas, we can prove the following proposition. It serves as the basis on which our decision algorithm is built. In the proof of this proposition, we will refer to the base cases established above by their encircled number.

\begin{proposition}\label{thismakesitwork}
    We can simplify any occurrences of consecutive $!$ and $?$ to a single connective, and any alternating sequence of length at least four can be simplified into a smaller one.
\end{proposition}

\begin{proof}
    We first prove the equivalence of modalities of consecutive symbols. We obtain $!!A\vdash{} !A$ by functoriality from $!A\vdash{} A$ \ding{192}, and have $!A\vdash{} !!A$ \ding{193}. Similarly, we obtain $?A\vdash{} ??A$ by functoriality from $A\vdash{} ?A$ \ding{192}, and have that $??A\vdash{} ?A$ \ding{193}.

    For the sequences of alternating connectives, we obtain $?!A\vdash{} ?!?!A$ from $!A\vdash{} !?!A$ \ding{194} through functoriality. We can also obtain $!?B\vdash{} !?!?B$ with $A \coloneq{} ?B$. Similarly, we obtain $!?!?A\vdash{} !?A$ from $?!?A\vdash{} ?A$ \ding{194} through functoriality and $?!?!B\vdash{} ?!B$ with $A \coloneq{} !B$.
\end{proof}

\begin{corollary}
    Every exponential modality is $\sim$-equivalent to one of the following exponential modalities: $\varepsilon, !, ?, !?, ?!, !?!, ?!?$.
\end{corollary}

To further characterize the order relation of the lattice, we observe that some order relations are not possible. They are listed in the following lemma.

\begin{lemma}\label{modarelatt}
    For atomic formulas $A$, we have $?A\not\vdash{} A$ and $A\not\vdash{}?!?A$.
\end{lemma}

The lattice of exponential modalities is now a simple observation that uses the results gained above. It provides us with the necessary structure to efficiently decide provability of \fragment{ELL}-sequents.

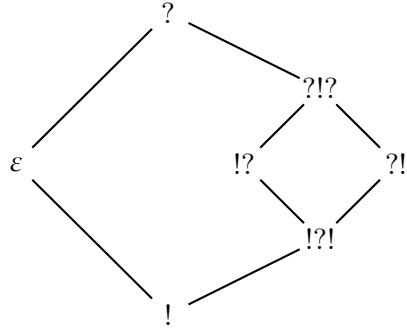
\begin{figure}[t]
    \centering
    \begin{tikzpicture}
        \node at (0,0) [] {$!$};
    
        \node at (2,1) [] {$!?!$};

        \node at (-2,2) [] {$\varepsilon$};
        \node at (1,2) [] {$!?$};
        \node at (3,2) [] {$?!$};

        \node at (2,3) [] {$?!?$};

        \node at (0,4) [] {$?$};

        \draw [thick, shorten >= 3mm, shorten <= 3mm] (0,0)  -- (-2,2);
        \draw [thick, shorten >= 3mm, shorten <= 3mm] (0,0)  -- (2,1);
            
        \draw [thick, shorten >= 3mm, shorten <= 3mm] (-2,2)  -- (0,4);
            
        \draw [thick, shorten >= 3mm, shorten <= 3mm] (2,1)  -- (1,2);
        \draw [thick, shorten >= 3mm, shorten <= 3mm] (2,1)  -- (3,2);
            
        \draw [thick, shorten >= 3mm, shorten <= 3mm] (1,2)  -- (2,3);
        \draw [thick, shorten >= 3mm, shorten <= 3mm] (3,2)  -- (2,3);

        \draw [thick, shorten >= 3mm, shorten <= 3mm] (2,3)  -- (0,4);
    \end{tikzpicture}
    \caption{Lattice of exponential modalities.}
    \label{latexpmod}
\end{figure}

\begin{lemma}[Lattice of exponential modalities]
    The equivalence classes of $\sim$ together with their order relation induced by $\lesssim$ form the lattice depicted in Figure~\ref{latexpmod}.
\end{lemma}

\begin{proof}
    To see this, observe that we have already shown $!A\vdash A$ and $!A\vdash{} !?!A$, and can deduce $!?!A\vdash{} !?A$ by functoriality and $!?!B\vdash{}?!B$ when $A\coloneq{} ?!B$. Furthermore, since we have $A\vdash{} B$ if and only if $B^\bot \vdash A^\bot$, the other relations follow.

    Next we show that no other relations are possible. First, from Lemma~\ref{modarelatt} and $A\vdash{} ?A$ we get $?A\not\vdash{}?!?A$. Since we have $\mu \lesssim{} \varepsilon$ or $\mu \lesssim{} ?!?$ for $\mu\in\{\,!, !?!, !?, ?!\,\}$, the modality $?$ can not be smaller than any other modality. The lemma and formula also give us that $\varepsilon$ cannot be smaller than $!$, $!?$, $?!$, $!?!$ or (by duality) $!?!$. This means that $\varepsilon$ and $?!?$ are both only smaller than $?$. Next, since we have $?!A\not\vdash{}!?A$ and $!?A\not\vdash{}?!A$, we have $?!\not\lesssim{} !?$ and $!?\not\lesssim{} ?!$. Functoriality gives us that $!?!\not\lesssim{} !$, so $!$ is the smallest element.
\end{proof}

\subsection{An Efficient Algorithm for \fragmenttxt{ELL}-\fragmenttxt{Provability}}

With the help of this lattice, the deterministic algorithm which decides the provability of \fragment{ELL} in polynomial time can be constructed by first reducing arbitrary modalities to their equivalent element in the lattice. New formulas can only be introduced via the identity axiom, which generates a dual to every formula in the sequent. A sequent is thus derivable in \fragment{ELL} if every formula in it occurs together with its dual. We can check this efficiently for sequents that have only formulas with modalities from the lattice. The algorithm which does this is given by Algorithm~\ref{ell}.

\begin{algorithm}\caption{Deterministic algorithm for \fragment{ELL}\textsf{-Provability}}\label{ell}
    \begin{algorithmic}[1]
        \Input{$\Gamma = \{\,\mu_1A_1, \mu_2A_2, \dots, \mu_nA_n\,\}$}
        \Output{Is $\Gamma$ provable in \fragment{ELL}?}

        \ForAll{$\mu_iA_i$, $1\leq i\leq n$}
            \Comment{Simplify the modalities}
            \While{$\mu_i\not\in\{\,\varepsilon, !, ?, !?, ?!, !?!, ?!?\,\}$}
                \State Simplify consecutive modalities
                \State Reduce every sequence of connectives with length $4$ according to Proposition~\ref{thismakesitwork}
            \EndWhile
        \EndFor
        \While{$\exists \mu_iA_i, \mu_jA_j \in\Gamma{}\ (\mu_iA_i = (\mu_jA_j)^\bot)$}
            \Comment{Delete dual formulas pairwise}
            \State Delete $\mu_iA_i$ and $\mu_jA_j$ from $\Gamma$
        \EndWhile
        \If{$\Gamma = \emptyset$}
            \State \Accept
        \Else
            \State \Reject
        \EndIf
    \end{algorithmic}
\end{algorithm}

When we take a closer look at the complexity of the algorithm, we see that for each modality in the list, the number of iterations of the while-loop reaching from line 2 to line 5 is bounded quadratically in the length of the modality. The complexity of pairwise deletion is also trivially bounded quadratically w.\,r.\,t. the input length. This gives us a runtime of $\mathcal{O}(n^2)$ for the algorithm. This trivially implies the following theorem.

\begin{theorem}[Complexity of \fragment{ELL}]
    $\fragment{ELL}$-$\mathsf{Provability}\in \class{P}$.
\end{theorem}

\section{Towards a Unified Lattice of Linear Logic Fragments}

Until now, we presented and extended the complexity-theoretic characterization of various fragments of linear logic. Each of these characterizations had a different underlying approach. We saw reductions from and to various machine models, other logics, and the exploitation of proof-theoretic properties of the various fragments. In the current state, presenting the characterization is very involved\footnote{In time and pages of this thesis.}. One approach to contain the complexity of the presentation is to follow a more structural approach. For this, we propose a lattice of fragments of linear logic, inspired by Post's lattice for propositional logic~\cite{10.2307/j.ctt1bgzb1r}.

Although first steps in this direction were already made by giving lattice presentations of the syntactic and Horn fragments, providing a unified view would be a huge undertaking, since it would rely on semantic, recursion-theoretic, and complexity-theoretic results not yet established for linear logic. We nevertheless believe that the benefits of such a presentation would be worth it because it would not only enable the establishment of dichotomy results like the ones given by~\textcite{DBLP:journals/mst/Lewis79},~\textcite{10.1145/800133.804350}, or, more recently, the conjecture by Feder and Vardi proven by~\textcite{10.1145/3292048.3292050}, but also give more insight in the relation of models that the various fragments can define.

\subsection{A Candidate for a Lattice}

We will now lay some groundwork to define such a unified lattice. First, we note that we can infer from the categorical semantics we have given for linear logic that the syntactic fragments admit distinct models. Next, we observe that the Horn, $(!, \with)$-Horn, $(\oplus, \with)$-Horn, and $(\oplus, !, \with)$-Horn fragments are restrictions of the \fragment{MLL}, \fragment{MELL}, \fragment{MALL}, and \fragment{LL} fragment, respectively. An interesting fact is that the ``undecidability barrier'' lays not just between the syntactical fragments, but between the Horn fragments as well.

The lattice we have constructed so far is not very symmetric, since the presence of the linear implication demands the presence of multiplicative connectives. To restore the symmetry, we define a dual set of fragments, the additive Horn fragments. Their underlying connective is the additive implication. We use Definition~\ref{addimpl}, keeping in mind that implications of the form $A\rightharpoonup B$ can be encoded as $(A\multimap 0) \oplus B$.

As we described above, additive linear implication has not achieved the same level of relevance as its multiplicative counterpart, due to the fact that it lacks a similar straightforward resource interpretation. From the additive implication, we can define the additive Horn fragments in an analogous way to the standard, multiplicative Horn fragments.

\begin{definition}[Generalized additive Horn sequents]
    The various variants of \define[Horn implication!generalized additive]{generalized additive Horn implications} are defined as follows:
    \begin{enumerate}
        \item An additive Horn implication is a formula of the form $(X\rightharpoonup Y)$,
        \item an additive $(\oplus)$-Horn implication is a formula of the form $(X\rightharpoonup (Y_1 \oplus Y_2))$,
        \item and an additive $(\with)$-Horn implication is a formula of the form $((X_1\rightharpoonup Y_1) \with (X_2\rightharpoonup Y_2))$.
    \end{enumerate}
    From these, \define[sequent!generalized additive Horn]{generalized additive Horn sequents} are defined by
    \begin{enumerate}
        \item For a multiset $\Gamma$ of additive Horn implications, a sequent of the form $W, \Gamma \vdash Z$ is called an additive Horn sequent, and a sequent of the form $W, !\Gamma \vdash Z$ is called an additive $!$-Horn sequent.
        \item Let $\diamondsuit \in \{\,\oplus, \with\,\}$. For a multiset $\Gamma$ of additive Horn and $(\diamondsuit)$-Horn implications, a sequent of the form $W, \Gamma\vdash Z$ is called an additive $(\diamondsuit)$-Horn sequent, and a sequent of the form $W, !\Gamma \vdash Z$ is called an additive $(!, \diamondsuit)$-Horn sequent.
        \item For a multiset $\Gamma$ of generalized additive Horn implications, a sequent of the form $W, \Gamma\vdash Z$ is called an additive $(\oplus, \with)$-Horn sequent.\qedhere
    \end{enumerate}
\end{definition}

Similar to the multiplicative case, we note that the additive Horn, additive $(\oplus, !)$-Horn, additive $(\oplus, \with)$-Horn, and additive $(\oplus, !, \with)$-Horn fragments are syntactical restrictions of \fragment{ALL}, \fragment{AELL}, \fragment{MALL}, and \fragment{LL} respectively. Unlike their multiplicative counterpart, a unified machine model is not yet known. Nevertheless, due to the simple encoding of additive implication into multiplicative implication, the complexity of the fragments should not differ much from their multiplicative counterparts.

\begin{figure}[t]
    \centering
    \includegraphics[width=\textwidth]{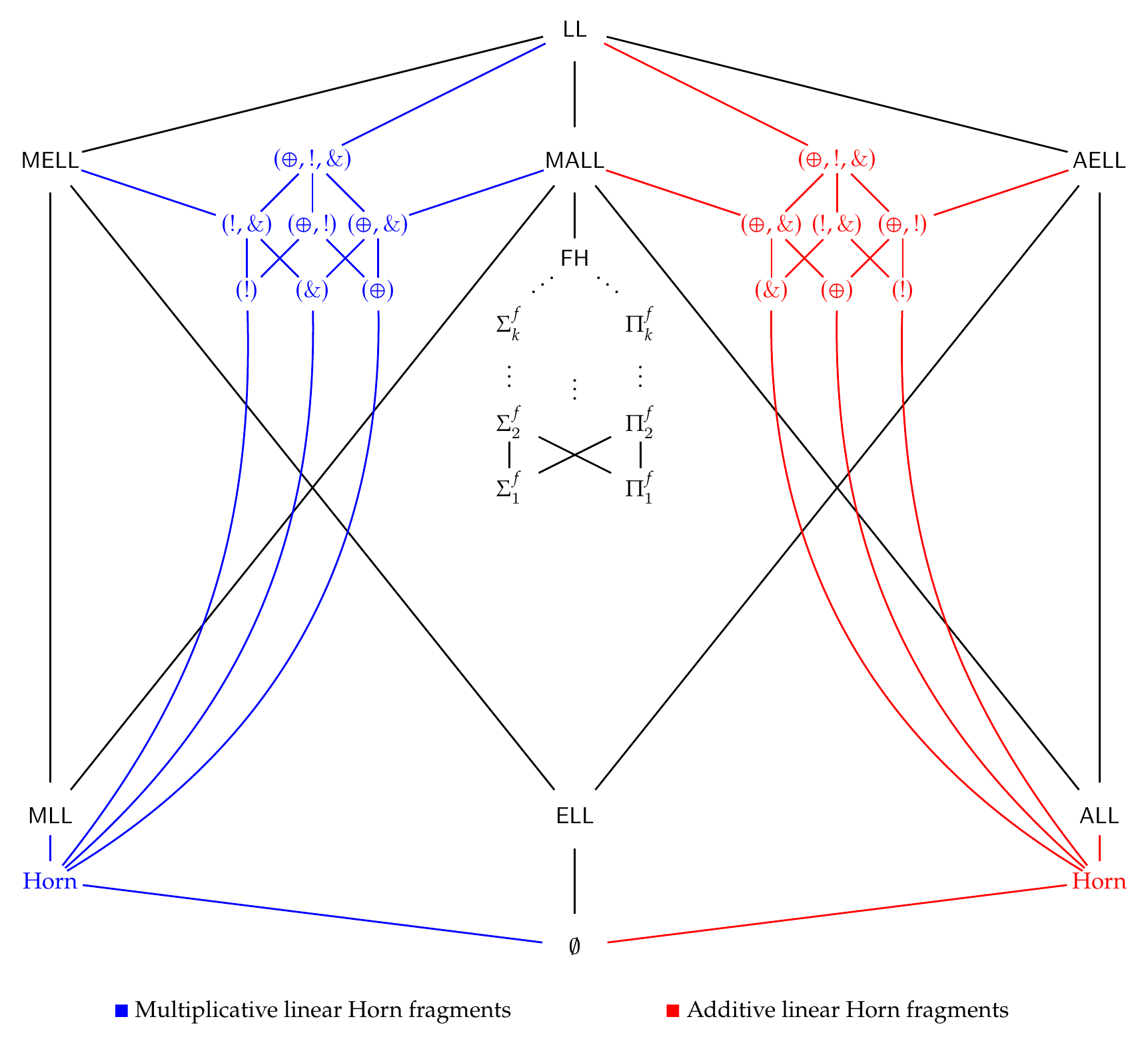}
    \caption{The lattice that encompasses the currently known relationships between the fragments we considered in this thesis.}
    \label{unilattice}
\end{figure}

We now also include the focussing hierarchy in the unified lattice, although due to the recency of its definition, there are no established lower fragments for the first levels of the hierarchy. The lattice that we have constructed is presented in Figure~\ref{unilattice}.

\subsection{Candidates for a Unified Semantics on the Lattice}

In this section, we shortly reiterate the different approaches to giving a semantics to linear logic, evaluating their fitness to be a candidate for a modular unified semantics for linear logic.

We start with the approaches that are likely not a good fit for an unified semantics: the phase semantics is quite simple to understand, but lacks modularity since the model of a phase space already incorporates all the connectives given by the full fragment. The game semantics incorporates the units on a very foundational level, and we would need to define new game rules for every fragment of linear logic, which would most likely lead to confusion. Sup lattices, vector spaces, and \reftodef[coherence space]{coherence spaces} are all incarnations of models for specific fragments of linear logic and do not generalize well to other fragments. This is also the case for Horn programs, which describe the Horn fragments very well, but it is not clear how to find computational models for other fragments, where the ``input to output'' relation is not quite as clear.

The most promising candidates seem to be the approach through Kripke semantics and categorical logic. Both feature a highly modular definition so that they can be made to correspond to different fragments. For Kripke semantics, new developments such as a canonical generalization of Kripke frames for substructural logics~\cite{COUMANS201450} could lead to a unified semantics that could encompass the main fragments. The categorical models are currently the approach that is best explored and that shows the most promise. There are already corresponding models for the syntactic fragments, and for the Horn fragments, it is feasible that new classes of categories could be defined, given the connection of category theory to programming language theory.

With its recent reformulation, another candidate that could provide an algebraic semantics for the whole lattice of linear logic is Geometry of Interaction. This approach, however, still being in its infancy itself, would need to be developed quite a lot further until a qualitative estimation in this regard can be made.
\chapter{Conclusion\strut }\label{ch:conclude}

In this chapter, we will review what we learned about linear logic and the complexity of provability in its fragments. We discuss the differences to classical logic and the insights the analysis of the logic provided. Last, we will give an outlook on the questions which are still open and the further research that can be pursued.

\section{Discussion}

In this thesis, we examined approaches to characterizing the complexity of deciding the provability problem of various fragments of linear logic. To do so, we first presented linear logic with its syntax and semantics, then conveyed the current state of research regarding the complexity of various fragments, and finally provided a new complexity characterization as well as ideas for a more structural approach to the complexity analysis of the provability problem in linear logic.

In the first part of the thesis, we saw that with a supposedly simple change to the sequent calculus of classical logic, the consideration of semantics becomes a non-trivial process. We saw that even to this day, many semantics are not fully fleshed out, although we made considerable progress throughout the years. We found linear logic to be applicable to many areas of mathematics and computer science, supplying a logic that admits as models widely used structures such as vector spaces or monoidal categories. Especially for computer science, the idea to provide implicit complexity bounds by constructing calculi that restrict the classes of functions they can represent is appealing and could provide, like descriptive complexity, a more structural approach to complexity theory. From a proof-theoretic standpoint, the categorical semantics is not only one of the most researched approaches for a semantics for linear logic, but linear logic is also one of the logics where categorical models are applied to the greatest success, which also helps to gain new insights in categorical logic. The practical use of linear logic will also be tested with the advent of different type systems which employ linear logic, tracking ownership and resources, or enforcing physical constraints of quantum mechanics through their formalism.

We found linear logic to be applicable to many areas of mathematics and computer science, supplying a logic that admits as models widely used structures such as vector spaces or monoidal categories. Especially for computer science, the idea to provide implicit complexity bounds by constructing calculi that restrict the classes of functions they can represent is appealing and could provide, like descriptive complexity, a more structural approach to complexity theory. From a proof-theoretic standpoint, the categorical semantics is not only one of the most researched approaches for a semantics for linear logic, but linear logic is also one of the logics where categorical models are applied to the greatest success, which also helps to gain new insights in categorical logic. The practical use of linear logic will also be tested with the advent of different type systems which employ linear logic, tracking ownership and resources, or enforcing physical constraints of quantum mechanics through their formalism.

When examining the complexity of the fragments of linear logic in the second part of the thesis, we saw that many different approaches are used, which are also focussed mainly on exploiting the syntactic properties of the fragments. The examination also showed that, compared to propositional logic, we gain a large increase in the complexity of the provability problem. But this is also accompanied by an increase in the expressibility of the logic so that many problems can be encoded as linear logic formulas. Even if the complexity characterizations put some fragments out of reach for practical use, they sometimes lead to interesting results for theory itself. For example, the problem of provability for \fragment{MLL}$_1$ stays \class{NP}-complete, and for \fragment{MALL}$_1$ it becomes \fragment{NEXP}-complete, but staying decidable. This shows that adding the quantifiers to classical propositional logic is not the sole source of the complexity increase, but rather the interaction of the quantifiers and the weakening and contraction rule. On the other hand, provability for both \fragment{MLL}$_2$ and \fragment{MALL}$_2$ is undecidable, see also Table~\ref{appendix:1}.

Lastly, we proved an original complexity result for \fragment{ELL} and gave ideas for a new approach to the complexity-theoretic examination of linear logic. The former provides a nice extension of the knowledge we have already gained for a more exotic fragment, exploiting the lattice of exponential modalities. The latter could be a starting point for deeper investigations to find a more structural representation of the fragments. While inspired by Post's lattice, many more semantical properties can not be taken over to linear logic, simply because linear logic is not necessarily two-valued. Without a doubt, other properties of linear logic formulas, besides being in Horn form or in a level of the focussing hierarchy, exist, and they can be used to extend the lattice in various directions.

\section{Future Work}

New research directions regarding linear logic and its complexity arise at every corner. We will give a short overview of problems which are of interest and some further ideas.

First of all, the expansion of semantics under all viewpoints given in this thesis is a vital step to gaining more insight into the inner workings of the logic. First and foremost, further research into Geometry of Interaction could bring forward new results in which linear logic is directly involved. New results regarding the finite model theory of linear logic could also help with open research questions about its complexity.

One of the most important open questions regarding the complexity of linear logic fragments is the decidability of \fragment{MELL} (cf. also~\cite{10.5555/212876.212887}). It is inter-reducible to decision problems for Petri nets and certain classes of counter machines~\cite{dG_04, LS_15}, and results regarding its complexity would also be beneficial for our understanding of these problems. There is also to this date no complexity bound on the \fragment{AELL} fragment whatsoever, and even results regarding its semantics are hard to come by, likely due to its more exotic behavior.

These are the most prominent research directions that are open right now, but we will give some further ideas that arose while we examined the problem of provability. One idea is the extension of the lattice we have defined in various directions. Of course, we can differentiate between the fragments at each level of higher-order logic. A number of hierarchies arise in this way, and the question is in which way they relate to other hierarchies of hypercomputation such as the arithmetic or analytic hierarchy. In the other direction, there are also open questions: can we define linear logic fragments which are weak enough to represent efficient complexity classes and hierarchies? In this thesis, the smallest class we considered was \class{P}, but it is possible to find even tighter bounds for \fragment{ALL} and \fragment{ELL}, and maybe even pendants to the circuit hierarchies $\fragment{AC}^i$, $\fragment{NC}^i$, and $\fragment{TC}^i$.

The idea of focussing plays a central part in the formulation of efficient proof search algorithms in linear logic and its derivatives. Closer examination of the base classes of the focussing hierarchy, that is, formulas which have proofs which are only nondeterministic or co-nondeterministic, could lead to new algorithms which put the provability problem for these classes into reach for efficient solving. The idea of the focussing hierarchy can also be adapted to the other syntactical fragments, yielding hierarchies which elements are in general less or more complex than their counterparts in \class{FH}.

The complexity characterizations we have presented so far can also all be considered as ``classical''. But advances in complexity theory also lead to many new techniques and viewpoints which have not yet been applied to linear logic. A small selection of new techniques includes methods from parameterized\footnote{Interestingly enough, a first result using parameterization was already given in 1997 by~\textcite{DBLP:conf/lfcs/Dudakov97}, even before parameterized complexity gained widespread popularity. He examined the concurrency complexity of the Horn fragment, and found parameterizations which put the respective problem in \class{FPT}, in todays terms.}, randomized, enumeration, incremental, as well as fine grained complexity.

Another door that is opened by the examination of linear logic is the complexity-theoretic treatments of various mathematical topics which have not yet experienced much of this. This could, for example, be research regarding category-theoretic or algebraic operations.

In conclusion, in this thesis we have answered one question about the complexity of linear logic, and asked countably many more. The options for further research are thus, quite literally, endless.

\appendix
\chapter{Appendix}\label{ch:appendix}

\section{Complexity Overview}

\begin{table}[ht!]
    \centering
    \caption{An overview of the various fragments along with their best known complexity bounds.}\label{appendix:1}
    \begin{tabular}{l l l l}
        \toprule
        \textbf{Fragment} & \textbf{Abbreviation} & \textbf{Complexity} & \textbf{Source}\\ \hline
        Linear Logic & \fragment{LL} & undecidable & \cite{LMSS_92}\\
        Multiplicative Linear Logic & \fragment{MLL} & \class{NP}-complete & \cite{K_92}\\
        Multiplicative Additive Linear Logic & \fragment{MALL} & \class{PSPACE}-complete & \cite{LMSS_92}\\
        Multiplicative Exponential Linear Logic & \fragment{MELL} & \class{TOWER}-hard & \cite{LS_15}\\ \hline
        Horn Linear Logic & & \class{NP}-complete & \cite{K_92}\\
        $(\oplus)$-Horn Linear Logic & & \class{NP}-complete & \cite{K_92}\\
        $(\with)$-Horn Linear Logic & & \class{NP}-complete & \cite{K_92}\\
        $(!)$-Horn Linear Logic & & decidable & \cite{K_92}\\
        $(\oplus, \with)$-Horn Linear Logic & & \class{PSPACE}-complete & \cite{K_92}\\
        $(\oplus, !)$-Horn Linear Logic & & undecidable & \cite{K_92}\\
        $(!, \with)$-Horn Linear Logic & & decidable & \cite{K_92}\\ \hline
        Constant Only \fragment{LL} & \fragment{COLL} & undecidable & \cite{K_17}\\
        Constant Only \fragment{MLL} & \fragment{COMLL} & \class{NP}-complete & \cite{DBLP:journals/tcs/LincolnW94}\\
        Constant Only \fragment{MALL} & \fragment{COMALL} & \class{PSPACE}-complete & \cite{K_17}\\ \hline
        Affine Linear Logic & \fragment{LLw} & \class{TOWER}-complete & \cite{LS_15}\\ 
        Affine \fragment{MELL} & \fragment{MELLw} & \class{TOWER}-hard & \cite{LS_15}\\ 
        Affine \fragment{MALL} & \fragment{MALLw} & \class{PSPACE}-complete & \cite{DBLP:journals/jar/Das20}\\ \hline
        Contractive Linear Logic & \fragment{LLc} & \class{ACKERMANN}-complete & \cite{LS_15}\\ 
        Contractive \fragment{MALL} & \fragment{MALLc} & \class{ACKERMANN}-hard & \cite{LS_15}\\ \hline
        Intuitionistic Linear Logic & \fragment{ILL} & undecidable & \cite{LMSS_92}\\ \hline
        First-Order Linear Logic & $\fragment{LL}_1$ & undecidable & \cite{Gir_87}\\ 
        First-Order \fragment{MALL} & $\fragment{MALL}_1$ & \class{NEXP}-complete & \cite{DBLP:conf/lics/LincolnS94}\\
        First-Order \fragment{MLL} & $\fragment{MLL}_1$ & \class{NP}-complete & \cite{DBLP:conf/lics/LincolnS94}\\
        Second-Order \fragment{IMLL} & $\fragment{ILL}_2$ & undecidable & \cite{DBLP:conf/lics/LincolnSS95}\\
        \bottomrule
    \end{tabular}
\end{table}

\section{Provable Formulas}

In this section, we list some interesting formulas which are provable in linear logic. Although we do not use them in the thesis, we also give rules for the quantifiers. The two tables are adapted from~\cite{llwiki}.

\subsection{Equivalences}

Two formulas $A$ and $B$ are linearly equivalent, written $A\multimapboth{} B$, if $A\multimap B$ and $B\multimap A$.

\begin{table}[ht!]
    \centering
    \caption{Important provable equivalences of linear logic.}\label{appendix:2}
    \begin{tabular}{l l}
        \toprule
        \textbf{Category} & \textbf{Formulas}\\
        \midrule
        Associativity & $A\otimes (B\otimes C) \multimapboth{} (A\otimes B)\otimes C$\phantom{xxx}$A\parr (B\parr C) \multimapboth{} (A\parr B) \parr C$\\
        & $A\oplus (B\oplus C)\multimapboth{} (A\oplus B)\oplus C$\phantom{xxx}$A\with (B\with C)\multimapboth{} (A\with B)\with C$\\\hline
        Commutativity & $A\otimes B\multimapboth{} B\otimes A$\phantom{xxx}$A\parr B\multimapboth{} B\parr A$\\
        & $A\oplus B\multimapboth{} B\oplus A$\phantom{xxx}$A\with B\multimapboth{} B\with A$\\\hline
        Neutrality & $A \otimes 1 \multimapboth{} A$\phantom{xxx}$A \parr \bot \multimapboth{} A$\\
        & $A \oplus 0 \multimapboth{} A$\phantom{xxx}$A \with \top \multimapboth{} A$\\\hline
        Idempotence of additives & $A\oplus A \multimapboth{} A$\\
        & $A\with A \multimapboth{} A$\\\hline
        Distributivity of multipli- & $A\otimes (B\oplus C)\multimapboth{} (A\otimes B)\oplus (A\otimes C)$\\
        catives over additives & $A\parr (B\with C)\multimapboth{} (A\parr B)\with (A\parr C)$\\
        & $A\otimes 0\multimapboth{} 0$\phantom{xxx}$A\parr \top \multimapboth{} \top$\\\hline
        Defining property of & $!(A\with B)\multimapboth{} !A\otimes{} !B$\phantom{xxx}$?(A\oplus B) \multimapboth{} ?A\parr{} ?B$\\
        exponentials & $!\top \multimapboth{} 1$\phantom{xxxxxxxxxxxxx}$?0\multimapboth{} \bot$\\\hline
        Monoidal structure of & $!A\otimes{} !A\multimapboth{} !A$\phantom{xxx}$?A\parr{} ?A\multimapboth{} ?A$\\
        exponentials & $!1 \multimapboth{} 1$\phantom{xxxxxxxxx}$?\bot \multimapboth{} \bot$\\\hline
        Digging & $!!A \multimapboth{} !A$\phantom{xxx}$??A\multimapboth{} ?A$\\\hline
        Other properties of & $!?!?A\multimapboth{} !?A$\phantom{xxx}$!?1\multimapboth{} 1$\\
        exponentials & $?!?!A\multimapboth{} ?!A$\phantom{xxx}$?!\bot\multimapboth{}\bot$\\\hline
        Commutation of quantifiers & $\exists\xi.\exists\psi.A \multimapboth \exists\psi.\exists\xi.A$\phantom{xxxxxxx}$\forall\xi.\forall\psi.A \multimapboth \forall\psi.\forall\xi.A$\\
        ($\zeta$ does not occur in $A$) & $\exists\xi.(A\oplus B)\multimapboth \exists\xi.A\oplus \exists\xi.B$\phantom{xx}$\forall\xi.(A\with B)\multimapboth \forall\xi.A\with \forall\xi.B$\\
        & $\exists\zeta.(A\otimes B)\multimapboth \exists\zeta.A\otimes \exists\zeta.B$\phantom{xx}$\forall\zeta.(A\parr B)\multimapboth \forall\zeta.A\parr \forall\zeta.B$\\
        & $\exists\zeta.A\multimapboth{} A$\phantom{xxxxxxxx\,xxxxxxxx}$\forall\zeta.A\multimapboth{} A$\\
        \bottomrule
    \end{tabular}
\end{table}

\clearpage
\subsection{Other}

\begin{table}[ht!]
    \centering
    \caption{Important provable formulas of linear logic.}\label{appendix:3}
    \begin{tabular}{l l}
        \toprule
        \textbf{Category} & \textbf{Formulas}\\
        \midrule
        Standard distributivities & $A \oplus (B\with C) \multimap (A\oplus B)\with (A\oplus C)$\\
        & $A\otimes (B\with C)\multimap (A\otimes B)\with (A\otimes C)$\\
        & $\exists \xi.(A\with B) \multimap (\exists \xi.A)\with (\exists.B)$\\\hline
        Linear distributivities & $A\otimes (B\parr C)\multimap (A\otimes B)\parr C$\\
        & $\exists \xi.(A\parr B)\multimap A\parr \exists \xi.B$ $(\xi\not\in A)$\\
        & $A\otimes \forall\xi.B \multimap \forall \xi .(A\otimes B)$ $(\xi\not\in A)$\\\hline
        Factorizations & $(A\with B)\oplus (A\with C)\multimap A\with (B\oplus C)$\\
        & $(A\parr B)\oplus (A\parr C) \multimap A\parr (B\oplus C)$\\
        & $(\forall \xi. A)\oplus (\forall\xi.B)\multimap \forall\xi.(A\oplus B)$\\\hline
        Identities & $1\multimap A^\bot \parr A$\\
        & $A\otimes A^\bot \multimap \bot$\\\hline
        Additive structure & $A\with B \multimap A$\phantom{xxx}$A\with B \multimap B$\\
        & $A \multimap A\oplus B$\phantom{xxx}$B \multimap A\oplus B$\\
        & $A \multimap \top$\phantom{xxxxxx\ \,}$0 \multimap A$\\
        \bottomrule
    \end{tabular}
\end{table}

\printindex

\printunsrtglossary[title={List of Symbols},type=symbols,style=long]

\cleardoublepage
\phantomsection\addcontentsline{toc}{chapter}{List of Definitions and Theorems}
\listoftheorems[title={List of Definitions and Theorems}, ignoreall, onlynamed={theorem,definition}]

\cleardoublepage
\phantomsection\addcontentsline{toc}{chapter}{List of Figures}
\listoffigures

\cleardoublepage
\phantomsection\addcontentsline{toc}{chapter}{List of Tables}
\listoftables

\cleardoublepage
\phantomsection\addcontentsline{toc}{chapter}{Bibliography}
\printbibliography

\cleardoublepage
\thispagestyle{empty}
\phantomsection\addcontentsline{toc}{chapter}{Erklärung der Selbstständigkeit}
\begin{center}
    {\huge \textbf{Erklärung der Selbstständigkeit}}
\end{center}
Hiermit versichere ich, dass ich die vorliegende Arbeit selbstständig verfasst und keine anderen als die angegebenen Quellen und Hilfsmittel benutzt habe, dass alle Stellen der Arbeit, die wörtlich oder sinngemäß aus anderen Quellen übernommen wurden, als solche kenntlich gemacht und dass die Arbeit in gleicher oder ähnlicher Form noch keiner Prüfungsbehörde vorgelegt wurde.

\vspace{3em}
\noindent Hannover, den 13.09.2021

\vspace{3em}
\noindent\rule{7cm}{0.4pt}

Florian Chudigiewitsch

\end{document}